%% file: DM.tex
\documentclass[twocolumn,floatfix,nofootinbib,british,balancelastpage]{revtex4-1}
\usepackage[pdftex]{graphicx}
\usepackage{rotate}
\usepackage{rotating}
\usepackage{relsize}
\usepackage{lineno}
\usepackage{slashed}
\usepackage{balance}
\usepackage{url}
\usepackage{amsmath,mathrsfs}
\usepackage{amssymb}
\usepackage{longtable}
\usepackage{multirow}
\usepackage{tabularx}
\usepackage{units}
\usepackage[usenames,dvipsnames]{xcolor}
\usepackage{subfigure}

\usepackage{hepunits}
\usepackage{hepnicenames}

\renewcommand{\GeV}{\text{Ge}\hspace{-0.05cm}\text{V}}

\begin{document}

\title{Illuminating Dark Matter at the ILC}%

\author{Herbert Dreiner }%
\email{dreiner@th.physik.uni-bonn.de}
\affiliation{Physikalisches
  Institut and Bethe
  Center for Theoretical Physics, University of Bonn, Bonn, Germany}
\author{Moritz Huck }%
\email{mhuck@physik.rwth-aachen.de}
\affiliation{Institute for
  Theoretical Particle Physics and Cosmology, RWTH Aachen University,
  Aachen, Germany}
\author{Michael Kr\"amer}%
\email{mkraemer@physik.rwth-aachen.de}
\affiliation{Institute for
  Theoretical Particle Physics and Cosmology, RWTH Aachen University,
  Aachen, Germany}
\author{Daniel Schmeier }%
\email{daschm@th.physik.uni-bonn.de}
\affiliation{{}Physikalisches
  Institut and Bethe
  Center for Theoretical Physics, University of Bonn, Bonn, Germany}
\author{Jamie Tattersall}%
\email{jamie@th.physik.uni-bonn.de}
\affiliation{{}Physikalisches
  Institut and Bethe
  Center for Theoretical Physics, University of Bonn, Bonn, Germany}

\begin{abstract}
  The \textsc{Wimp} (weakly interacting massive particle) paradigm for dark matter is currently being probed via many
  different experiments. Direct detection, indirect detection and
  collider searches are all hoping to catch a glimpse of these elusive
  particles. Here, we examine the potential of the \textsc{Ilc} (International Linear Collider) to
  shed light on the origin of dark matter. By using an effective field
  theory approach we are also able to compare the reach of the
  \textsc{Ilc} with that of the other searches. We find that for low
  mass dark matter ($<10$~GeV), the \textsc{Ilc} offers a unique
  opportunity to search for \textsc{Wimps} beyond any other experiment. In
  addition, if dark matter happens to only couple to leptons or via a
  spin dependent interaction, the \textsc{Ilc} can give an unrivalled
  window to these models. We improve on previous \textsc{Ilc} studies by
  constructing a comprehensive list of effective theories that allows
  us to move beyond the non-relativistic approximation.
\end{abstract}


\date{November 9th, 2012}%
\maketitle

\input{introduction}

\input{models}

\input{astrophysics}
\input{ilc}
\input{conclusions}

\section*{Acknowledgments}
\noindent 
We would especially like to thank Jenny List for many detailed
discussions regarding the \textsc{Ilc} analysis.  The work has been supported
by the Helmholtz Alliance ``Physics at the Terascale'', the \textsc{Dfg
Sfb/Tr9} ``Computational Particle Physics'', and the \textsc{Dfb Sfb/Tr33} (`The
Dark Universe'). H.K.D. would like to thank the Aspen Center for
Physics where part of this work was completed.

\onecolumngrid
\pagebreak
\twocolumngrid

\pagebreak
\appendix

\input{astrophysics_appendix.tex}

\input{appendix_ilc.tex}

\bibliographystyle{utphys}
\bibliography{DM}

\end{document}

%% file: introduction.tex
\section{Introduction}

Weakly interacting massive particles (\textsc{Wimp}s) are one of the leading candidates to solve the dark matter puzzle \cite{Bertone:2004pz}. Primarily this is due to the fact that a neutral particle that interacts with roughly the strength of the weak force, naturally gives the correct relic abundance. In addition many theoretical models predict that the masses of these 
states should exist around the scale of electroweak symmetry breaking, e.g.\ Supersymmetry (\textsc{Susy}) \cite{Martin:1997ns,Drees:2004jm}, Universal Extra Dimensions (\textsc{Ued}) \cite{Appelquist:2000nn}, Little Higgs \cite{ArkaniHamed:2001nc} etc.

Currently, this \textsc{Wimp} paradigm is being actively explored in a number of different ways. Perhaps the most well known 
are the direct detection searches that aim to observe interactions between the dark matter and an atomic nucleus \cite{Goodman:1984dc}. As these are extremely low rate experiments, the detectors are typically placed deep underground to reduce 
background. The annihilation of dark matter into Standard Model particles in high density regions of our universe offers another potential method to see a signal e.g.\ \cite{Bouquet:1989sr}.

In particle colliders here on Earth the same interactions may be probed in the production of dark matter.  Unfortunately, the fact that \textsc{Wimp}s are neutral and only weakly interacting means that they cannot be detected directly in these experiments. Therefore collider based searches must rely on particles produced in combination with the dark matter candidates. If dark matter is produced directly, one possibility is to use initial state radiation (\textsc{Isr}), such as gluon jets, or photons, that will recoil against the \textsc{Wimp}s.

This idea was first explored in a model independent approach for the International Linear Collider (\textsc{Ilc}) using mono-photons in a non-relativistic approximation \cite{Birkedal:2004xn,Konar:2009ae}. Later, detailed detector studies have been performed to understand the full capabilities of the \textsc{Ilc} for such a signature \cite{Bartels:2007cv,Bartels:2009fa,Bartels:2010qv,Bartels:2012ex,Bernal:2008zk}. Furthermore the same signature has been considered in the case of \textsc{Susy} \cite{Dreiner:2006sb,Dreiner:2007vm}. At the \textsc{Lhc} (Large Hadron Collider) and Tevatron similar signals have also been studied but with a mono-jet signal  \cite{Cao:2009uw,Bai:2010hh,Fox:2011pm,Goodman:2010ku,Goodman:2010yf,Beltran:2010ww,Rajaraman:2011wf,Bai:2012he,Cheung:2012gi}. All of these papers used the idea of parameterising the dark matter interactions in the form of effective operators. This has the advantage that the bounds can be compared with those coming from direct detection and also that a non-relativistic approximation is not required to compare with the relic density measurement. These methods have now been used by the \textsc{Lhc} experiments to set bounds on different effective operators that are competitive with other methods \cite{Chatrchyan:2012pa,ATLAS-CONF-2012-084}. In addition, \textsc{Lep} (Large Electron-Positron Collider) data has been re-interpreted to determine corresponding constraints \cite{Fox:2011fx}.

In this paper we take the effective field theory approach to dark matter and apply this to an \textsc{Ilc} search 
\cite{Kurylov:2003ra,Beltran:2008xg,Agrawal:2010fh}. To apply the effective field theory in a consistent way we 
assume that the dark matter particles can only interact with the Standard Model fields via a heavy mediator. 
The mediator is always assumed to be too heavy to be produced directly at the \textsc{Ilc} and thus can be 
integrated out. For our model choices we consider the possibility that the dark matter candidate could be a scalar, 
a Dirac (or Majorana) fermion or a vector particle. The same choices are taken for the heavy mediator 
and all combinations are considered. The collider phenomenology can vary significantly, depending on 
whether the mediator is exchanged in the $s$- or $t$-channel and consequently we examine both. In addition, we also study the different ways in which the mediator can couple to both the dark matter and Standard Model particles. We note that using the effective field theory approach allows us to move away from the non-relativistic approximation that had previously been used in \textsc{Ilc} studies. This can be especially important if the dark matter candidate happens to be light.\footnote{The mass determination of a light neutralino dark matter candidate at the \textsc{Ilc} has been discussed in Ref.~\cite{Conley:2010jk}.}

For all models we compare the reach of the \textsc{Ilc} with the bounds derived from direct and indirect detection. We also calculate the couplings expected to lead to the correct relic density and see whether the \textsc{Ilc} can probe these regions of parameter space. We also note that an \textsc{Ilc} search is complementary to that at the \textsc{Lhc} thanks to the different initial state.

The paper is laid out as follows. We begin in Sec.~\ref{sec:models} by explaining how we derive the effective field theories for the dark matter interactions and we explicitly give the Lagrangian for both the full and effective theory. We also describe the benchmark models that we use throughout the study. In Sec.~\ref{sec:astro} we describe the various astrophysical constraints on our effective theories. We begin with the calculation of the relic density abundance before moving on to explain the bounds from direct and indirect detection.

Section~\ref{sec:ilc} describes in detail the potential search for dark matter at the \textsc{Ilc}. Here we explain the calculation of the signal rate and the dominant backgrounds that were considered. In addition we detail how the \textsc{Ilc} detectors are modeled to account for relevant experimental effects. We find that the polarisation of incoming beams is particularly important for many models of dark matter to discriminate the signal and background. We also investigate the advantage of a doubling of the \textsc{Ilc} energy to $\sqrt{s}=1$~TeV. 

In Sec.~\ref{sec:results} we present the results of the paper. We begin by examining the potential bounds of the \textsc{Ilc} on the effective coupling of the dark matter model at the collider. Afterwards, we combine these results with those from direct and indirect detection to understand for which models and mass ranges the \textsc{Ilc} presents a unique opportunity to discover dark matter. Finally in Sec.~\ref{sec:conclusions} we conclude and summarise the main results of our work.

%% file: models.tex
\section{Models}
\label{sec:models}

\subsection{General Motivation}
\noindent
The idea of parametrising the interaction of a dark matter particle
with Standard Model particles by using effective operators is not new,
see for example Refs.~\cite{Agrawal:2010fh, Beltran:2008xg,
  Zheng:2010js, Yu:2011by, Fox:2011fx, Bai:2010hh}.  Many authors
construct a list of effective 4--particle-interactions with Lorentz--invariant
combinations of $\gamma^\mu$, $\partial_\mu$ and
spinor--/vector--indices up to mass dimension 5 or 6. In many cases
there is no explanation how those operators may arise in an underlying
fundamental theory. That makes it difficult to judge how exhaustive
the lists of operators are, whether interference between different
operators should be taken into account and how the effective model is
connected to realistic fundamental theories and their couplings.

We follow the effective approach introduced in \cite{Agrawal:2010fh} by
starting from different fundamental theories with given renormalisable
interactions between Standard Model fermions and the
  hypothesized dark matter particles that are mediated by a very massive
particle. From these theories we deduce effective 4--particle--vertices for
energies significantly smaller than the mass of the mediator. Working
with these effective operators, one can deduce information about the
effective coupling and propagate this information to the parameters of
the corresponding underlying fundamental theory. The effective 
approach allows us to reduce the dimensionality of the parameter space
and more easily compare the different experimental searches.

\subsection{Deriving Effective Lagrangians}
We start with a list of fundamental Lagrangians taken from
\cite{Agrawal:2010fh}. However we do not perform a non--relativistic
approximation, since we are interested in the phenomenology of
  this Lagrangian at a high energy
experiment and therefore the results for our effective operators
differ. We also use a different method to evaluate the effective
vertices, motivated in Ref.~\cite{Haba:2011vi}, which uses the path
integral formalism.

We give one explicit example for the derivation of the effective
operators and only mention specific peculiarities for the other cases,
which are apart from that calculated similarly. Let $\psi$ be a
Standard Model fermion and $\chi$ a complex scalar field representing
the dark matter candidate. For our example, we assume the mediator to
be a real scalar field, $\phi$, with mass $M_\Omega$ (we will keep
this notation for the mediator mass throughout). The relevant terms in
the UV completed Lagrangian are then given by,
\begin{align}
\mathscr{L_{\text{UV}}} &= \frac{1}{2} \left[\partial_\mu \phi(x) \right]^2 -
\frac{1}{2} M_{\Omega}^2 \phi^2(x) - g_\chi \chi^\dagger(x) \chi(x) \phi(x)  \nonumber \\
& \quad - \bar{\psi}(x) \left( g_s +
 i g_p \gamma^5 \right) \psi(x) \phi(x)\;, \\
&\equiv - \frac{1}{2} \phi(x) \square_x \phi(x) - \frac{1}{2}
M_{\Omega}^2 \phi^2(x) - F(x) \phi(x)\;. 
\end{align}
where the function $F(x)$ is given by,
\begin{align}
F(x) & \equiv g_\chi \chi^\dagger(x) \chi(x) + \bar{\psi}(x) \left( g_s + i
  g_p \gamma^5 \right) \psi(x) \,.
\end{align}
We have not included the kinetic terms for $\chi,\,\psi$, as they are
not relevant for the computation of the effective Lagrangian. In this
particular example, $g_s$, $g_p$ are dimensionless couplings and
$g_\chi$ is a dimension one parameter but these definitions can change
depending upon the precise model studied and we shall use this notation throughout. 
We have included the kinetic term
for $\phi$, the heavy mediator field. After integrating out $\phi$, we
obtain the effective Lagrangian,
\begin{align}
\mathscr{L}_\text{eff} &= \frac{1}{2 M_\Omega^2} F^2 \supset \frac{g_\chi}{M_\Omega^2} \chi^\dagger
 \chi \bar{\psi} \left(g_s + i g_p \gamma^5 \right) \psi \;.
\end{align}
\begin{table*}
\centering
\renewcommand{\arraystretch}{1.525}
\begin{tabular}{c c c l}
\hline
\multirow{2}{*}{DM} & \multirow{2}{*}{Med.} & \multirow{2}{*}{Diagram} &
$-\mathscr{L}_{\text{UV}}$ \\
& & & $-\mathscr{L}_{\text{eff}}$ \\
\hline \hline
\multirow{2}{*}{S} & \multirow{2}{*}{S} & \multirow{2}{*}{\raisebox{-0\height}{\includegraphics[width=0.1\textwidth]{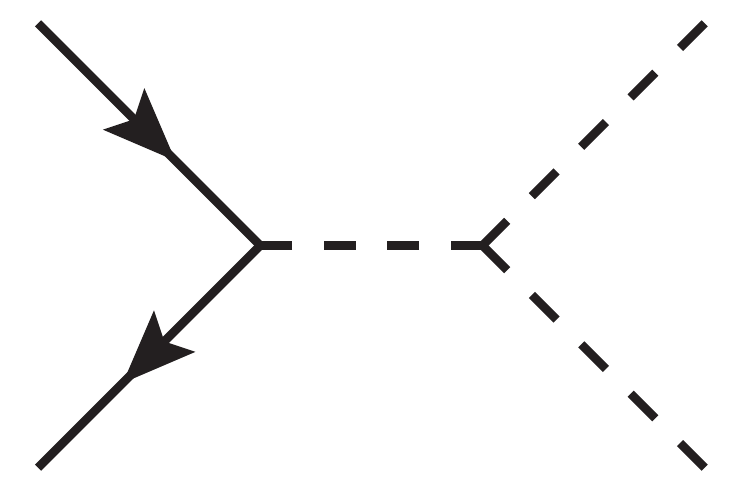}}} & $g_\chi \chi^\dagger \chi \phi + \bar{\psi} (g_s + i g_p \gamma^5) \psi \phi$ \\
& & &  $\displaystyle \frac{g_\chi}{M_\Omega^2} \chi^\dagger \chi \bar{\psi} (g_s + i g_p \gamma^5) \psi$ \\
\hline
\multirow{2}{*}{S}       & \multirow{2}{*}{F} &
       \multirow{2}{*}{\raisebox{-0\height}{\includegraphics[width=0.1\textwidth]{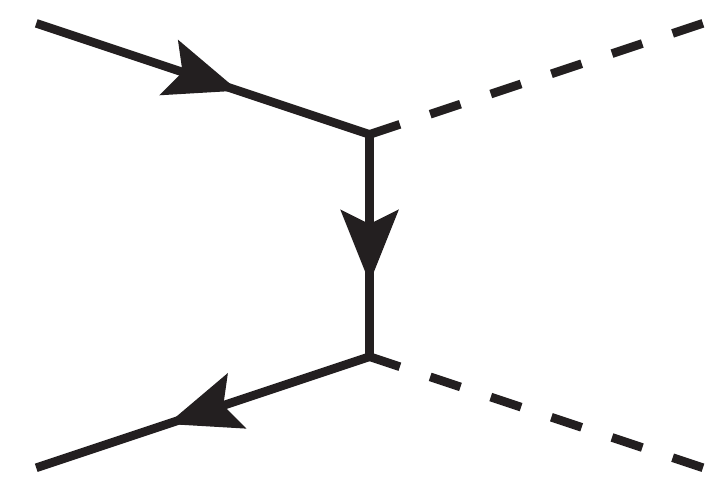}}}&
       $ \bar{\eta} (g_s + g_p \gamma^5 ) \psi \chi +   \bar{\psi} (g_s - g_p
       \gamma^5 ) \eta \chi^\dagger$ \\
& & &  $ \displaystyle   \frac{1}{M_\Omega} \left[ (g_s^2 - g_p^2) \bar{\psi}
  \psi \chi^\dagger \chi + \frac{i}{M_\Omega} \chi^\dagger \bar{\psi} \left(g_s^2 + g_p^2 - 2     g_s g_p \gamma^5 \right)\gamma^\mu  \partial_\mu \left( \psi \chi \right) \right]$ \\
\hline 
 \multirow{2}{*}{S}      & \multirow{2}{*}{V} &
       \multirow{2}{*}{\raisebox{-0\height}{\includegraphics[width=0.1\textwidth]{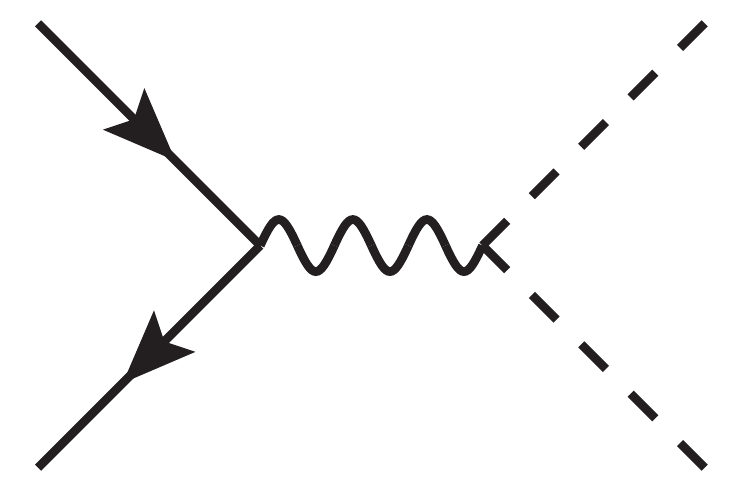}}}&$
       g_\chi (\chi^\dagger \partial_\mu \chi - \chi \partial_\mu
       \chi^\dagger) Z^\mu + \bar{\psi} \gamma^\mu (g_l P_L + g_r P_R) \psi
       Z_\mu$ \\
& & & $\displaystyle \frac{g_\chi}{M_\Omega^2} \bar{\psi} \gamma^\mu \left( g_l P_L + g_r P_R \right) \psi \left( \phi^\dagger \partial_\mu \phi - \phi \partial_\mu \phi^\dagger \right)$\\
\hline \hline
\multirow{2}{*}{F} & \multirow{2}{*}{S} &
\multirow{2}{*}{\raisebox{-0\height}{\includegraphics[width=0.1\textwidth]{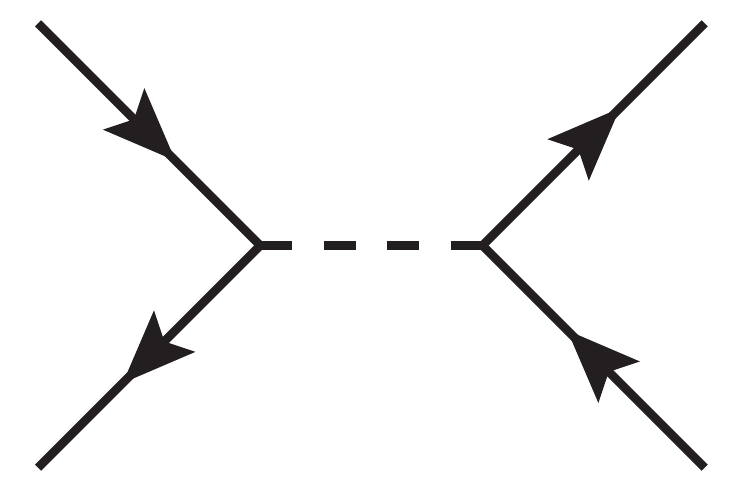}}}&
$\bar{\chi} \left(g_s + g_p \gamma^5 \right) \chi \phi + \bar{\psi} \left( g_s
  + g_p \gamma^5 \right) \psi \phi$ \\
& & & $\displaystyle \frac{1}{M_\Omega^2} \bar{\chi} \left(g_s + i g_p \gamma^5 \right) \chi \bar{\psi} \left( g_s + i g_p \gamma^5 \right)\psi$\\
\hline
\multirow{2}{*}{F}        & \multirow{2}{*}{V} &
        \multirow{2}{*}{\raisebox{-0\height}{\includegraphics[width=0.1\textwidth]{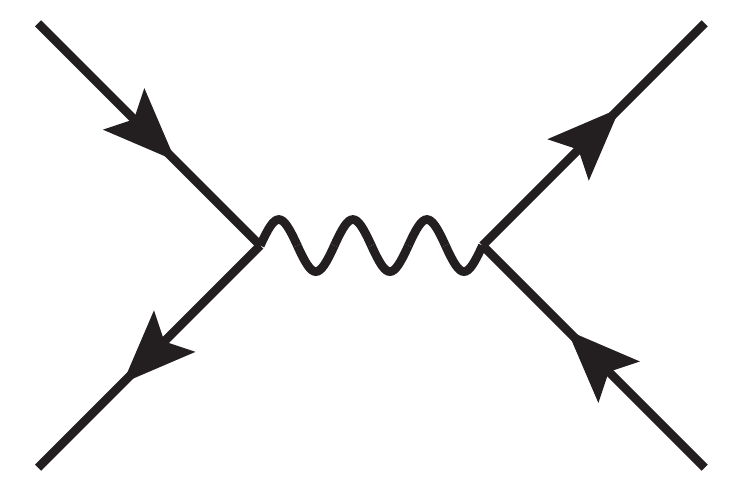}}}&
        $\bar{\psi} \gamma^\mu (g_l P_L + g_r P_R) \psi Z_\mu + \bar{\chi}
        \gamma^\mu (g_l P_L + g_r P_R) \chi Z_\mu $ \\
& & & $\displaystyle \frac{1}{M_\Omega^2} \bar{\psi} \gamma^\mu \left(  g_l P_L + g_r P_R \right) \psi \ \bar{\chi} \gamma_\mu \left(g_l P_L + g_r P_R    \right) \chi$\\
\hline
\multirow{2}{*}{F}        & \multirow{2}{*}{tS} &
        \multirow{2}{*}{\raisebox{-0\height}{\includegraphics[width=0.1\textwidth]{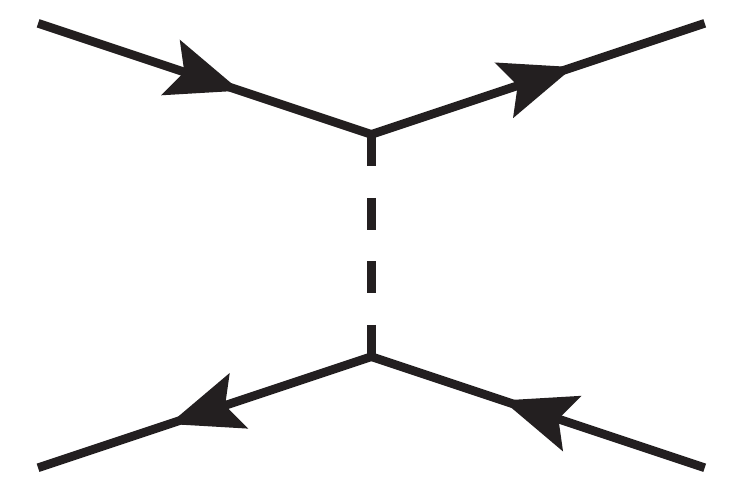}}}&$\bar{\chi}
        \left(g_s + g_p \gamma^5 \right) \psi \phi + \bar{\psi} \left( g_s +
          g_p \gamma^5 \right) \chi \phi$ \\
&&& $\displaystyle \frac{1}{M_\Omega^2} \bar{\psi} \left(g_s - g_p \gamma^5 \right) \chi \bar{\chi} \left( g_s + g_p \gamma^5 \right)\psi $ \\
\hline
\multirow{2}{*}{F}        & \multirow{2}{*}{tV} &
        \multirow{2}{*}{\raisebox{-0\height}{\includegraphics[width=0.1\textwidth]{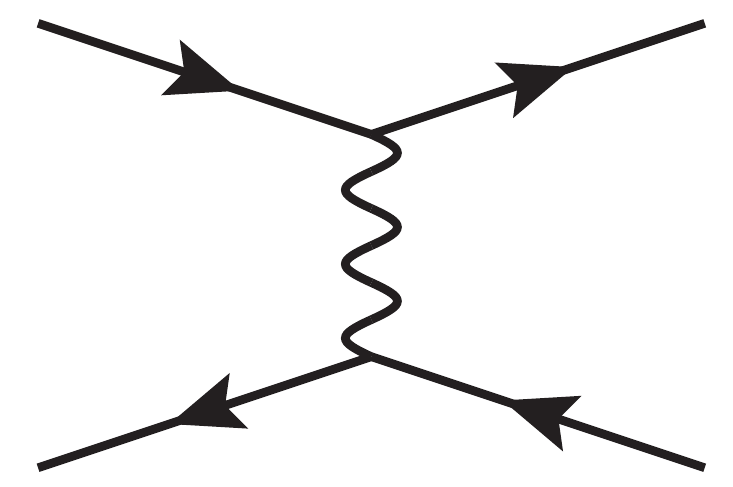}}}&$\bar{\psi}
        \gamma^\mu (g_l P_L + g_r P_R) \chi Z_\mu + \bar{\chi} \gamma^\mu (g_l
        P_L + g_r P_R) \psi Z_\mu $ \\
&&& $ \displaystyle \frac{1}{M_\Omega^2} \bar{\psi}
        \gamma^\mu \left(g_l P_L + g_r P_R \right)   \chi  \bar{\chi} \gamma_\mu \left(g_l P_L + g_r P_R  \right) \psi $ \\
\hline \hline
\multirow{2}{*}{V} & \multirow{2}{*}{S} &
\multirow{2}{*}{\raisebox{-0\height}{\includegraphics[width=0.1\textwidth]{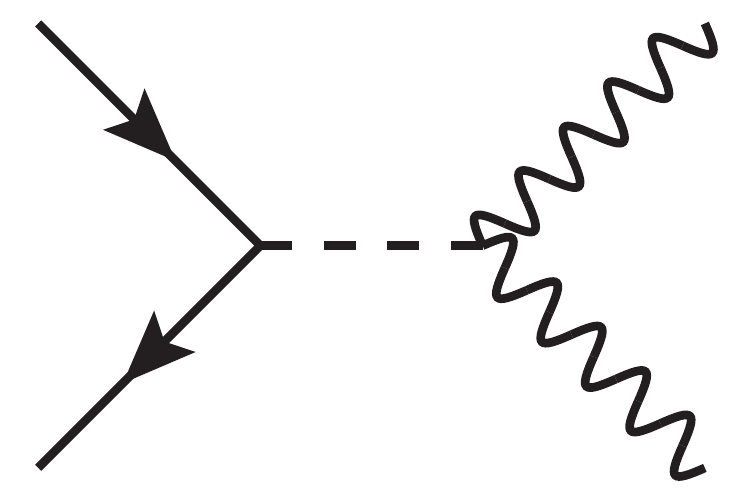}}}
& $-\chi^\mu \chi_\mu \phi + \bar{\psi} (g_s + i g_p \gamma^5)\psi \phi$ \\
&&& $\displaystyle - \frac{g_\chi}{M_\Omega^2} \chi^\mu \chi_\mu \bar{\psi} \left( g_s + i g_p \gamma^5 \right)\psi$\\
\hline
\multirow{2}{*}{V}       & \multirow{2}{*}{F} &
       \multirow{2}{*}{\raisebox{-0\height}{\includegraphics[width=0.1\textwidth]{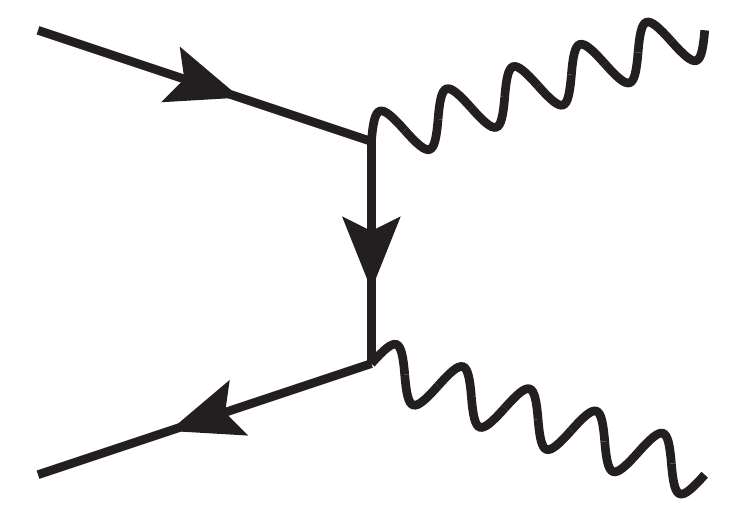}}}&
       $ - \bar{\eta} \gamma^\mu (g_l P_L + g_r P_R ) \chi_\mu
         + \bar{\psi} \gamma^\mu (g_l P_L + g_r P_R) \eta \chi_\mu^\dagger$      \\
&&& $\displaystyle \frac{1 }{M_\Omega} \left[g_l g_r \bar{\psi} \gamma^\nu
  \gamma^\rho \psi \ \chi^\dagger_\nu\chi_\rho  +  \frac{i}{M_\Omega} \chi^\dagger_\nu  \bar{\psi}
       \gamma^\nu\gamma^\mu \gamma^\rho \left(g_l^2 P_L + g_r^2P_R
       \right) \partial_\mu \left( \psi \chi_\rho \right) +\right]$\\
\hline
\multirow{2}{*}{V}       & \multirow{2}{*}{V} &
       \multirow{2}{*}{\raisebox{-0\height}{\includegraphics[width=0.1\textwidth]{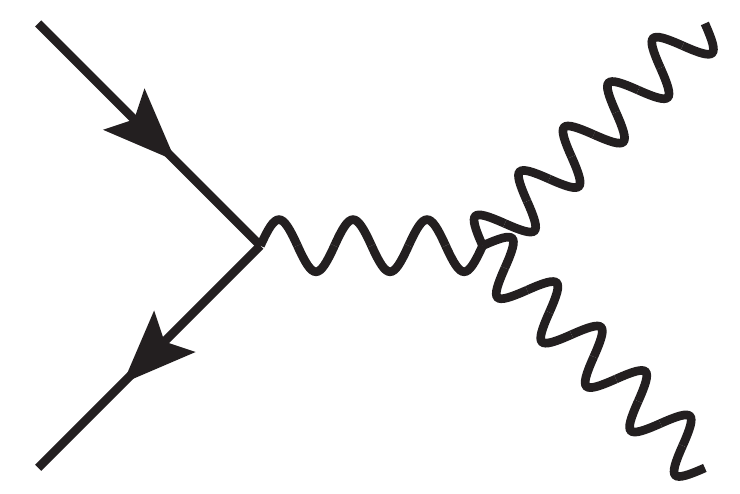}}}&
       $i g_\chi \left[ Z_\mu \chi^\dagger_\nu {\partial \chi}^{\mu \nu} + Z_\mu
         \chi_\nu  \partial \chi^{\mu \nu} + \chi^\dagger_\mu
         \chi_\nu \partial Z^{\mu \nu} \right]  + \bar{\psi} \gamma_\mu (g_l P_L + g_r P_R ) \psi$ \\
&&& $\displaystyle \frac{i g_\chi}{M_\Omega^2} \bar{\psi} \gamma^\mu \left( g_l P_L + g_r
  P_R \right) \psi \left[\chi^\nu \partial \chi^\dagger_{\mu \nu} - \chi^{\dagger, \nu} \partial \chi_{\mu \nu} + \partial^\nu \left(\chi^\dagger_\nu\chi_\mu - \chi^\dagger_\mu \chi_\nu \right) \right]$\\
\hline
\end{tabular}
\caption{List of interaction vertices for S(calar), F(ermion) and 
V(ector) dark matter, $\chi$,  before and after integrating out the 
heavy mediator scalar field $\phi$, spinor field $\eta$ or vector field
$Z^\mu$ with mass $M_{\Omega}$. $\psi$ denotes the Standard Model fermion.
$\partial X^{\mu \nu} \equiv \partial^\mu X^\nu  - \partial^\nu X^\mu$.
tS and tV denote cases where the mediator is exchanged in the $t$-channel.}
\label{tbl:allmodels}
\end{table*}
Cases with different spin for the dark matter or the mediator particle
are evaluated similarly. We only want to give some special remarks:
\begin{itemize}
\item For spin--$1/2$ mediators, the Dirac propagator
has only one power of $M_\Omega$ in the denominator,
\begin{align}
\frac{1}{\slashed{p} - M_\Omega} \approx -\frac{1}{M_\Omega} - \frac{\slashed{p}}{M_\Omega^2}. 
\end{align}
We therefore get two effective vertices after expanding the Lagrangian up to order $1/M_\Omega^2$.

\item Some effective operators give derivatives on the Standard Model fermion
  fields. These are not negligible, since they only vanish if the
  Dirac equation $i \slashed{\partial} \psi = m \psi$ can be used and
  the fermion mass $m$ is small. This is not the case for e.g.\ heavy
  quark contributions in the annihilation sector and processes with
  off--shell fermions.

\item We use the same list of effective operators for the cases of real scalar
  ($\chi = \chi^\dagger$), real vector ($\chi_\mu = \chi^\dagger_\mu$)
  or Majorana fermion \cite{Denner:1992vza} dark matter
  fields. However, we would like to mention that for consistency we do
  not introduce additional factors of $\nicefrac{1}{2}$ in the
  couplings as is often done in the case of real fields.

\end{itemize}

The full list of models with their respective fundamental and effective
Lagrangians is given in Table~\ref{tbl:allmodels}. Note that all Lagrangians are
hermitian by construction.

\subsection{Benchmark Models}

The effective operators described above have multiple
independent parameters to describe the effective coupling, for example $g_\chi, g_l, g_r$ and $M_
\Omega$  in the scalar dark matter, vector mediator (SV) case or
$g_s, g_p$ and $M_\Omega$ in the fermion dark matter, scalar mediator (FS) case in Table~\ref{tbl:allmodels}. 
\begin{table}
\centering
\begin{tabular}{l@{\quad}l@{\quad}l}
\hline
Operators & Definition & Name \\
\hline \hline
SS, VS, FS, & $g_p = 0$ & scalar \\
  FtS, FtSr:       & $g_s = 0$ & pseudoscalar \\
\hline
SF, SFr: & $g_p = 0, M_\Omega = \unit{1}{\TeV}$ & scalar\_low \\
        & $g_p = 0, M_\Omega = \unit{10}{\TeV}$ & scalar\_high \\
        & $g_s = 0, M_\Omega = \unit{1}{\TeV}$ & pseudoscalar\_low \\
        & $g_s = 0, M_\Omega = \unit{10}{\TeV}$ & pseudoscalar\_high \\
\hline
SV, FV, FtV,  & $g_l = g_r$ & vector \\
  FtVr, VV:      & $g_l = -g_r$ & axialvector \\
        & $g_l = 0$ & right--handed \\
\hline
VF, VFr: &  $g_l = g_r, M_\Omega = \unit{1}{\TeV}$ & vector\_low \\
        & $g_l = -g_r, M_\Omega = \unit{10}{\TeV}$ & vector\_high \\
        &  $g_l = g_r, M_\Omega = \unit{1}{\TeV}$ & axialvector\_low \\
        & $g_l = -g_r, M_\Omega = \unit{10}{\TeV}$ & axialvector\_high \\
\hline
FVr : & $g_l = 0$ & right--handed \\
\hline
\hline
\end{tabular}
\caption{Benchmark models with specific values for the coupling constants shown in  Table \ref{tbl:allmodels}.}
\label{tbl:constraints}
\end{table}
Considering the full range of parameters would lead to a
plethora of scenarios, well beyond the scope of this paper.  Thus we restrict our 
analysis to specific benchmark models (see
Table~\ref{tbl:constraints}) with constraints on the individual
couplings such that only one overall multiplicative factor
remains. The effective coupling constant $G$ for each model is then
defined as $G \equiv g_ig_j/ M_\Omega^2$.
For models with fermionic
mediators, the leading term has only a $1/M_\Omega$ dependence, which is
why we define $G \equiv g_ig_j/ M_\Omega$ for these.  We also choose two
possible values for $M_\Omega$ to represent different
suppression scales of the respective second order terms. Models with
real fields that are trivially connected to the corresponding complex
cases by multiplicative prefactors are not taken into account
separately. We also omit models with left--handed couplings that are
related to the respective right--coupled cases. Information on these
can easily be extracted from the related models by rescaling the
corresponding result accordingly.

%% file: astrophysics.tex
\section{Astrophysical constraints}
\label{sec:astro}

Any model which aims to describe dark matter, for example through a
\textsc{Wimp}, has to agree with present data. It has to give the correct relic
abundance, and must be consistent with the bounds from direct and
indirect detection
searches \cite{Komatsu:2010fb,Aprile:2012nq,Adriani:2008zr}.
\allowdisplaybreaks
\subsection{The Relic Abundance}
We first consider the best measurement of the relic abundance from
\textsc{Wmap}-7 \cite{Komatsu:2010fb},
\begin{equation}
\Omega^{\text{DM}} h^2 = 0.1099 \pm 0.0056\,.
\end{equation}
We employ the solution of the model dependent Boltzmann equation obtained
in \cite{Beltran:2008xg},
\begin{subequations}
\begin{align}
\Omega^\text{DM}_0 h^2 \approx 1.04\cdot10^9 \,\GeV^{-1} \frac{x_f}
{m_\text{Pl} \sqrt{g_*(x_f)} (a + 3b/x_f)}\;. \label{eqn:Omega0}
\end{align}
Here $m_{\mathrm{Pl}}$ is the Planck mass. $x_f=M_\chi/T_f$ is the
inverse freeze--out temperature, $T_f$, rescaled by the \textsc{Wimp} mass,
$M_\chi$. It is implicitly given by the equation,
\begin{align}
x_f = \text{ln} \left[ c (c+2) \sqrt{\frac{45}{8}} \frac{1}{2 \pi^3} 
\frac{g\ m_\text{pl} M_\chi (a + 6b/x_f)}{\sqrt{x_f} \sqrt{g_*(x_f)}} 
\right]. \label{eqn:x0}
\end{align}
\end{subequations}
$g_*(x_f)$ denotes the relativistic degrees of freedom in equilibrium
at freeze-out and is given in Ref.~\cite{Coleman:2003hs}. $a$ and $b$
are the first two coefficients of the non-relativistic
expansion of the thermally averaged annihilation cross section,
\begin{equation}
  \langle\sigma v \rangle \approx a + b v^{2} + O(v^{4}),
\end{equation}
where $v$ is the relative velocity of the colliding particles. Here the center-of-mass energy squared is approximated by \cite{Zheng:2010js,Yu:2011by},
\begin{equation}
    s \approx 4 M_{\chi}^2+ M_{\chi}^2 v^2 +3/4 ~ M_{\chi}^2 v^4.
\end{equation}
$g$ are the internal degrees of freedom of the \textsc{Wimp}. $c$ is an order unity parameter which is determined numerically in the solution of the Boltzmann equation and we set this parameter to 0.5.

Instead of testing all the models presented in
Table~\ref{tbl:allmodels}, we shall focus on a few exemplary cases.
First, the relic density depends on the possible Standard Model
particles, $f$, the \textsc{Wimp}s can annihilate into $\chi \overline{\chi}
\rightarrow f \overline{f}$. We shall consider two cases for the set of particles $f$:
(i) all leptons, (ii) all SM fermions.
Second, two variants of couplings are tested. In one scenario all SM
particles couple via the mediator to the \textsc{Wimp} with the same strength;
this is called \textit{universal coupling}. In the other they have a
coupling proportional to their mass, which we call \textit{Yukawa-like
coupling}. In the cases where we have the same effective operator our
results agree with Refs.~\cite{Zheng:2010js,Yu:2011by}, up to the
normalisation (see Appendix \ref{sec:sigmarelic}).

In order to set constraints, we must determine the total relic
density, which is the sum of the relic density of the particle and the
anti-particle (if the latter exists). This means the relic density for
a complex particle-pair is two times the density of a real particle.
If we consider the \textsc{Wmap} result as an upper bound on the relic density,
i.e.\ allowing for other dark matter, then this corresponds to
a lower bound on the effective coupling of the \textsc{Wimp} to the SM
particles. If we require our \textsc{Wimp} to be the only dark matter, we shall also obtain an
upper bound on the effective coupling.

The strict interpretation that our model only contains a heavy
mediator and a single \textsc{Wimp} ensures that there are no
resonances or co-annihilations.  However we also note that in many
full theories that contain dark matter, a `co-annihilation' regime can
exist that can significantly alter the relic density in the
universe. Whilst the co-annihilation mechanism cannot be incorporated
into the strict definition of our model, it may actually have no
observable effect on the collider based phenomenology.  An example of
such a feature could be stau co-annihilation in \textsc{Susy} that would not
change the \textsc{Ilc} production process of the lightest supersymmetric
particle. Another example is that a more complicated model may contain
resonant annihilations. Both of these examples can significantly
weaken the relic abundance bounds.

\subsection{Direct Detection}
We shall also impose bounds on our operators from the direct detection
searches for \textsc{Wimp} dark matter. The experiments are designed to measure
the recoil energy from the scattering between a (dark matter halo)
\textsc{Wimp} and the target nucleus. The interactions are difficult to detect
since the energy deposited is quite small, 1 to \unit{100}{\keV},
\cite{Bertone:2004pz}. These experiments give an upper limit for the 
cross section between the dark matter and the nucleus of the
target. One drawback is that in the cases where the \textsc{Wimp} does not
couple to quarks, the coupling can only occur through loop diagrams.

The direct detection experiments give a much stronger bound on spin
independent (SI) interactions than on spin dependent (SD). The reason
is that in the SI case the interaction with all nucleons add
coherently which enhances the corresponding cross section by the
atomic number squared. However, the spins of the nucleons cancel if
they are paired. Thus SD interactions are only enhanced for very
special nuclei.

The SI interactions are scalar or vector interactions in the
$s$-channel, the axialvector and tensor interactions in the
$s$-channel give a SD interaction. Note that due to the low
kinetic energy of the \textsc{Wimp}s the cross section should be
computed in the non-relativistic limit. In that case the pseudoscalar
interaction, $\overline{\psi} \gamma ^5 \psi$, vanishes.

The $t/u$--channel diagrams are cast into a sum of $s$--channel diagrams
via the Fierz identities. From this only the SI parts are employed, since any SD
contribution is negligibly small.
Tensor interactions occur only via the Fierz identities, since we do
not consider fundamental tensor interactions. However, since Fierz
  identities will always give at least one SI contribution, tensor
  terms can be dropped. 

For the SI interactions we shall consider the limits set by the \textsc{Xenon100} experiment \cite{Aprile:2012nq}. These are the most recent and set
the strictest limits over a broad parameter range. For the SD
interactions we consider the \textsc{Xenon}10 data \cite{Angle:2008we} since
\textsc{Xenon}100 gives no statement on SD interactions. The smaller data 
set along with the physical reasons mentioned above lead to a bound 
that is $\sim 10^6$ times weaker than for the SI interactions. The 
calculations for the \textsc{Wimp}--nucleus cross sections follow 
Ref.~\cite{Agrawal:2010fh} and for identical models
we find the same results. See Appendix \ref{sec:directdetect} for the
complete list of cross sections.

\subsection{Indirect Detection}

We also consider the indirect detection searches for dark
matter. These are much more model dependent, as the dark matter is
seen via an agent, for example neutrinos, which could also be produced
via other means. Specifically we shall consider the \textsc{Pamela} experiment
\cite{Adriani:2008zr} which measured an excess of positrons. These
could potentially originate from dark matter annihilation. To
implement this we need to compute the propagation of the produced
positrons and electrons from the source to the earth. This is
described by the diffusion--loss equation \cite{Baltz:1998xv},
\begin{equation}
\frac{\partial \psi}{\partial t} - \nabla [K(\textbf{x},E) \nabla 
\psi ] - \frac{\partial}{\partial E} [b(E) \psi] = q(\textbf{x},E).
\label{diff_loss}
\end{equation}
Here $\psi(x,E) = \mathrm{d} n_{e+}/ \mathrm{d}E$ is the positron
density per energy. $K(x,E)$ is the diffusion coefficient which
describes the interaction with the galactic magnetic field. $b(E)$
denotes the energy loss due to synchrotron emission and inverse
Compton scattering. $q(x,E)$ is the source term due to dark matter
annihilation. We note that convection and re--acceleration terms are
ignored as these do not apply to positrons \cite{Delahaye:2008ua}.

We use the conventional formalism
\cite{Delahaye:2007fr,Perelstein:2010fq} to derive a solution of
Eq.~(\ref{diff_loss}). It is also possible to use the so-called
extended formalism that takes the corrections from sources in the free
propagation zone into account as well as those from the diffusion
zone. However, this increases the runtime of the calculation
considerably while only giving a small correction that is less than
the measurement error. To perform the numerical comparison we use the
cored isothermal dark matter density profile \cite{Bahcall:1980fb} and
the galactic propagation model M2 \cite{Delahaye:2007fr}.

The above choices result in the following positron flux,
\begin{align}
\Phi_{e^+}(E)&=\frac{\beta_{e+}}{4 \pi} \psi(r_{\odot},z_{\odot},E), \\
\psi(r,z,E) &=\frac{\tau_{E}}{\epsilon^2} \int^{\epsilon_{max}}_{\epsilon} d \epsilon_S f(\epsilon_S) I(r,z,\epsilon,\epsilon_S), \\
I(r,z,\epsilon,\epsilon_S) &= \sum_i \sum_n J_0(\frac{\alpha_i r}{R}) \sin{\frac{n \pi (z+L)}{2L}} \nonumber \\
&\hspace{1.8cm}\times \exp{(-\omega_{i,n} (t-t_S))}	 R_{i,n}, \\
\omega_{i,n}&= K_0 [ (\frac{\alpha_i}{R})^2+ (\frac{n \pi}{2L})^2].
\end{align}

Here $\tau_E,~R,~K_0,~L$ are parameters which describe the
M2 propagation model.  They are set to the standard choices \cite{Delahaye:2007fr,Perelstein:2010fq}
$\tau_E = \unit{\power{10}{16}}{\second}$, $R=\unit{20}{\kilo \text{pc}}$ as well as to the M2 propagation model
$L=\unit{1}{\kilo \text{pc}}$, $K_0 = \unit{0.00595}{\kilo \text{pc$^2$/Myr}}$, $\delta=0.55$. $f(\epsilon)$ is
the energy distribution of the positrons from the annihilation and is
generated with \textsc{Pythia}8 \cite{Sjostrand:2007gs}.
$R_{i,n}$ are the coefficients of the Bessel-Fourier expansion of $R(r,z)$,
\begin{align}
R(r,z)&\equiv \eta \langle\sigma v \rangle \left[\frac{\rho(r,z)}{M_{\chi}}\right]^2, \\
\rho(r,z) &= \rho_{\odot}(\frac{r_{\odot}}{r})^{\gamma}\left[\frac{1+(r_{\odot}/r_s)^{\alpha}}{1+(r_{\odot}/r)^{\alpha}}\right] ^{(\beta-\gamma)/\alpha}.
\end{align}
Here $\langle \sigma v \rangle$ is the thermally averaged annihilation
cross section. We include all possible final states, not just those
resulting in positrons. Furthermore $\eta= 1/2$ for real
particles and 1/4 for complex particles. $ r_{\odot}=\unit{8.5}{\kilo \text{pc}}$ is the
distance of the solar system from the galactic center. $\rho_{\odot}=
\unit{0.3}{\GeV\per\centi\meter\cubed}$ is the local dark matter density and $\alpha=\beta=2,
~\gamma=0, ~ r_S =\unit{5}{\kilo \text{pc}}$ are chosen according to the cored isothermal
dark matter density distribution \cite{Delahaye:2007fr,Perelstein:2010fq}.

\textsc{Pamela} measures the ratio $\Phi_{e^+}/(\Phi_{e^-}+\Phi_{e^+})$,
where the fluxes, $\Phi_{e^\pm}$, contain the flux from dark matter
annihilation and from any astrophysical background. The background we
take is \cite{Baltz:1998xv},
\begin{subequations}
\begin{align}
\dfrac{d \Phi_{e^- bg}}{dE} &=\left(\frac{0.16 \epsilon^{-1.1}}{1+11 \epsilon^{0.9}+3.2 \epsilon^{2.15}}\right. \nonumber \\
&\hspace{-1.1cm}+\left.\frac{0.7 \epsilon^{0.7}}{1+110 \epsilon^{1.5}+600 \epsilon^{2.9}+580 \epsilon^{4.2}}\right)
\mathrm{GeV}^{-1} \mathrm{cm}^{-2} \mathrm{s}^{-1} \mathrm{sr}^{-1}, \\
\dfrac{d \Phi_{e^+ bg}}{dE} &=\frac{4.5 \epsilon^{0.7}}{1+650 \epsilon^{2.3}+1500 \epsilon^{4.2}} 
\mathrm{\GeV}^{-1} \mathrm{cm}^{-2} \mathrm{s}^{-1} \mathrm{sr}^{-1}, \\[3mm]
\epsilon &\equiv E/ \mathrm{GeV}. \nonumber
\end{align}
\end{subequations}
The quantity we compare to \textsc{Pamela} is,
\begin{equation}
\frac{\Phi_{e^+}}{\Phi_{e^+}+\Phi_{e^-}} = \frac{\Phi_{e^+ \chi} +\Phi_{e^+ bg}}{\Phi_{e^+ bg} + \Phi_{e^+ \chi} + \Phi_{e^- \chi} +\Phi_{e^- bg}},
\end{equation} 
and we note that  $\Phi_{e^+ \chi }= \Phi_{e^- \chi}$. 

We find an upper bound on the annihilation cross section by assuming that all of the excess comes from dark matter. However, it is possible that other background sources contribute and thus we also allow models that produce a flux smaller than the one seen.

We also note that for dark matter masses above $\sim$\unit{1}{\TeV}, the \textsc{Fermi--Lat} \cite{Atwood:2009ez} experiment may provide competitive bounds from inverse Compton scattering \cite{Cirelli:2009vg,Bernal:2010ip}. However, since we are only interested in models that can be probed at the \textsc{Ilc} we ignore them here.

The \textsc{IceCube} collaboration also sets limits on heavier dark matter masses via annihilations into neutrino final states \cite{Abbasi:2012ws,IceCube:2011aj}. In addition these bounds may be competitive for spin dependent interactions but we do not consider the limits in this study.

%


%% file: ilc.tex

\newcommand{\myinvfb}{\femto\reciprocal\barn}

\section{Dark matter search at the \textsc{Ilc}}
\label{sec:ilc}
\subsection{Radiative Production of Dark Matter}

\begin{figure}
\centering
\includegraphics[width=0.49\columnwidth]{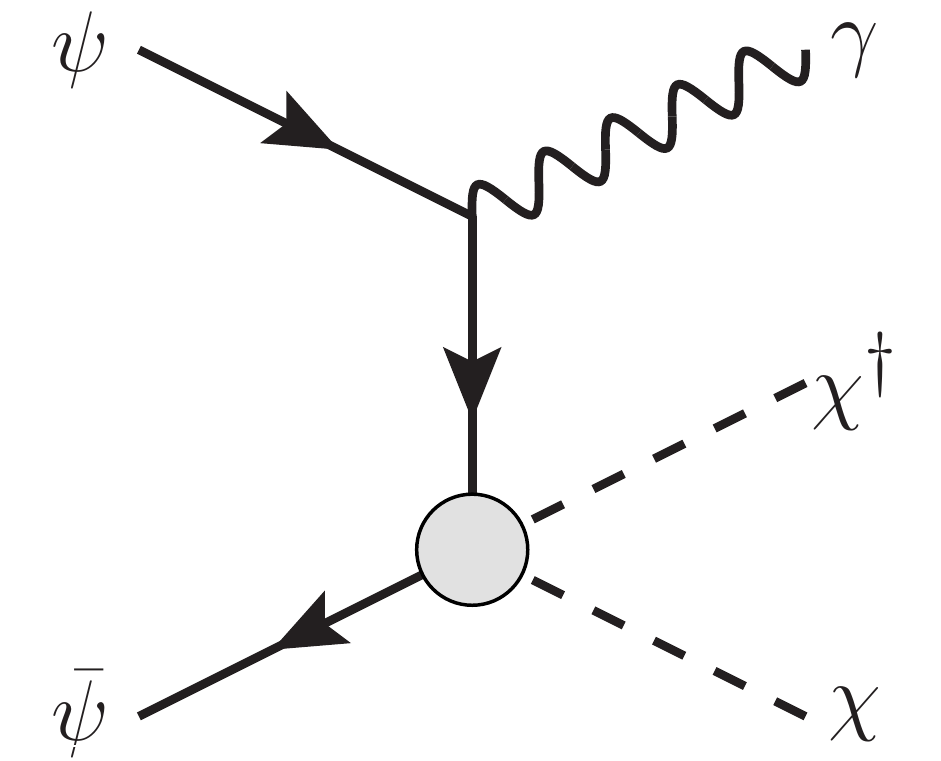} \hfill
\includegraphics[width=0.49\columnwidth]{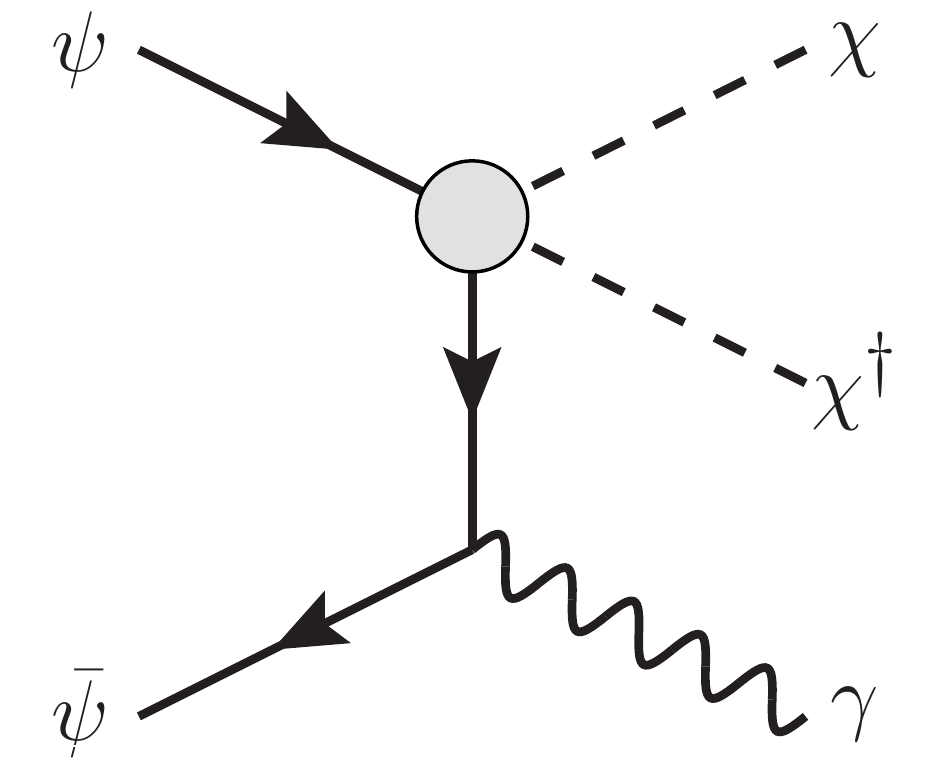}
\caption{Diagrams for radiative pair production of dark
  matter. Terms in which the heavy mediator can emit a photon are neglected.}
\label{fig:signalprocess_feynmandiagrams}
\end{figure}

\begin{figure*}
\centering
\includegraphics[width=0.49\textwidth]{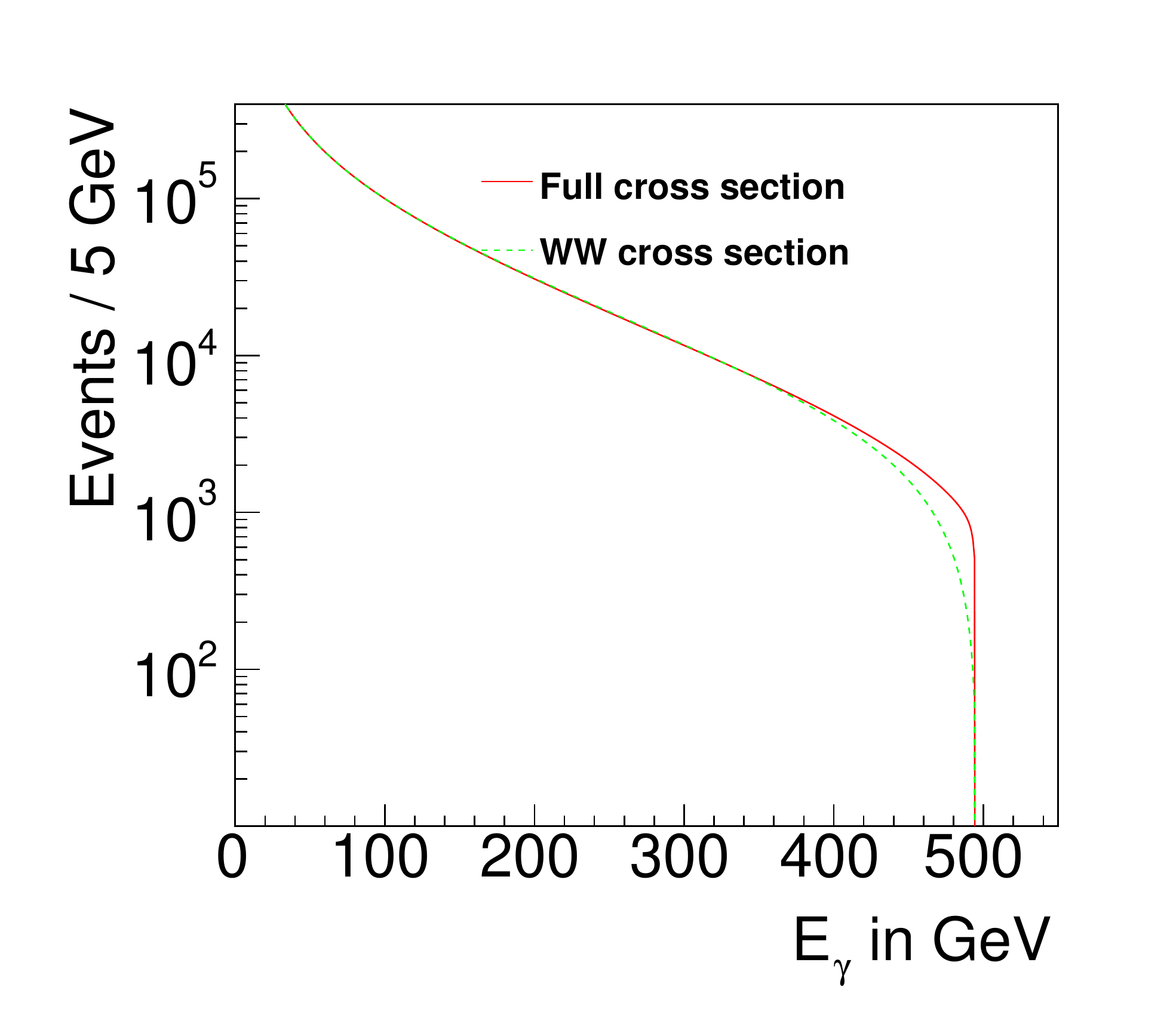}
\put(-45,180){(a)} 
\hfill
\includegraphics[width=0.49\textwidth]{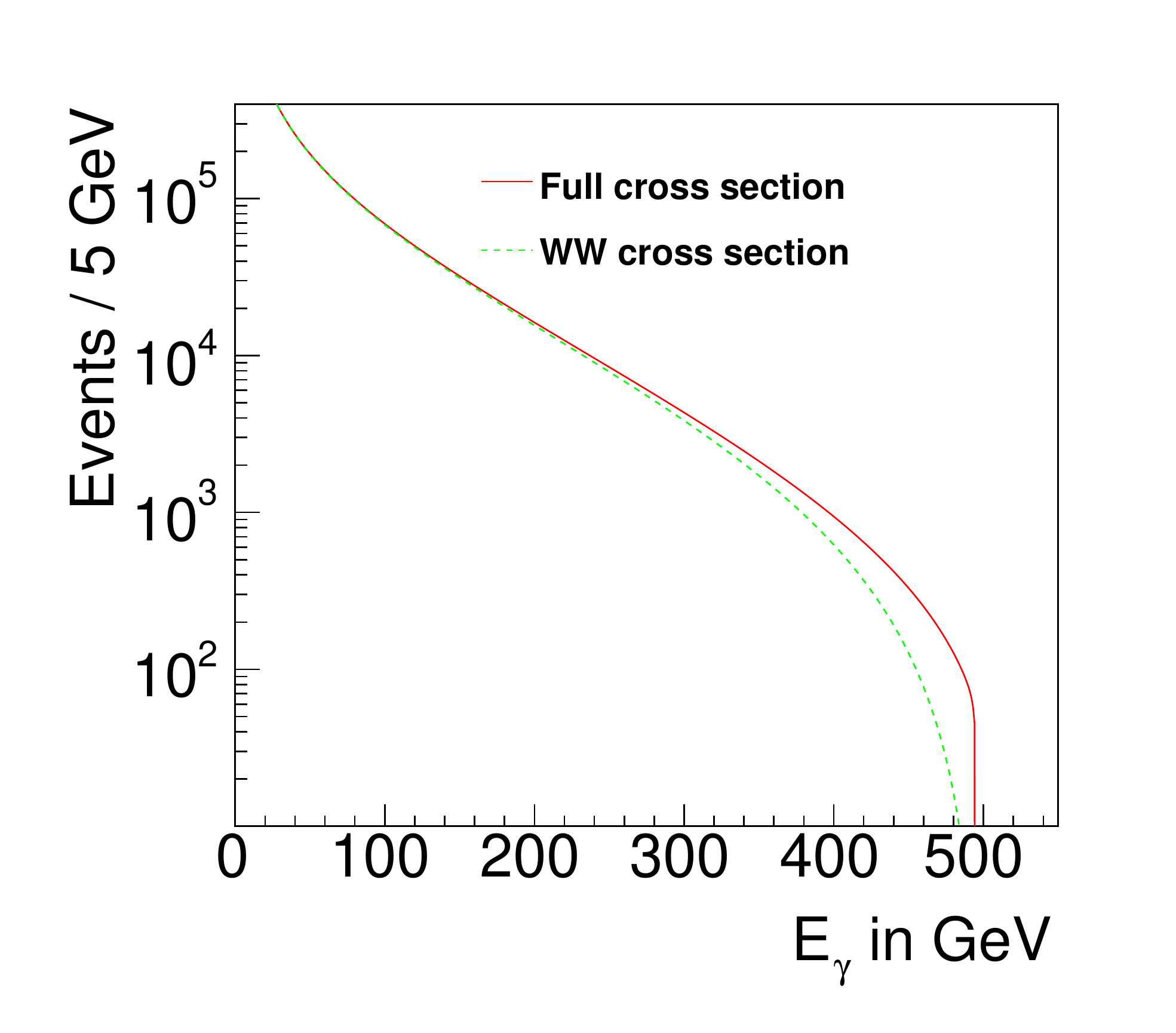} 
\put(-45,180){(b)} 
\caption{Comparison of tree level photon energy distributions in the
  WW--approximation and the analytical solution for $M_{\chi} = \unit{50}{\GeV}$,
  $|\cos \theta_\gamma|_\text{max} = 0.98$ and
  $\sqrt{s} = \unit{1}{\TeV}$. (a) SV, (b) FtS.}
\label{img:comparison}
\end{figure*}

For the \textsc{Ilc} search, we look at the process $\Ppositron
\Pelectron \rightarrow \chi \chi\Pphoton$ with a hard photon being the
only detected particle in the final state, Fig.~\ref{fig:signalprocess_feynmandiagrams}. We determine the polarized
differential cross section for this process with respect to the
relative photon energy $x \equiv 2 E_\gamma / \sqrt{s}$ and its polar
angle $\theta$ by integrating over the full phase space of the
final state dark matter particles.  The results for this
calculation are given in Table~\ref{tbl:2t3crosssections}, with
further explanation of the abbreviations used given in
Appendix~\ref{app:xsectterms}.  Previous \textsc{Ilc} studies,
e.g.\ \cite{Bartels:2012ex, Birkedal:2004xn}, have used the
Weizs\"acker--Williams approximation for soft photons. This formula
relates the differential photon cross section to the total pair
production cross section $\Ppositron\Pelectron \rightarrow \chi \chi$
with a reduced center of mass energy $s \rightarrow \hat{s} \equiv
s(1-x)$ and multiplied by the kinematical function $F_{x \theta}$,
\begin{align}
\frac{\mathrm{d} \sigma \left[\Ppositron \Pelectron  \rightarrow \bar{\chi} \chi
    \gamma \right] }{\mathrm{d} x \ \mathrm{d}
   \cos \theta_\gamma} \approx F_{x \theta}\ 
 \hat{\sigma} \left[\Ppositron \Pelectron \rightarrow \bar{\chi} \chi
 \right].
\label{weiz-will}
\end{align}
\begin{table*}
\begin{tabular}{r@{\quad}l}
\hline
Model & $ \displaystyle \frac{ \mathrm{d}\sigma}{ \mathrm{d} x \
  \mathrm{d}\cos \theta} $\\ [1.5ex]
\hline
\hline
SS & $ \displaystyle \frac{\hat{\beta}  F_{x \theta}}{32 \pi M_\Omega^4}
G_{s+p}  g_\chi^2 C_s $\\[1.5ex]
SF & $\displaystyle \frac{  \hat{\beta}  F_{x \theta}}{32 \pi M_\Omega^2} 
\left[  G_{s-p}^2 C_s  + \frac{\hat{\beta}^2 \hat{s}}{12 M_\Omega^2} 
    \boldsymbol{ V_{x \theta}} \left[(g_s+g_p)^4 C_R + (g_s-g_p)^4 C_L \right]
    + \boldsymbol{A_{\text{SF}}} \right] $ \\[1.5ex]
SFr & $\displaystyle \frac{  \hat{\beta}}{16 \pi M_\Omega^2} \left[ F_{x \theta}
G_{s-p}^2 C_s  + \boldsymbol{A_{\text{SFr}}}\right]$ \\[1.5ex]
SV & $ \displaystyle \frac{\hat{s} \hat{\beta}^3 F_{x \theta}}{96 \pi M_\Omega^4}  \boldsymbol{V_{x \theta}} 
 \left[ g_l^2C_L + g_r^2C_R \right]  g_\chi^2 $ \\[1.5ex]
\hline
\hline
FS &  $ \displaystyle \frac{\hat{s} \hat{\beta}  F_{x \theta}}{16 \pi M_\Omega^4} 
G_{s+p}  C_s \left[ g_{s}^2 \hat{\beta^2} + g_{p}^2 \right] $ \\[1.5ex]
FV & $ \displaystyle \frac{ \hat{\beta} F_{x \theta} }{48 \pi
  M_\Omega^4}  \boldsymbol{V_{x \theta}} \left[
G_{l+r} \hat{s} \hat{\beta}^2 + 3 \left(g_l + g_r \right)^2 M_\chi^2 \right] \left[ g_l^2C_L + g_r^2C_R \right] 
$ \\[1.5ex]
FVr & $ \displaystyle \frac{ \hat{s} \hat{\beta^3} F_{x \theta} }{48 \pi
  M_\Omega^4}  \boldsymbol{V_{x \theta}} \left(g_l - g_r \right)^2 \left[ g_l^2C_L + g_r^2C_R \right] $\\[1.5ex]
FtS &  $\displaystyle \frac{  F_{x \theta} \hat{\beta}}{48 \pi M_\Omega^4} G_{s+p}^2
\left[\boldsymbol{V_{x \theta}}(\hat{s}-M_\chi^2) + \boldsymbol{A_{\text{FtS}}} \right]$ \\[1.5ex]
FtSr & $\displaystyle \frac{ \hat{\beta} F_{x \theta}}{192 \pi M_\Omega^4}
G_{s+p}^2 \left[ 3 (\hat{s}-2 M_\chi^2) C_P + \boldsymbol{V_{x \theta}} 2 (\hat{s}
  - 4 M_\chi^2) C_V \right] $\\[1.5ex]
FtV & $\displaystyle \frac{\hat{\beta} F_{x \theta}}{48 \pi M_\Omega^4} \left[ 
  6 G_{lr}^2 C_s
  (\hat{s} - 2 M_\chi^2 ) +  (\hat{s} - M_\chi^2) \boldsymbol{V_{x \theta}} (g_l^4 C_L + g_r^4 C_R) \right] $ \\[1.5ex]
FtVr & $\displaystyle \frac{\hat{\beta} F_{x \theta}}{48 \pi M_\Omega^4} \left[ 
  12 G_{lr}^2 C_s   (\hat{s} - 2 M_\chi^2 ) + (\hat{s} - 4 M_\chi^2) \boldsymbol{V_{x \theta}} (g_l^4 C_L + g_r^4 C_R) \right] $\\[1.5ex]
\hline
\hline
VS &  $ \displaystyle \frac{ \hat{\beta} F_{x \theta}} {128 \pi M_\chi^4 M_\Omega^4} 
G_{s+p}  g_\chi^2 C_s (12 M_\chi^4-4M_\chi^2\hat{s}+\hat{s}^2) $ \\[1.5ex]
VF & $\displaystyle \frac{\hat{\beta}F_{x \theta}}{3840 \pi M_\chi^4 M_\Omega^2} 
\Big[ 40 G_{lr}^2 C_s (7 M_\chi^4 - 2 M_\chi^2 \hat{s} + \hat{s}^2)   +\frac{1}{M_\Omega^2}\left(g_l^4C_L+g_r^4C_R\right)
  (40 M_\chi^6-22M_\chi^4\hat{s}+56 M_\chi^2 \hat{s}^2 + 3 \hat{s}^3)    + \boldsymbol{A_{\text{VF}}}\Big]$\\[1.5ex]
VFr &  $\displaystyle \frac{\hat{\beta}F_{x \theta}}{3840 \pi M_\chi^4 M_\Omega^2} 
\Big[ 60 G_{lr}^2 C_s (12 M_\chi^4 - 4 M_\chi^2 \hat{s} + \hat{s}^2) + \frac{1}{M_\Omega^2}\left(g_l^4C_L+g_r^4C_R\right)
  (320 M_\chi^6-104^4\hat{s}+32 M_\chi^2 \hat{s}^2 + \hat{s}^3) + \boldsymbol{A_{\text{VFr}}}\Big]$\\
VV & $ \displaystyle \frac{ \hat{s} \hat{\beta}^3 F_{x \theta} \boldsymbol{V_{x \theta}}}{3840 \pi M_\chi^4 M_\Omega^4}\left[ g_l^2C_L + g_r^2C_R \right]  g_\chi^2 
( M_\chi^4 + 20 M_\chi^2 \hat{s} + \hat{s}^2)$ \\[1.5ex]
\hline
\end{tabular}
\caption{Analytical differential cross sections for the process $\Ppositron
  \Pelectron \rightarrow \chi \chi \gamma$ in the various effective
  models. Terms in bold do not appear in the Weizs\"acker--Williams
  approach and are given in Appendix~\ref{app:xsectterms} where we also
    define all used abbreviations. Models with a suffix `r' correspond to the case of real particles. 
  Cross sections for SSr, FSr and VSr are twice as large as
  in the complex case while SV and VV vanish completely for real particles.}
\label{tbl:2t3crosssections}
\end{table*}
Due to the soft collinear approximation used, we expect that the above equation will perform poorly for large angle and high $p_T$ photons. We compare the analytical result to this approximation to test
the reliability. In Table~\ref{tbl:2t3crosssections} we put terms in
bold, which are purely caused by our analytical
treatment. The corrections are either of the form of an additional
kinematical factor $V_{x \theta}$, mostly appearing in models with
vector mediators, or completely new terms that typically appear in
t--channel interactions. Since $\lim_{x \rightarrow 0} V_{x \theta} =
1$ and $\lim_{x \rightarrow 0} (A_i) = 0$, the WW--approximation is in
agreement with our full result for small energies. In
Fig.~\ref{img:comparison} we show the respective photon energy
distributions for different models in both the WW--approximation and
the full analytical treatment. The curves behave quite congruently
with differences visible in the high energy sector. Since most of the
signal events lie in the low energy part, the approximation gives
accurate results for counting experiments. A shape dependent analysis
would need to use the analytical result to estimate the correct
threshold behaviour for high energies. Our subsequent analysis
is performed with the full analytical cross section.

When we restrict the various couplings in our model according to the benchmark scenarios, Table~\ref{tbl:constraints}, most of the cross sections simplify and have only one polarisation dependent term $C_i$. To
determine the polarisation leading to the best signal to
background ratio, we only need to consider cases with different $C_i$.
We therefore classify our models as follows:
\begin{align}
\text{Scalar--like}: \sigma_{\text{pol}} &= C_S
\sigma_{\text{unpol}}, \label{eqn:polclasses} \\
\text{Vector--like}: \sigma_{\text{pol}} &= C_V \sigma_{\text{unpol}}, \nonumber
\\
\text{Right--like}: \sigma_{\text{pol}} &= C_R \sigma_{\text{unpol}}, \nonumber \\
\text{Left--like}: \sigma_{\text{pol}} &= C_L \sigma_{\text{unpol}}. \nonumber
\end{align}

Models with t--channel mediators usually have multiple terms with
different polarisation behaviour and do not fall into one of the basic
polarisation classes given in Eq.~(\ref{eqn:polclasses}). We choose
the following polarisation settings for those:
\begin{itemize}
\item Models with fermionic mediators are classified according to their
  leading term, which is always scalar--like. 
\item All other models have both scalar--like and vector--like parts of about
  the same size. We analyse them in a vector--like scenario that naturally leads to a
  better background suppression.
\end{itemize}

\subsection{Standard Model Background for Monophotons}
We consider the two leading dominant Standard Model background contributions
after selection, determined with a full \textsc{Ild} (International Linear Detector concept) detector simulation
\cite{BartelsThesis, Bartels:2012ex}. All numbers here and in the following
paragraphs refer to the nominal \textsc{Ilc} center of mass energy of
\unit{500}{\GeV} \cite{Phinney:2007gp}. We also consider the case of an
increased energy of \unit{1}{\text{TeV}} and mention the differences later.
\begin{itemize}
\item Neutrinos from $\Ppositron \Pelectron \rightarrow
  \Pnu \APnu \Pphoton (\Pphoton)$ form a polarisation dependent background. The leading
  contribution comes from t--channel $\PW$--exchange, which only couples to
  left--chiral leptons. Additional smaller contributions come from s--channel
  $\PZ$--diagrams with both left-- and right--chiral couplings. We also
  consider the case of one additional undetected photon, which contributes
  with a size of roughly $\unit{10}{\%}$. 
\item Bhabha scattering of leptons with an additional hard photon, $\Ppositron \Pelectron \rightarrow
  \Ppositron \Pelectron \Pphoton$ has a large cross section but a very small
  selection efficiency, since both final state leptons must be undetected. It has been determined to give a contribution of the same order
  of magnitude as the neutrino background, after application of all selection
  criteria. It is mostly polarisation independent \cite{BartelsThesis, Bartels:2012ex}.
\end{itemize}
 Other background sources contribute with less
than \unit{1}{\%} compared to the neutrino background and are therefore omitted. 

\subsection{Data Modeling}
To evade the use of a full detector simulation, we build on the results of
Refs.~\cite{BartelsThesis, Bartels:2012ex}. For the signal and monophoton neutrino
background, we generate the events by ourselves with the given phase space criteria. 
We then apply the \textsc{Ild} estimates for the
energy resolution as well as the reconstruction and selection
efficiencies\footnote{From here on, the expression `efficiency' abbreviates
  `reconstruction and selection efficiencies'.}  and compare the final energy
distributions. For the diphoton neutrino and Bhabha background, we model the
final distributions directly from the given results performed with a full detector simulation \cite{BartelsThesis, Bartels:2012ex}.

\begin{figure*}
\centering
\includegraphics[width=0.49\textwidth]{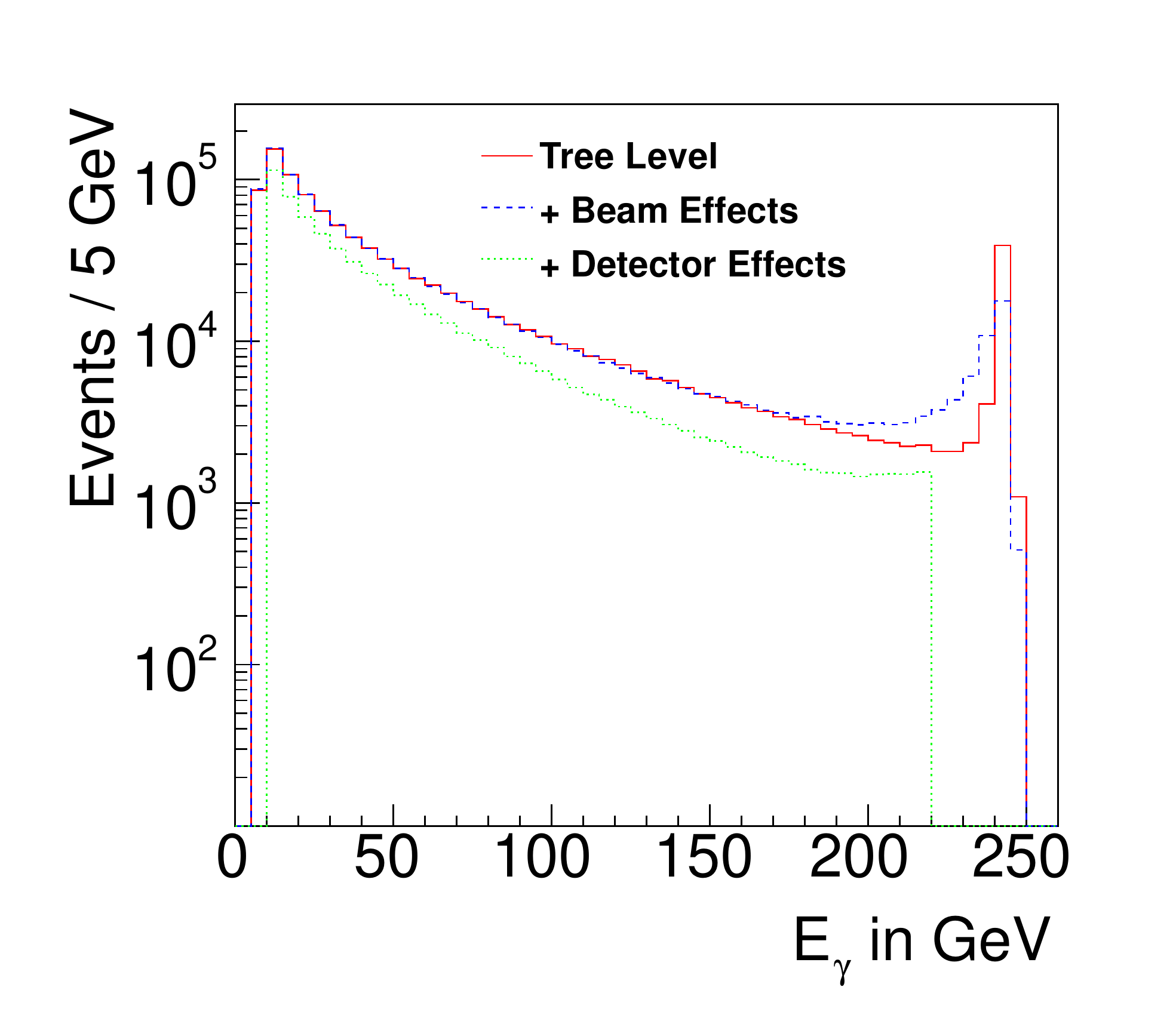} 
\put(-45,180){(a)}
\hfill
\includegraphics[width=0.49\textwidth]{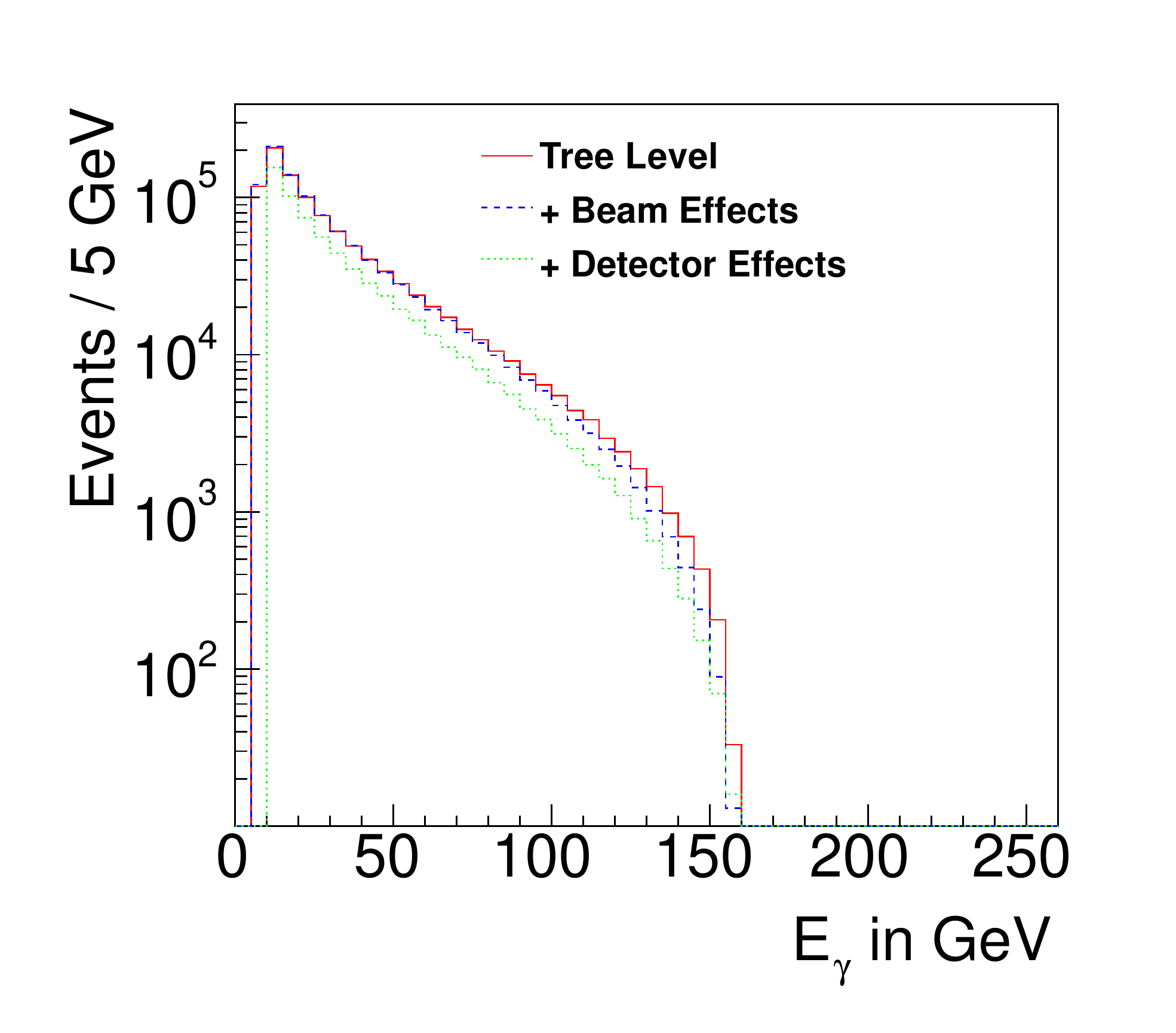}
\put(-45,180){(b)}
\caption{Photon energy distribution before and after application of beam effects
  (\textsc{Isr} + beamstrahlung) and detector effects (resolution +
  efficiency) for a) unpolarised neutrino background and b) unpolarised FS
  scalar signal with $M_\chi = \unit{150}{\GeV}$. Distributions are normalised to
  $10^6$ tree level events.}
\label{img:photonenergy}
\end{figure*}

\begin{figure}
\centering
\vspace{-0.55cm}
\includegraphics[width=0.49\textwidth]{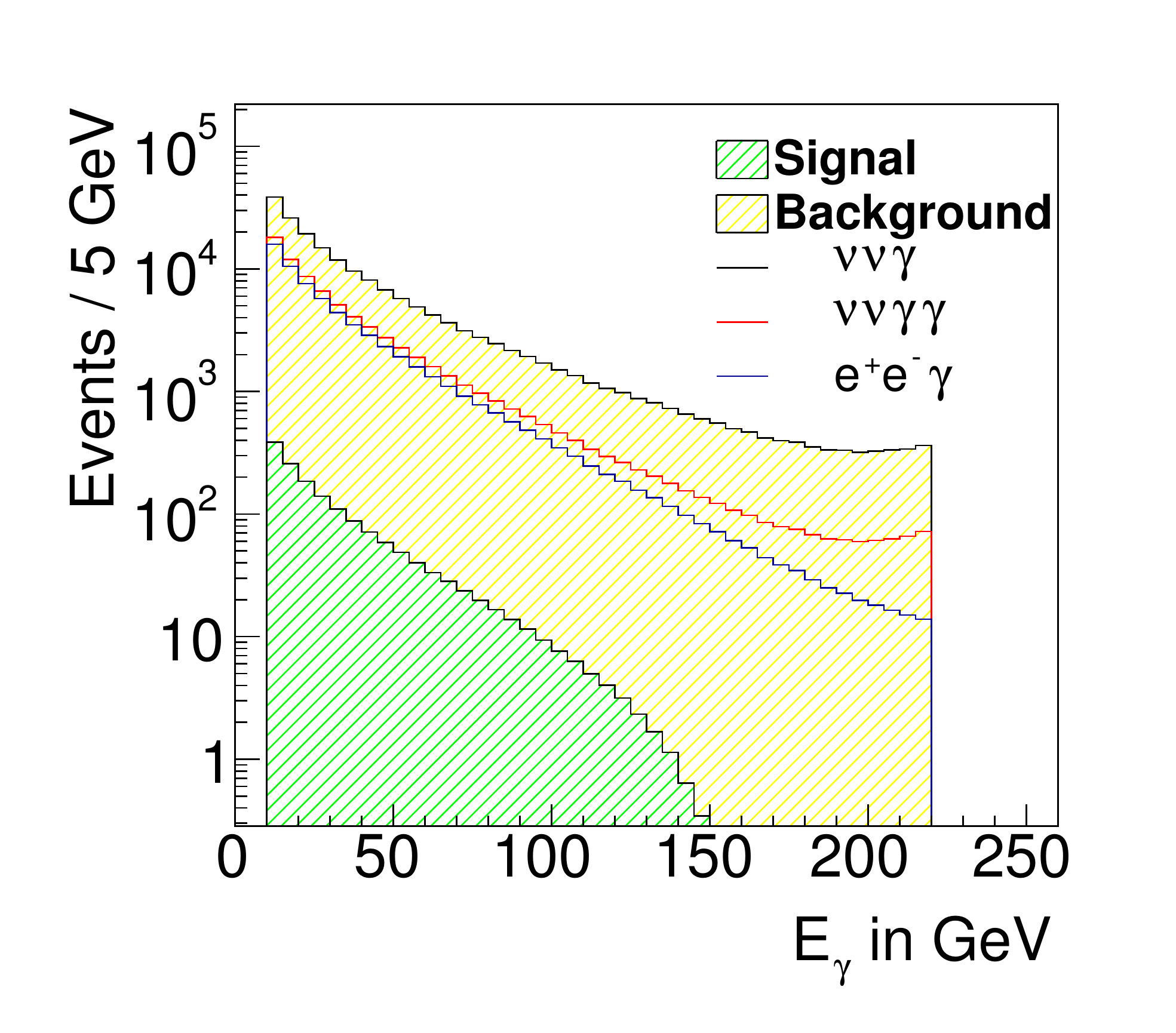}
\vspace{-0.8cm}
\caption{Photon energy distributions of the most dominant background
  contributions (stacked) compared to an example signal (FS Scalar, $M_\chi =
  \unit{150}{\GeV}$) with a total cross section of \unit{100}{\femto\barn}. All spectra are taken after selection for an unpolarised
  initial state.}
\label{img:backgroundcontributions}
\vspace{0.775cm}
\begin{tabular}{r@{\qquad}crr c@{\quad} rr c@{\quad}  rr }
\hline
$P^-/P^+$ && \multicolumn{2}{c}{$\Pnu \Pnu \Pphoton$} && \multicolumn{2}{c}{$\Pnu \Pnu \Pphoton \Pphoton$} &&\multicolumn{2}{c}{$\Ppositron \Pelectron \Pphoton$} \\
\hline
\hline
$0/0$ && 2257 & (2240) && 226 & (228) && 1218 & (1229) \\
$+0.8/-0.3$ && 493 & (438) && 49 & (43) && 1218 & (1204) \\
$-0.8/+0.3$ && 5104 & (5116) && 510 & (523) && 1218 & (1227) \\
\hline
\end{tabular}
\caption{Total number of events in the different background
  sources after application of all selection criteria. The numbers are
  given for an integrated luminosity of \unit{1}{\myinvfb} in different
  polarisation settings. Numbers in brackets are taken from
  Ref.~\cite{BartelsThesis} which employed a proper detector
    simulation.}
\label{tbl:neventsperscenario}
\end{figure}

For the generation of signal and monophoton neutrino events we use
\textsc{C}alc\textsc{hep} \cite{Pukhov:1999gg}. We produce signal events for all benchmark
scenarios with dark matter masses ranging from \unit{1}{\GeV} to
\unit{240}{\GeV}. To avoid collinear and infrared divergences,
we limit phase space in the event generation to $E_\gamma \in \left[\unit{8}{\GeV}, \unit{250}{\GeV}
    \right]$ and $\cos \theta_\gamma \in \left[ -0.995, 0.995
    \right]$. Initial State Radiation (\textsc{Isr})
    and beamstrahlung significantly change the width and position of the neutrino
    \PZzero--resonance, Fig.~\ref{img:photonenergy}a), and are taken into account. We
    set the accessible parameters in \textsc{C}alc\textsc{hep} according to the \textsc{Ild}
    Letter of Intent \cite{Abe:2010aa} to \unit{645.7}{\nm} for the bunch size, \unit{0.3}{\mm}
    for the bunch length and a total number of particles per bunch of $2 \cdot
    10^{10}$.

The finite resolution of the detector components and the use of 
selection criteria to reduce beam--induced background are taken into account by
applying the following steps to both signal and background data. First we shift the photon
energy, given in \GeV, according to a Gaussian distribution by taking into account the estimated
  resolution of the \textsc{Ild} detector components \cite{Abe:2010aa},
\begin{align}
\frac{\Delta E_\gamma}{E_\gamma} =
\frac{\unit{16.6}{\%}}{\sqrt{E_\gamma \text{ in \GeV}}} \oplus \unit{1.1}{\%}.
\end{align}
Afterwards we further limit the phase space to reduce background processes in
  the \PZzero resonance peak at \unit{242}{\GeV} and additional collinear photons from \textsc{Isr},
\begin{align}
E_\gamma &\in \left[ \unit{10}{\GeV}, \unit{220}{\GeV} \right], \\
\cos \theta_\gamma &\in \left[ -0.98, 0.98 \right]. \nonumber
\end{align}
The additional angular cut ensures a good photon reconstruction within
the detector.  Finally a random elimination of events is used
to simulate the efficiency factor for reconstruction and selection
determined in Ref.~\cite{BartelsThesis}. The efficiency consists of an
energy dependent part $\epsilon_1$ and a constant part $\epsilon_2$
that are applied successively,
\begin{align}
\epsilon_1 &= \unit{97.22}{\%} - (E_\gamma \text{ in \GeV}) \cdot\unit{0.1336}{\%}, \label{eqn:efficiency1}\\
\epsilon_2 &= \unit{96.8}{\%}.  \nonumber
\end{align}
Fig.~\ref{img:photonenergy} shows how these settings affect the signal and
background spectrum and Fig.~\ref{img:backgroundcontributions} shows a stacked histogram of the dominant background processes along with a example dark matter signal.


\begin{table*}
\centering
\begin{tabular}{r@{\quad}r@{\qquad}c rr@{\qquad}c rr@{\qquad}c rrrr}
\hline 
&&&&&&&&&&&& \\ [-2.ex]  
$P^-/P^+$ & $N_{\text{B}}$ && $\Delta^\text{stat}_{50}$  & $\displaystyle\tilde{\Delta}^\text{stat}_{500}$ && $\Delta^\text{sys}_{P}$
& $\displaystyle\Delta^\text{sys}_{\tilde{P}}$  && $\displaystyle \Delta^\text{tot}_{50 P}$ &
$\Delta^\text{tot}_{50 \tilde{P}}$ & $\displaystyle \tilde{\Delta}^\text{tot}_{500 P}$
& $\displaystyle \tilde{\Delta}^\text{tot}_{500 \tilde{P}}$ \\ [1ex]
\hline\hline
0/0 & 184998 &&  &  && &  &&  &  &  &  \\
\hline
$+0.8$/$+0.3$ & 97568 && 312 & 99 && 312 & 125 && 441 & 336 & 327 & 159 \\
$+0.8$/$+0.6$ & 102365 && 320 & 101 && 385 & 154 && 500 & 355 & 398 & 184 \\
\hline
$+0.8$/$-0.3$ & 87974 && 297 & 94 && 169 & 68 && 341 & 304 & 193 & 116 \\
$+0.8$/$-0.6$ & 83177 && 288 & 91 && 104 & 42 && 307 & 291 & 138 & 100 \\
\hline
$-0.8$/$+0.3$ & 341597 && 584 & 185 && 351 & 140 && 682 & 601 & 396 & 232 \\
$-0.8$/$+0.6$ & 404970 && 637 & 201 && 501 & 200 && 811 & 668 & 546 & 284 \\
\hline
$-0.8$/$-0.3$ & 212851 && 461 & 156 && 233 & 93 && 517 & 471 & 275 & 173 \\
$-0.8$/$-0.6$ & 148478 && 385 & 122 && 337 & 135 && 512 & 408 & 359 & 182 \\
\hline
\end{tabular}
\caption{Total amount of background events, $N_{\text{B}}$, with statistical
  error, $\Delta^\text{stat}$, systematic error, $\Delta^\text{sys}$, and the total error,
  $\Delta^\text{tot}$. The subscripts 50 and 500 denote the integrated
  luminosity in inverse femtobarn. In case of a ten times larger luminosity,
  one will get ten times as many events in all channels; to better compare to the error
  of the low luminosity case,
  we give $\tilde{\Delta}^\text{stat}_{500} \equiv \Delta^\text{stat}_{50}/\sqrt{10}$. The polarisation uncertainties are set to
  \unit{0.25}{\%} ($P$) and \unit{0.1}{\%}
  ($\tilde{P}$).}
\label{tbl:bkgevts}
\end{table*}

\subsection{Analysis}

We are interested in determining the lower bound on the effective coupling constants that the
\textsc{Ilc} can find for each individual model under the assumption that no signal
events are measured. We perform a counting experiment by using the
\textsc{TRolke} \cite{Rolke:2004mj} statistical test. We determine the total
number of background events along with its statistical and systematic fluctuation
$\Delta N_{\text{B}}$ and exclude coupling constants which would lead to a
larger number of signal events than the \unit{90}{\%} confidence interval of the background--only
assumption.

\subsection{Systematic Uncertainties}

Systematic uncertainties play an important role in determining the
total error on the background, $\Delta N_B$, and for estimating the
bounds on the effective couplings. There are two dominant
contributions, motivated in Ref.~\cite{BartelsThesis} which we now discuss.

The experimental efficiency given in Eq.~(\ref{eqn:efficiency1}) will
be determined at the real experiment by measuring the
$\PZzero$--resonance peak, which is theoretically known to a very good
accuracy. Systematic uncertainties on that value are given by the
finite statistics of this measurement and further broadening of the
peak by unknown beam effects. These errors can be extrapolated down to
the dark matter signal region at small photon energies and, since the
same efficiency factor is used for signal and background, is highly
correlated between those two. This global uncertainty will therefore
approximately cancel in the determination of the maximum coupling
$G_\text{eff}$.

Cancellation will not take place for model dependent effects
however. This is due to the fact that the signal energy distribution 
depends on the unknown mass of the dark matter particle and the
underlying interaction model. Therefore, the correct function
$\epsilon(E_\gamma)$ for the signal will be different from the used
neutrino background efficiency given in
Eq.~(\ref{eqn:efficiency1}). Since we do not know the model a priori,
we use the same value for both and introduce an error on the
determination of the signal events, $N_S$. Compared to Ref.~\cite{BartelsThesis}, we use a
conservative value of $\Delta \epsilon = \unit{2}{\%}$.

Since the neutrino spectrum depends on the incoming lepton's
polarisation $P^\pm$, any fluctuation within those parameters will
give additional systematic uncertainties on the number of expected
background events. One can not use the information from measuring the
$\PZzero$--resonance in this case to infer information in the low
energy signal range because of the polarisation dependence of the
shape itself. Given the assumed accuracy of at least $\Delta P/P =
\unit{0.25}{\%}$ \cite{Abe:2010aa} with a possible improvement to
\unit{0.1}{\%} at the \textsc{Ilc}, we can derive the corresponding
error on the polarised number of background events. As an example we
show the left handed background,
\begin{align}
N_{\text{pol}} &= (1+P^+)(1-P^-)N_{\text{unpol}}, \nonumber \\
\Delta N_{\text{pol}} &= \sqrt{\left[P^- (1+P^+) \right]^2 + \left[P^+ (1-P^-)
  \right]^2} \  \frac{\Delta P}{P} \ N_{\text{unpol}}. \label{eqn:deltap}
\end{align}
From the numbers in Table~\ref{tbl:neventsperscenario}, we assume an identical
polarisation dependence for $\Pnu \Pnu \Pphoton$ and $\Pnu \Pnu \Pphoton
\Pphoton$ events and no dependence for the Bhabha background. 

\begin{table}[b]
\centering
\begin{tabular}{rr@{\quad}rrrrrr}
\hline
IA type & $P^-/P^+$ & $N_{\text{S}}$ && $r_{50 P}$ & $r_{50 \tilde{P}}$ & $r_{500 P}$ &
$r_{500 \tilde{P}}$ \\
\hline
Scalar &$+0.8/+0.3$   &  620  &&  1.41  &  1.85  &  \textbf{1.90}  &  3.90 \\
 &$+0.8/+0.6$   &  740  &&  \textbf{1.48}  &  \textbf{2.08}  &  1.86  &  \textbf{4.02} \\
\hline
Vector &$+0.8/-0.3$   & 620  &&  1.82  &  2.04  &  3.21  &  5.34 \\
& $+0.8/-0.6$   &  740  &&  \textbf{2.41}  &  \textbf{2.54}  &  \textbf{5.36}
&  \textbf{7.40} \\
\hline
Left & $-0.8/+0.3$   &  1170  &&  1.72 &  1.95  &  \textbf{2.95}  &  5.04 \\
& $-0.8/+0.6$   &  1440  &&  \textbf{1.78}  &  \textbf{2.16}  &  2.64  &  \textbf{5.07} \\
\hline
Right &$+0.8/-0.3$   &  1170  &&  3.43  &  3.85  &  6.06  &  10.09 \\
& $+0.8/-0.6$   &  1440  &&  \textbf{4.69}  &  \textbf{4.95}  &  \textbf{10.43}  &  \textbf{14.4} \\
\hline
\end{tabular}
\caption{Determination of the best ratio $r \equiv
  {N_{\text{S}}}/{\Delta N_{\text{B}}}$ with $\Delta N_{\text{B}}$ given by the different
  total errors determined in Table~\ref{tbl:bkgevts}. $N_{\text{S}}$ describes the
number of polarised signal events for the different classes described in Sec.~\ref{sec:models}
with a common reference value of 500 unpolarised events for an integrated
luminosity of $\unit{50}{\myinvfb}$. We only show the polarisation signs with the largest ratios. We mark the
numbers which lead to the best signal to background ratio in bold.}
\label{tbl:sigoverbkgestimate}
\end{table}

\begin{table*}
\begin{tabular}{l@{\qquad}r@{\quad}rr@{\quad}rr@{\quad}rrrr}
\hline
 &&&&&&&&& \\ [-2.ex]  
$P^-/P^+$ & $N_{\text{B}}$ & $\Delta^\text{S}_{50}$  & $\displaystyle\tilde{\Delta}^\text{S}_{500}$ & $\delta^\text{P}_{P}$
& $\delta^\text{P}_{\tilde{P}}$  & $\Delta^\text{tot}_{50 P}$ & $\Delta^\text{tot}_{50 \tilde{P}}$ & $\displaystyle\tilde{\Delta}^\text{tot}_{500 P}$
& $\displaystyle \tilde{\Delta}^\text{tot}_{500 \tilde{P}}$ \\ [1ex]
\hline
\hline
$0/0$ & 162437 &  &  &  &  &  &  &  &  \\
\hline
$+0.8$/$+0.3$ & 54649 & 234 & 74 & 380 & 152 & 446 & 279 & 387 & 169 \\
$+0.8$/$+0.6$ & 62791 & 251 & 79 & 469 & 188 & 531 & 314 & 476 & 203 \\
\hline
$+0.8$/$-0.3$ & 38365 & 196 & 62 & 201 & 82 & 281 & 212 & 210 & 102 \\
$+0.8$/$-0.6$ & 30223 & 174 & 55 & 125 & 50 & 214 & 181 & 137 & 74 \\
\hline
$-0.8$/$+0.3$ & 357173 & 598 & 189 & 428 & 171 & 735 & 622 & 468 & 255 \\
$-0.8$/$+0.6$ & 435979 & 660 & 209 & 612 & 245 & 900 & 704 & 647 & 322 \\
\hline
$-0.8$/$-0.3$ & 199561 & 447 & 141 & 284 & 114 & 530 & 461 & 317 & 181 \\
$-0.8$/$-0.6$ & 120755 & 348 & 110 & 411 & 165 & 538 & 385 & 425 & 198 \\
\hline
\end{tabular}
\caption{Total amount of background events ($N_{\text{B}}$) and different
  error sources (see Table~\ref{tbl:bkgevts}) for $\sqrt{s} = \unit{1}{\TeV}$.}
\label{tbl:bkgevts_1tev}
\end{table*}

\begin{table}[b]
\begin{tabular}{r@{\qquad}r@{\quad}r@{\quad}r@{\quad}}
\hline
$P^-/P^+$ & $\Pnu \Pnu \Pphoton$ & $\Pnu \Pnu \Pphoton \Pphoton$ & $\Ppositron \Pelectron$ \\
\hline
\hline
$0/0$ & 2677  & 268 & 304  \\
$+0.8/-0.3$ & 421 & 42  & 304  \\
$-0.8/+0.3$ & 6217 & 622  & 304  \\
\hline
\end{tabular}
\caption{Simulated and modeled number of events in the different background
  sources after application of all selection criteria for $\sqrt{s} = \unit{1}{\TeV}$. The numbers are
  calculated for an integrated luminosity of \unit{1}{\myinvfb} in different
  polarisation settings.}
\label{tbl:neventsperscenario_1tev}
\end{table}

\begin{table}[b]
\begin{tabular}{r@{\qquad}rr@{\quad}rrrr}
\hline
Model & $P^-/P^+$ & $N_{\text{S}}$ & $r_{50 P}$ & $r_{50 \tilde{P}}$ & $r_{500 P}$ &
$r_{500 \tilde{P}}$ \\
\hline
\hline
Scalar &$+0.8$/$+0.3$   &  620  &  1.39  &  2.22  &  \textbf{1.60}  &  \textbf{3.7} \\
 &$+0.8$/$+0.6$   &  740  &  \textbf{1.39}  &  \textbf{2.36}  &  1.55  &  3.65 \\
\hline
Vector &$+0.8$/$-0.3$   &  620  &  2.21  &  2.92  &  2.95  &  6.08 \\
& $+0.8$/$-0.6$   &  740  &  \textbf{3.46}  &  \textbf{4.09}  &  \textbf{5.40}
&  \textbf{10.00} \\
\hline
Left & $-0.8$/$+0.3$   &  1170  &  1.59 &  1.88  &  2.50  &  \textbf{4.59} \\
& $-0.8$/$+0.6$   &  1440  &  \textbf{1.60}  &  \textbf{2.05}  &  2.23  &  4.47 \\
\hline
Right &$+0.8$/$-0.3$   &  1170  &  4.16  &  5.52  &  5.57  &  11.47 \\
&$+0.8$/$-0.6$   &  1440  &  \textbf{6.73}  &  \textbf{7.96}  &  \textbf{10.51}  &  \textbf{19.46} \\
\hline
\end{tabular}
\caption{Determination of the best ratio $r \equiv
  {N_{\text{S}}}/{\Delta N_{\text{B}}}$ (see
  Table~\ref{tbl:sigoverbkgestimate}) for $\sqrt{s} =
  \unit{1}{\TeV}$. }
\label{tbl:sigoverbkgestimate_1tev}
\end{table}
\subsection{Polarisation Settings}
Polarisation can be used to significantly increase the number of
signal events according to Eq.~(\ref{eqn:polclasses}) but also
increases the systematical contribution to the total background error,
$\Delta N_\text{B}$ via Eq.~(\ref{eqn:deltap}). We are interested in
the settings for each individual model that leads to the largest
$N_\text{S}/\Delta N_\text{B}$ ratio allowing for the strictest bounds
on $G_\text{eff}$. In Table~\ref{tbl:bkgevts} we give the total number
of background events in different polarisation settings $P^-$ = $\pm
0.8$ and $P^+$ = $\pm 0.3$/$\pm 0.6$ that are feasible at the
\textsc{Ilc} \cite{Phinney:2007gp}. We give the statistical
fluctuation for integrated luminosities of
\unit{50}{\femto\reciprocal\barn} as well as for
\unit{500}{\myinvfb}. Since the latter will give ten times as much events in
all channels, we reduce the statistical error accordingly to give a value
comparable to the small luminosity case.
We also give the systematic error that is dominated by the polarisation uncertainty for two estimates of the
polarisation error  $\delta P/P = \unit{0.25}{\%} $ and $
\unit{0.1}{\%}$ \cite{Helebrant:2008qz}. Finally we give the total errors adding all combinations of
individual errors in quadrature.

On the signal side, we look at the different classes derived in Sec.~\ref{sec:models} with respect to their polarisation dependence. For comparison,
we use a common reference value of 500 unpolarised events for an integrated
luminosity of $\unit{50}{\myinvfb}$ and derive the corresponding number of
events for polarised input.

We look for the maximum ratio $r \equiv N_{\text{S}} / \Delta N_{\text{B}}$ and the results for the best settings are displayed in Table~\ref{tbl:sigoverbkgestimate}. In most cases the largest possible polarisation for the incoming leptons enhances the result. For high statistics and a non--reduced polarisation error, the systematic uncertainty
from increased polarisation may outweigh the gain in the number of signal events though. In those
cases, which appear only in scalar-- and left--coupling models, less polarised
beams lead to better results.

\subsection{Increasing \boldsymbol{$\sqrt{s}$} to \unit{1}{\text{TeV}}}

We also consider the possibility of a doubled center of mass
energy. This changes the previous analysis as follows:
\begin{itemize}
\item We generate events in a larger photon energy range $E_\gamma \in
  \left[\unit{8}{\GeV}, \unit{500}{\GeV}\right]$ and reduce it to the interval $\left[
      \unit{10}{\GeV}, \unit{450}{\GeV} \right]$ after performing the energy resolution
    shift $\Delta E / E$. This again reduces background events from the
    $\PZzero$--resonance, which now is positioned at \unit{496}{\GeV}.
\item Dark matter signal processes can now be produced with masses up to \unit{490}{\GeV}.
\item We use our previously modeled distribution for the Bhabha background
  and rescale it by a factor of 1/4, taking into account that the full cross
  section for that process is approximately proportional to $1/s$. 
\item We use, as a rough approximation, the same \textsc{Isr}-- and beamstrahlung parameters in
\textsc{C}alc\textsc{hep}, efficiency factors and systematic error estimates. 
\end{itemize}

Tables~\ref{tbl:bkgevts_1tev}-\ref{tbl:sigoverbkgestimate_1tev}
summarise again the number of background events per background
scenario, the individual error sources and the determination of the
best polarisation setting for the increased center of mass energy. In
contrast to the Bhabha cross section that falls mainly according to
$\sigma \propto 1/s$, the neutrino background gets significant
contributions from t--channel $\PW^{\pm}$s, which give $s / m_W^4$
--terms in the evaluation of the total cross section. The left--handed
neutrino contribution therefore gets enhanced whereas the Bhabha
background becomes less dominant in some polarisation channels. This
leads to a larger relative polarisation error and therefore a larger
impact on the size of the background fluctuation. In the end, vector--
and right--coupling models receive stronger enhancement for polarised
input than in the $\sqrt{s} = \unit{500}{\GeV}$ case, whereas the
other models suffer from the larger impact of polarisation on the
total error and prefer smaller polarisation.

\begin{figure*}
\centering
\includegraphics[width=0.45\textwidth]{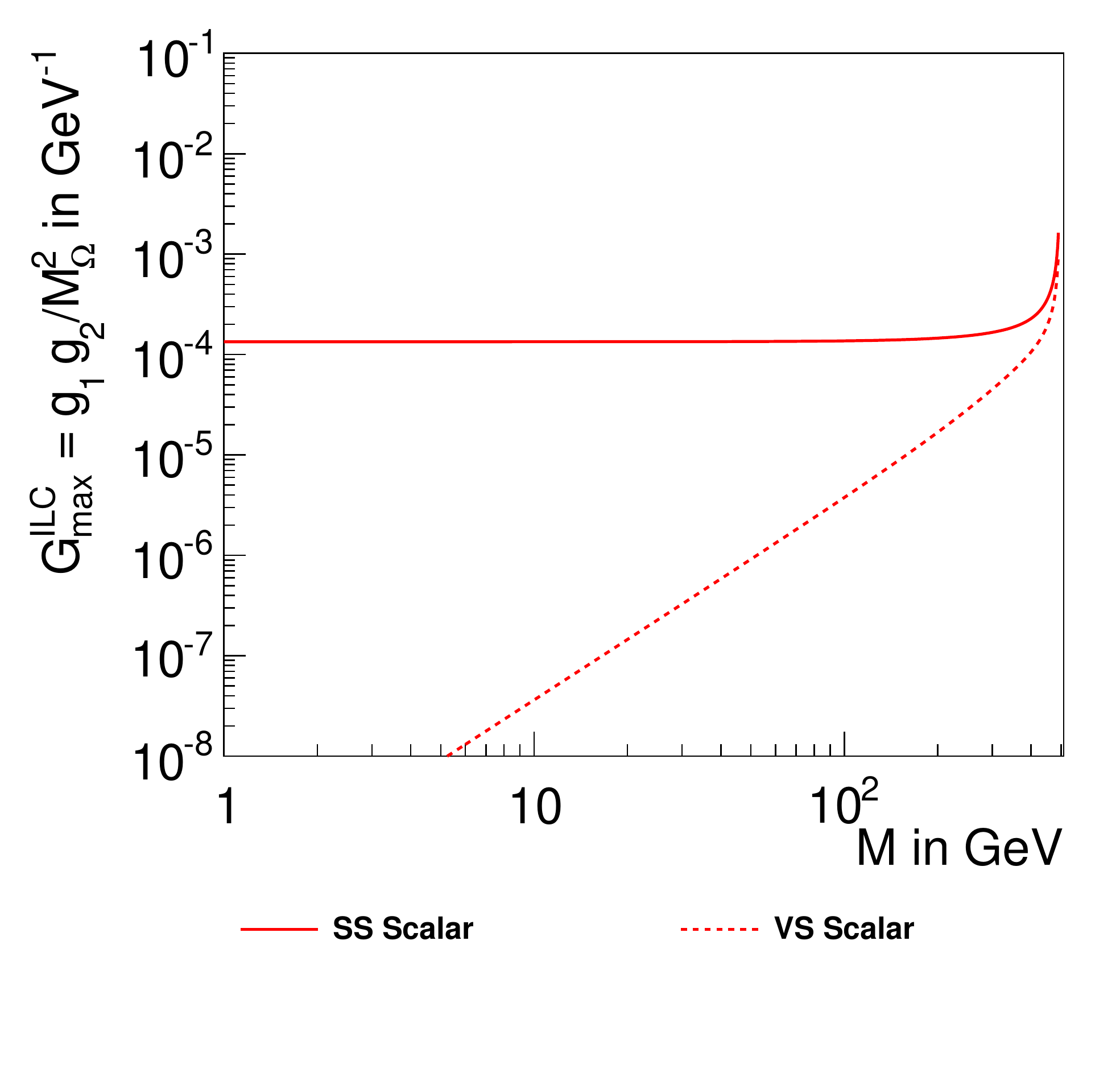} \hfill
\includegraphics[width=0.45\textwidth]{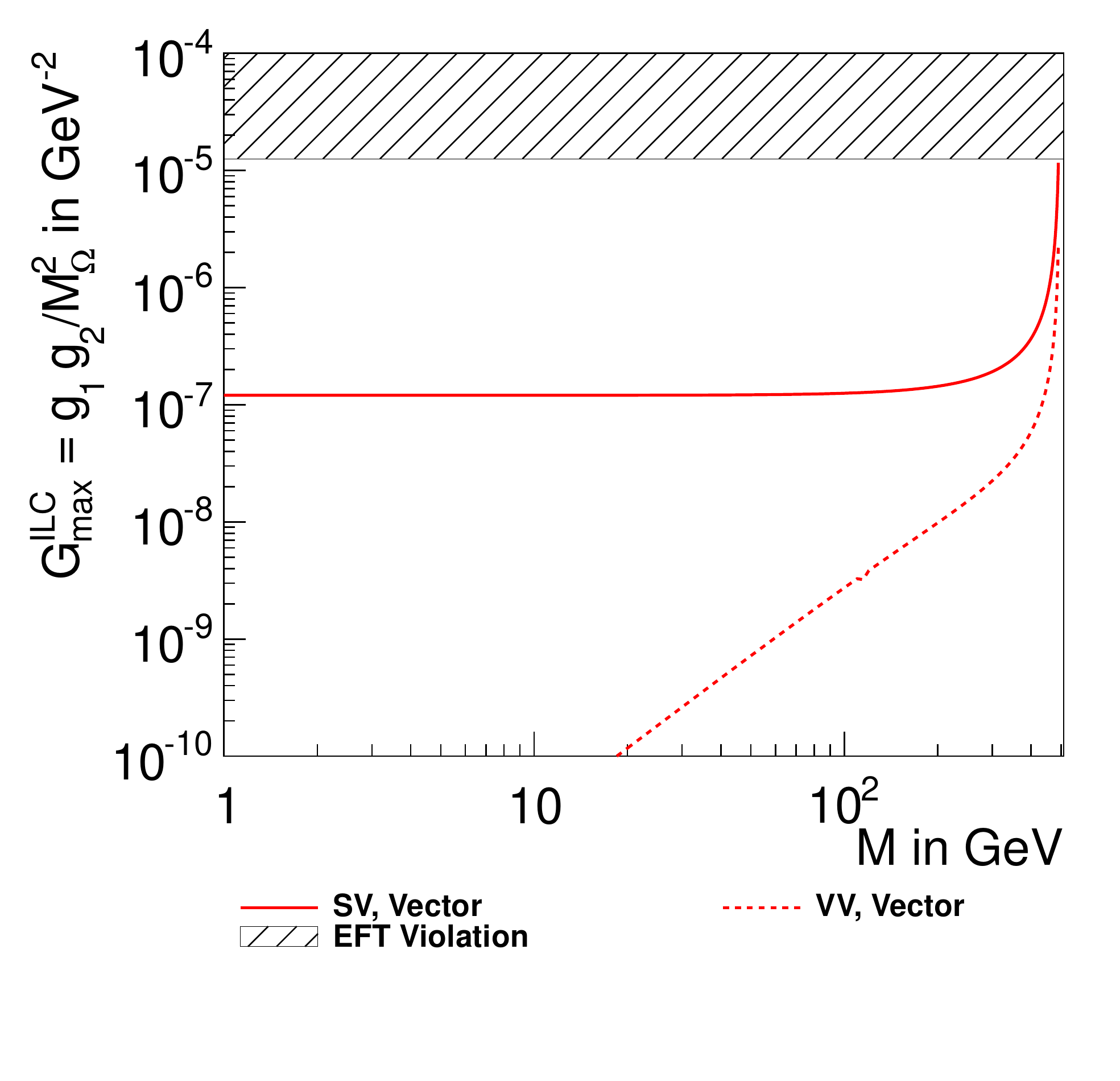} \\
\includegraphics[width=0.45\textwidth]{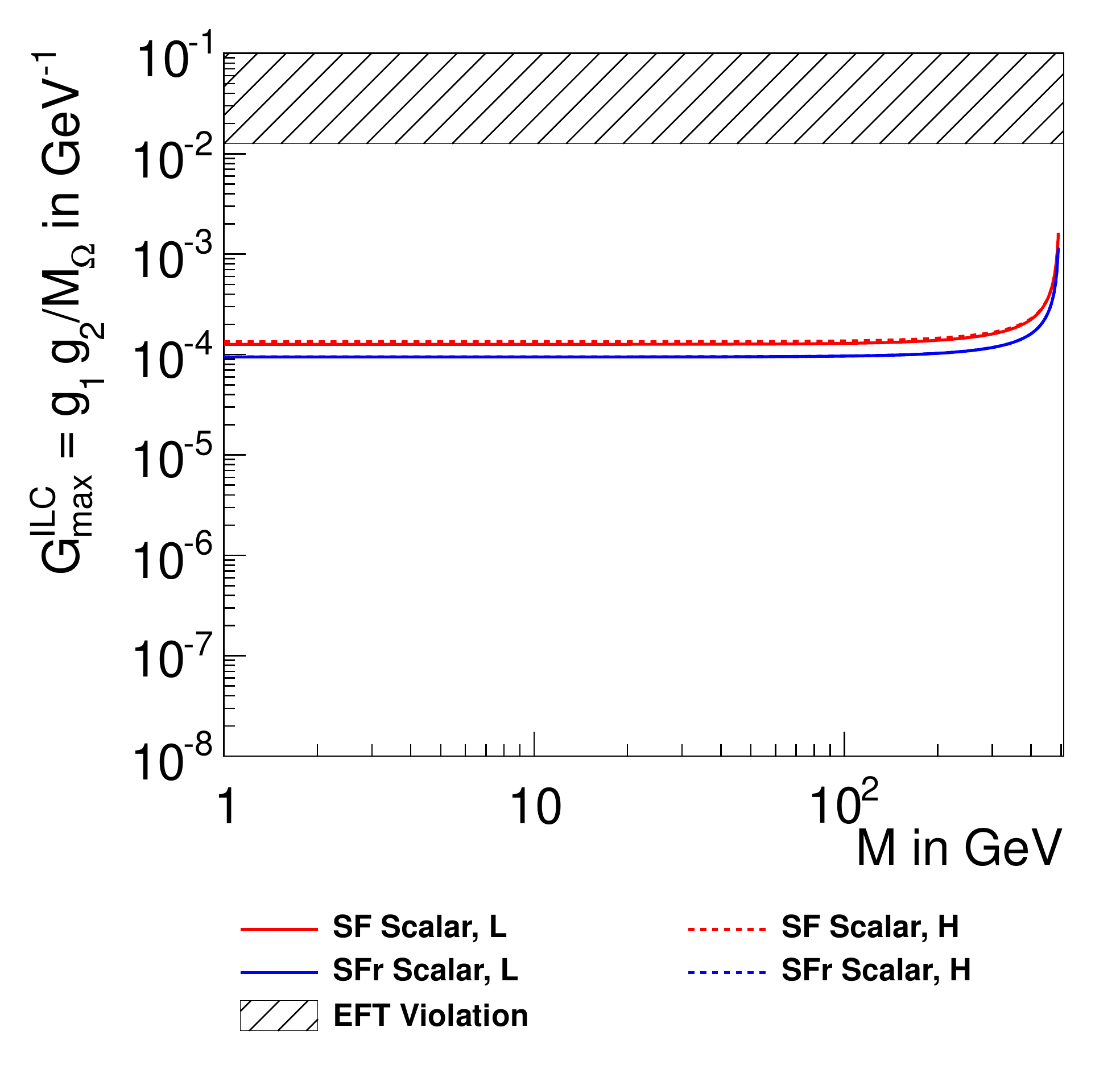} \hfill
\includegraphics[width=0.45\textwidth]{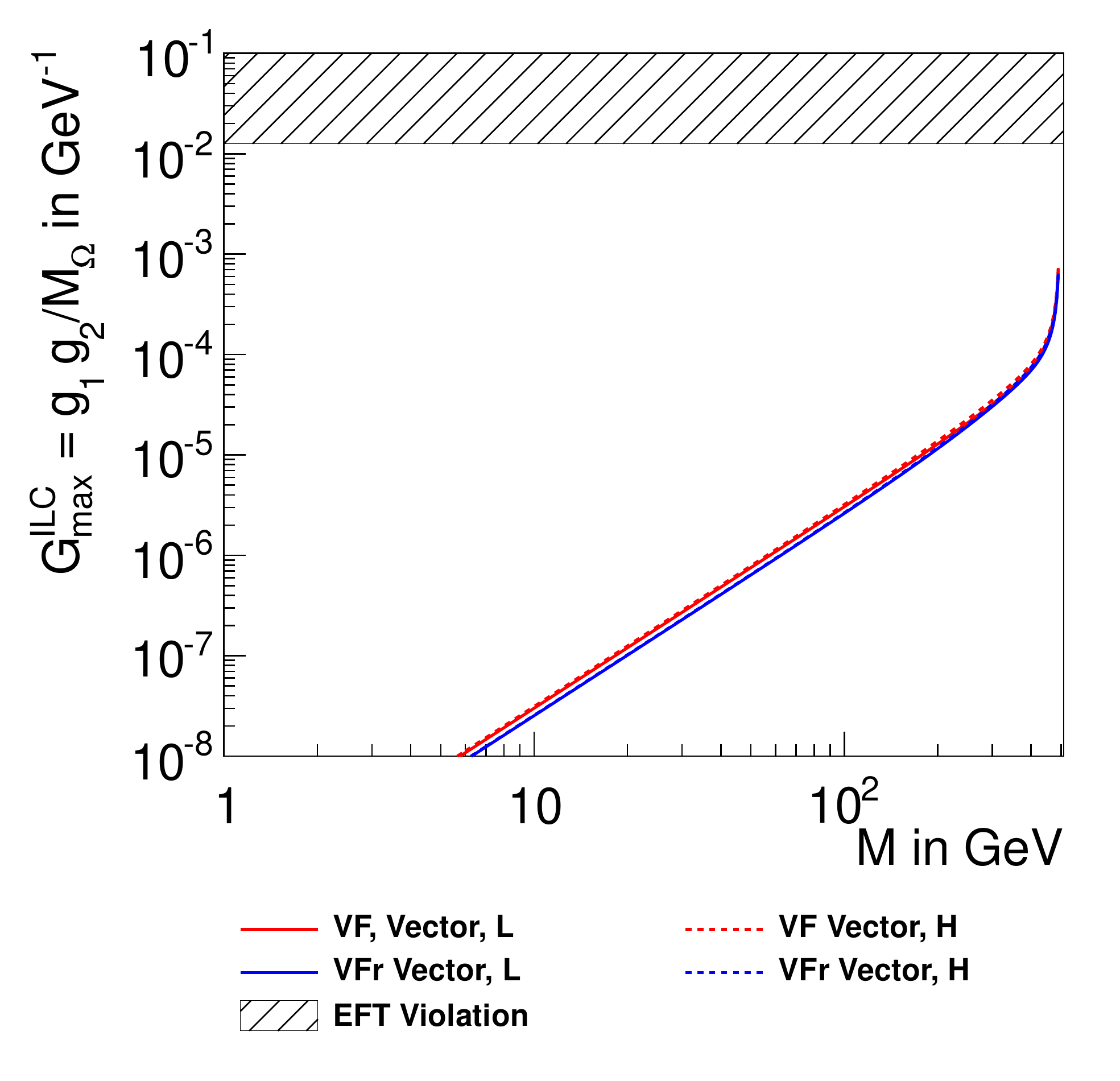} \\
\includegraphics[width=0.45\textwidth]{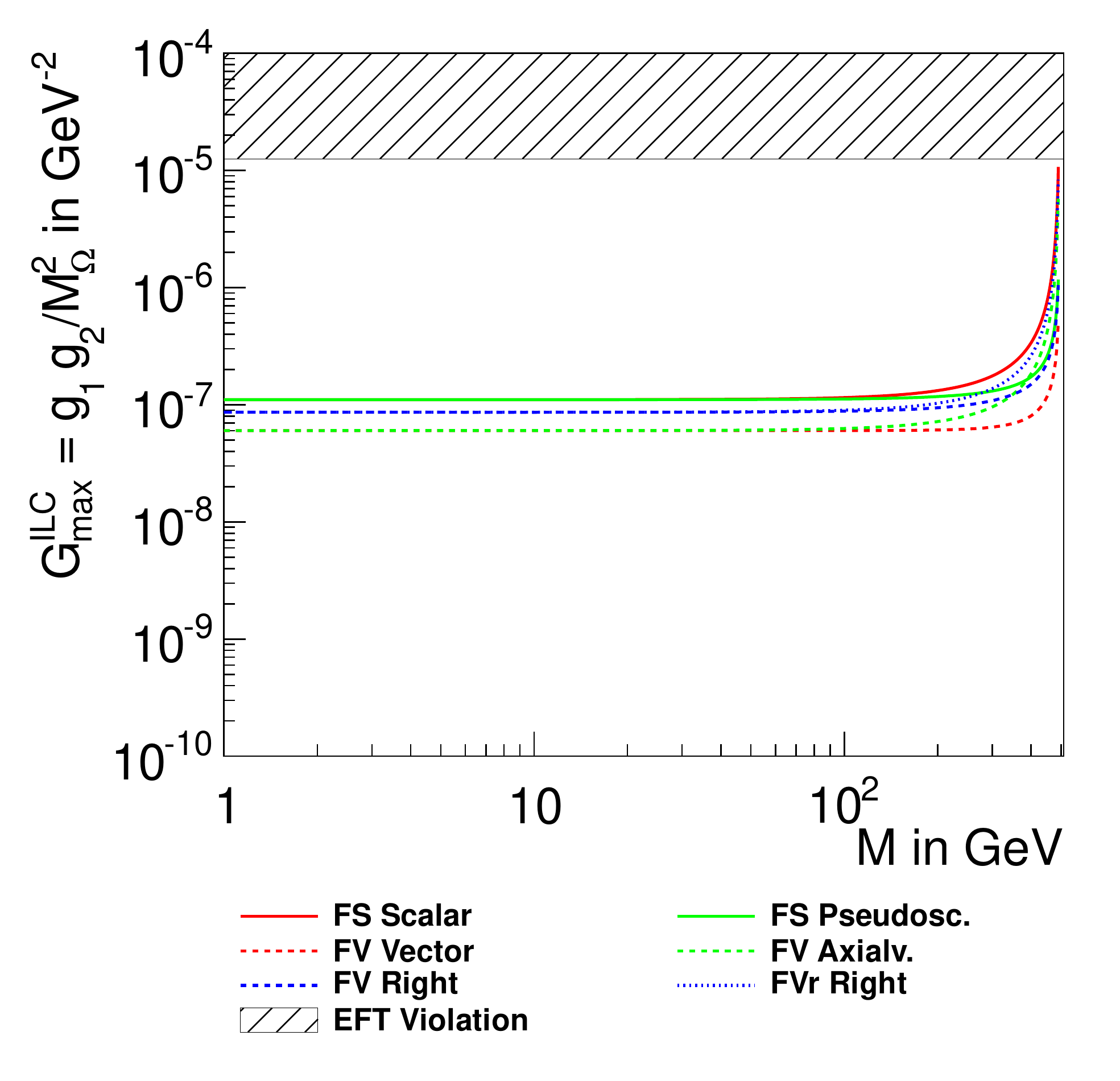} \hfill
\includegraphics[width=0.45\textwidth]{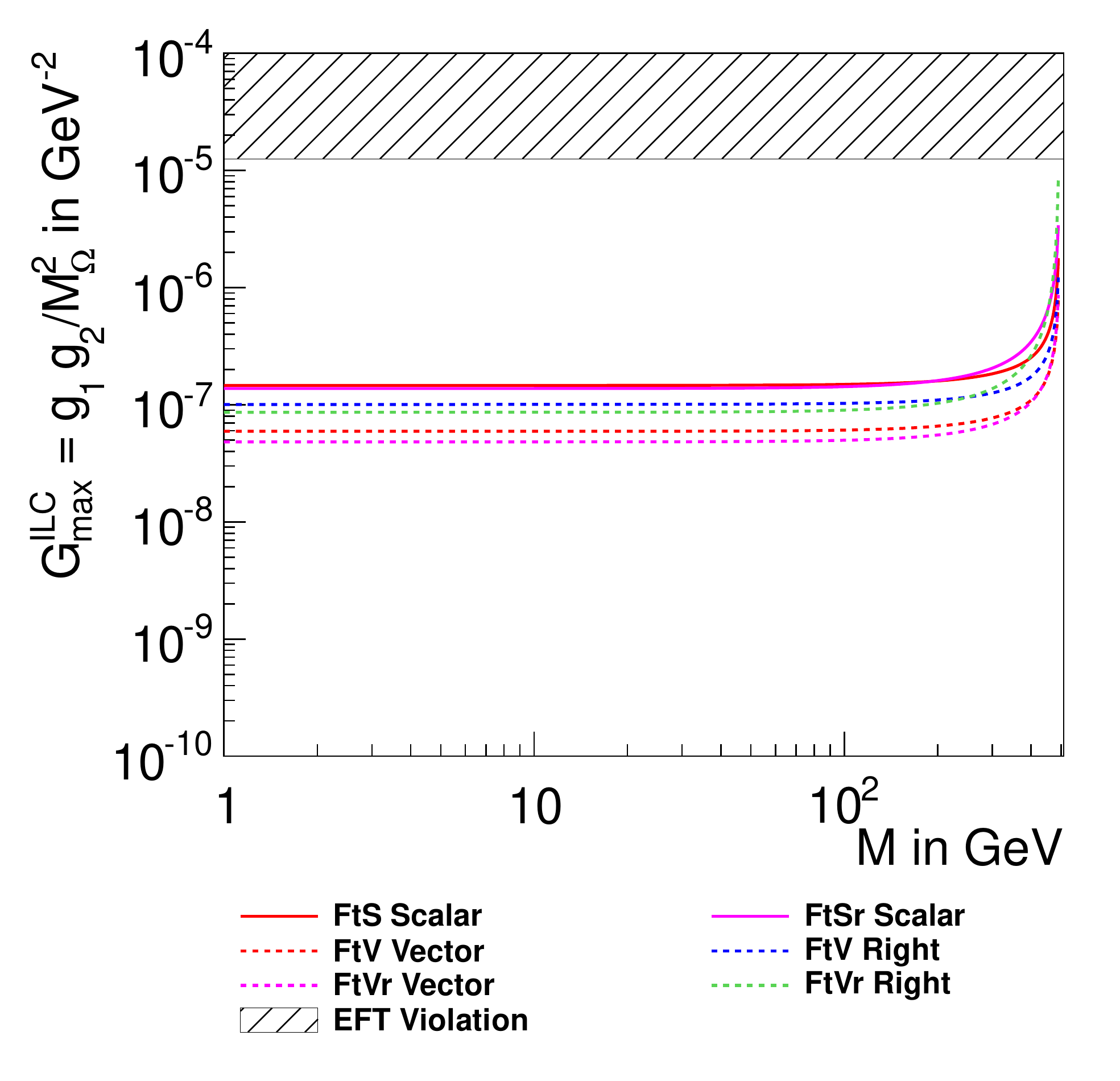} \\
\caption{\unit{90}{\%} exclusion limits on the effective couplings accessible at the \textsc{Ilc} with $\sqrt{s} =
  \unit{1}{\TeV}$. We only give effectively allowed regions for models with
  dimensionless fundamental couplings $g$. }
\label{img:ilcbounds}
\end{figure*}

%% file: conclusions.tex
\section{Results}
\label{sec:results}
We begin by presenting the reach at the \textsc{Ilc} in terms of the
effective coupling constant in Sec.~\ref{sec:ilc_bounds}. We then
compare these potential bounds with the couplings
predicted by the cosmological relic density and the bounds coming from
direct and indirect detection experiments. Of course we would also like to discover a dark matter at the \textsc{Ilc} and the bounds provide an estimate of the potential sensitivity of the collider. 

\subsection{ILC Bounds}
\label{sec:ilc_bounds}
We determine the \unit{90}{\%} exclusion bound for the effective
coupling constant in each benchmark model for the best case
scenario. The integrated luminosity is set to \unit{500}{\myinvfb} and
the systematic polarisation error to $\Delta P/P =
\unit{0.1}{\%}$. For each benchmark model we choose the polarisation
setting that leads to the best signal to background ratio for the
corresponding polarisation behaviour according to
Tables~\ref{tbl:sigoverbkgestimate} and
\ref{tbl:sigoverbkgestimate_1tev}. Results for different polarisation
settings can be found by rescaling the bound on the coupling according
to $G^\prime = G \sqrt{r^\prime / r}$ with $r$ denoting the ratio
$N_\text{S} / \Delta N_\text{B}$ given in
Table~\ref{tbl:sigoverbkgestimate_1tev}. We choose to present all of
the results for an \textsc{Ilc} with a center of mass energy of
\unit{1}{\TeV} due to the increased range of dark matter masses that
this option can probe. In addition, smaller effective couplings can be
probed, mainly due to the falling Bhabha background.

In Fig.~\ref{img:ilcbounds} we show the derived bounds on the coupling
constants for an \textsc{Ilc} center of mass energy of
\unit{1}{\TeV}. The hashed area denotes the region that either
violates the tree level approach with a too large dimensionless
coupling constant $g^2 > 4 \pi$, or by having a too small mediator
mass $M_\Omega < \unit{1}{\TeV}$, for the effective approach to be
valid. Note that the leading order in models with fermionic mediators
has a different mass dimension and therefore gives a different
definition for the effective coupling constant $G_\text{eff}$. If a
model has no separate `pseudoscalar' or `axialvector' results, it is
identical to the corresponding `scalar'/ `vector' line due to
identical cross section formulas. For masses away from the threshold,
the \textsc{Ilc} is able to exclude coupling constants down to the order of \unit{\power{10}{-7}}{\GeV \rpsquared}
or \unit{\power{10}{-4}}{\reciprocal\GeV}, depending on the mass
dimension. This corresponds to a total cross section (for the given
phase space criteria) of about \unit{0.3}{\femto\barn}. Exceptions
however arise for models with vector dark matter that tend to have
very strong exclusion limits for small masses. This is caused by the
$1/M_\chi^4$ dependence in the photon cross section, which leads to
divergences for very small vector boson masses. It has been shown
\cite{Cornwall:1974km} that only spontaneously broken gauge theories
can lead to models with massive vector particles that are not
divergent. Therefore, our initial fundamental model cannot be the full
theory for all energies. In our effective approach, we restrict the
energy to a maximum and in that case one can still receive
perturbative valid results for mass ranges that do not violate unitary
bounds. However, the perturbatively allowed mass range cannot be given
in this model independent approach, since such an analysis needs more
information about the size of the individual couplings and the
relation between the mass of the mediator and the dark matter mass
itself. In summary, a more detailed fundamental theory is needed to
evaluate the breakdown of perturbation theory in this scenario.

We note that in models with fermionic operators, the sub-leading
order has a negligible effect, as can be seen from the nearly identical lines
for fermionic mediators with different masses. 

\subsection{Combined Results}
The combined maximum
exclusion limits for spin independent DM--proton interaction at \textsc{Pamela}, \textsc{Wmap}
and the \textsc{Ilc} are shown in
Figs.~\ref{img:totalbounds1}-\ref{img:totalbounds3}. We choose a subset of
models that couple to all Standard Model fermions and give an overview
of the bounds that we can expect. Other models behave similarly and are
therefore not shown again separately. We can give the following statements about
the comparison of the \textsc{Ilc} exclusion bound with the current \textsc{Xenon} limits:
\begin{itemize}
\item We have sensitivity to spin independent proton cross sections for, as an example, the
  FV Vector model down to \unit{\power{10}{-42}}{\cm\rpsquared} or
  equivalently \unit{\power{10}{-3}}{\femto\barn}, which
  is an improvement of about four orders of magnitude compared to
  current \textsc{Lep} \cite{Fox:2011fx} and two orders of magnitude
  compared to current Tevatron \cite{Bai:2010hh} and \textsc{Cms}
  \cite{Chatrchyan:2012pa} results.
\item An increased center of mass energy can lead to stronger bounds
  by up to one order of magnitude. It also allows a larger dark matter
  mass range to be probed. 
\item \textsc{Ilc} bounds get significantly weakened if the interaction is
  Yukawa--like. At the \textsc{Ilc} the mediator must couple to
    electrons, which have a suppressed Yukawa coupling. The production cross section
is thus small, leading to weaker bounds.
\item Models with scalar mediators give weaker bounds than models with
  vector interactions. For fermionic dark matter we observe a
  difference of about two orders of magnitude, which is in agreement
  with previously mentioned results from e.g.\
  \textsc{Lep}. For scalar and vector dark matter the difference is
  mass--dependent and can increase to up to six orders of magnitude,
  which is due to the different mass dimension of the
  couplings.
\item The \textsc{Wmap} bounds are for many effective models very
  constraining, Figs.\ref{img:totalbounds1}--\ref{img:totalbounds4}. However, we would like to point out that these can be
  highly dependent on the full theory whilst not affecting the
  \textsc{Ilc} or direct detection phenomenology. For example,
  annihilation can occur via some resonance or as in some \textsc{Susy} models,
  co-annihilation with staus or stops.
\end{itemize}

In Fig.~\ref{img:totalbounds4} we show some models which allow for
lepton couplings only. In that case, dark matter can only interact
with protons via photons through a fermion loop, \textit{cf.}
Appendix.~\ref{app:PhotonLoop}. The loop factor significantly lowers
the cross section and therefore increases the bound in the case of
vector coupled models. Other models allow quark couplings only at the
two--loop level or theoretically completely forbid them
\cite{Fox:2011fx}. In all cases, the \textsc{Ilc} would give the
strongest exclusion bounds for dark matter lepton couplings.
For models with fermionic mediators there is an extra subtlety when
comparing the bounds. In particular the exclusion limit at the
\textsc{Ilc} is mainly given by the leading term in the operator
expansion, which is scalar like. Loop couplings can only happen for
vector currents, which in the case of a fermionic mediator is only
given by the sub-leading order and has an additional factor of
$1/M_\Omega^2$. In that case, when translating any exclusion limits
into bounds on the {\textsc{Wimp}--proton cross section, we need to know
the exact mass of the mediator. We show this in
Fig.~\ref{img:totalbounds4} for the two different chosen suppression
scales `Low' ($M_{\Omega} =$\unit{1}{\TeV}) and `High' ($M_{\Omega} =$\unit{10}{\TeV}), Table~\ref{tbl:constraints}.

In Fig.~\ref{img:totalbounds5} we show the exclusion limits for
the spin--dependent interaction. In our
case, only the model with fermionic dark matter, a vector mediator and
an axial--vector coupling leads to such an interaction. In that case,
we compare with data from the previous \textsc{Xenon} experiment
(\textsc{Xenon}10), since no results for the \textsc{Xenon}100 phase
were available when this study was completed. Since in this scenario
dark matter only couples to a single nucleon on average because of the
natural spin anti--alignment in nuclei, the \textsc{Xenon} bounds are
not coherently enhanced by the atomic number and therefore strongly
lose sensitivity. The \textsc{Ilc} would also give strongest exclusion
bounds over the whole accessible mass range here.
\onecolumngrid
\twocolumngrid
\begin{figure*}
\centering
\includegraphics[width=0.45\textwidth]{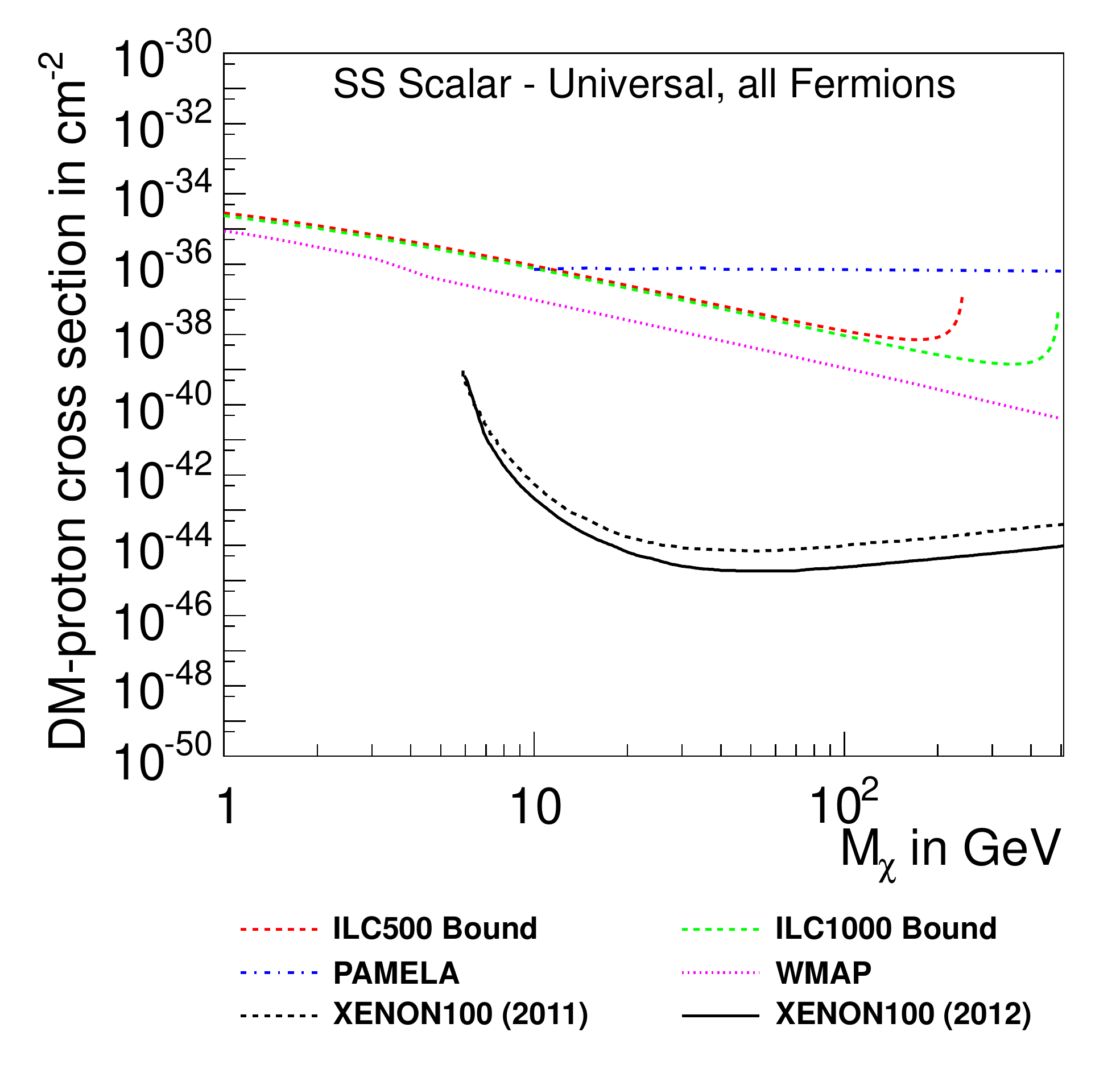} \hfill
\includegraphics[width=0.45\textwidth]{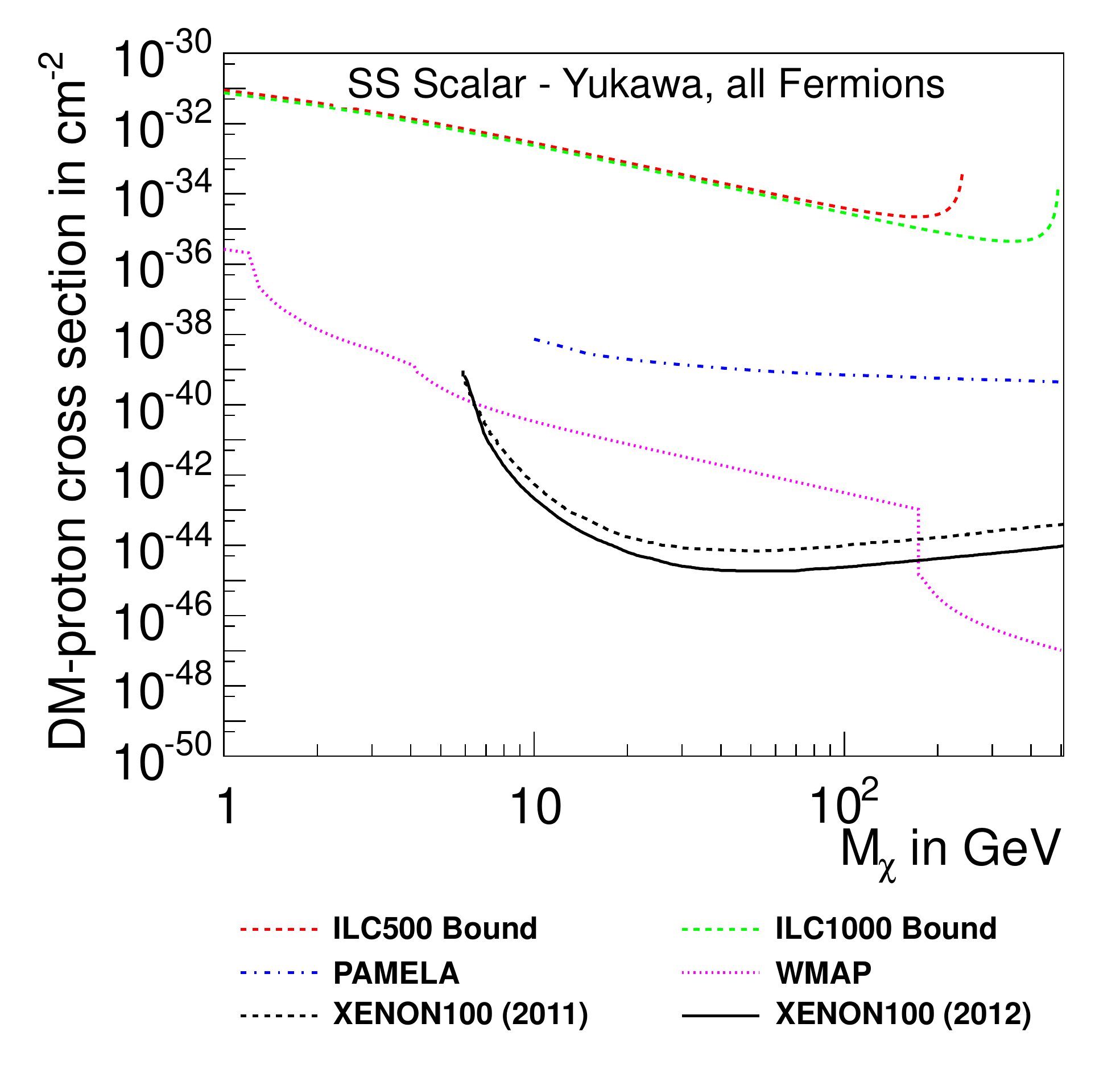}\\
\includegraphics[width=0.45\textwidth]{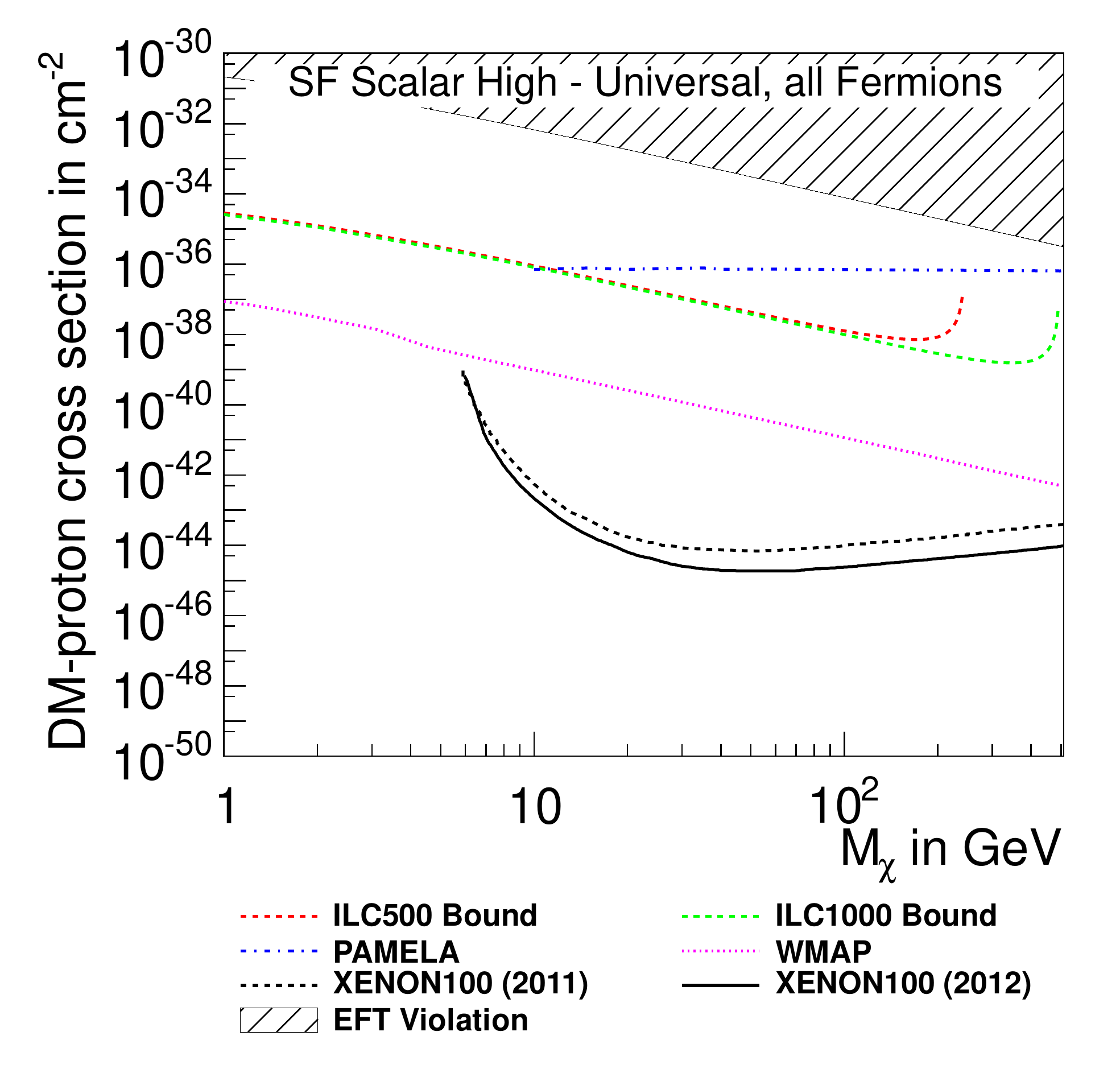} \hfill
\includegraphics[width=0.45\textwidth]{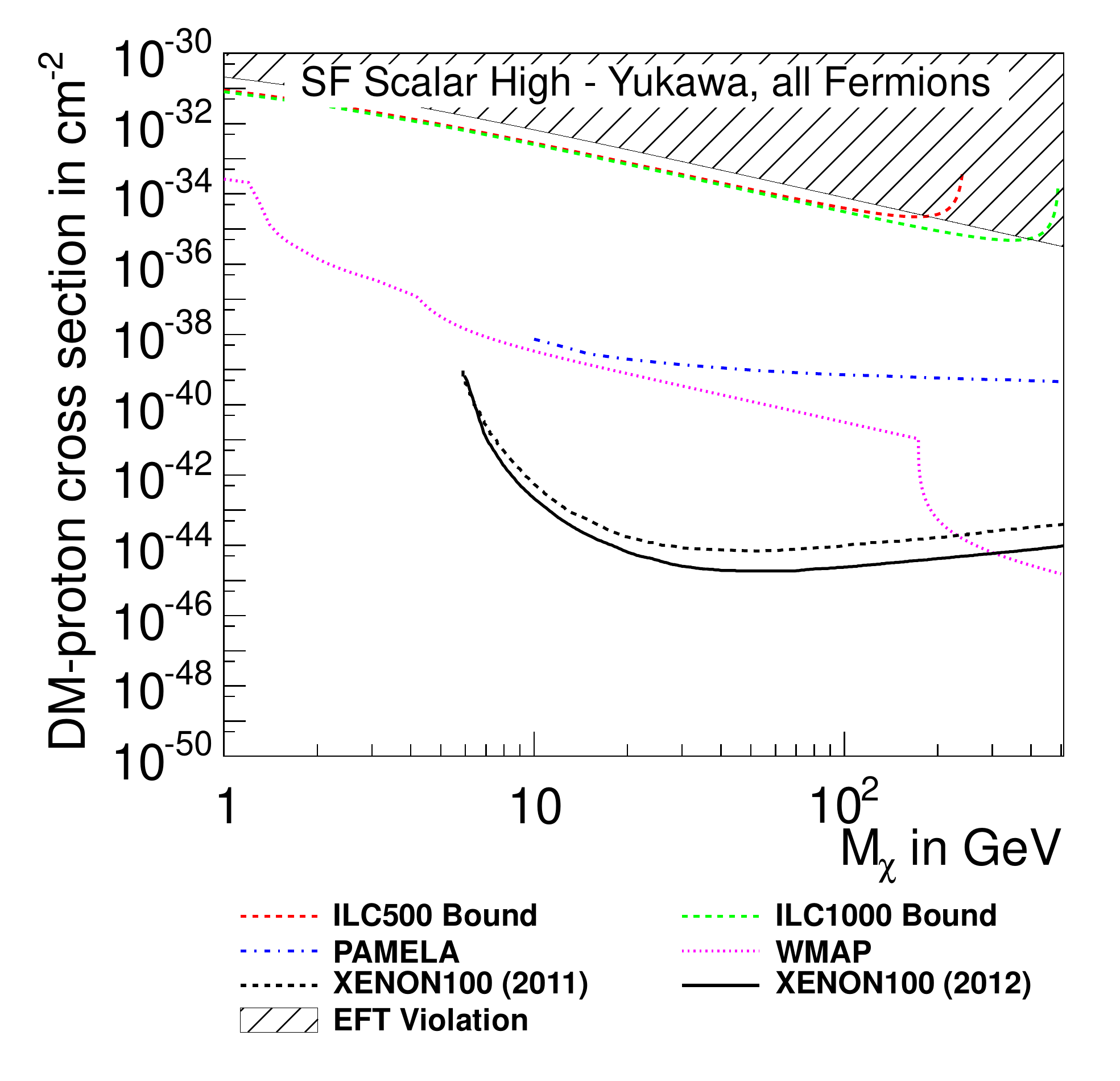}\\
\includegraphics[width=0.45\textwidth]{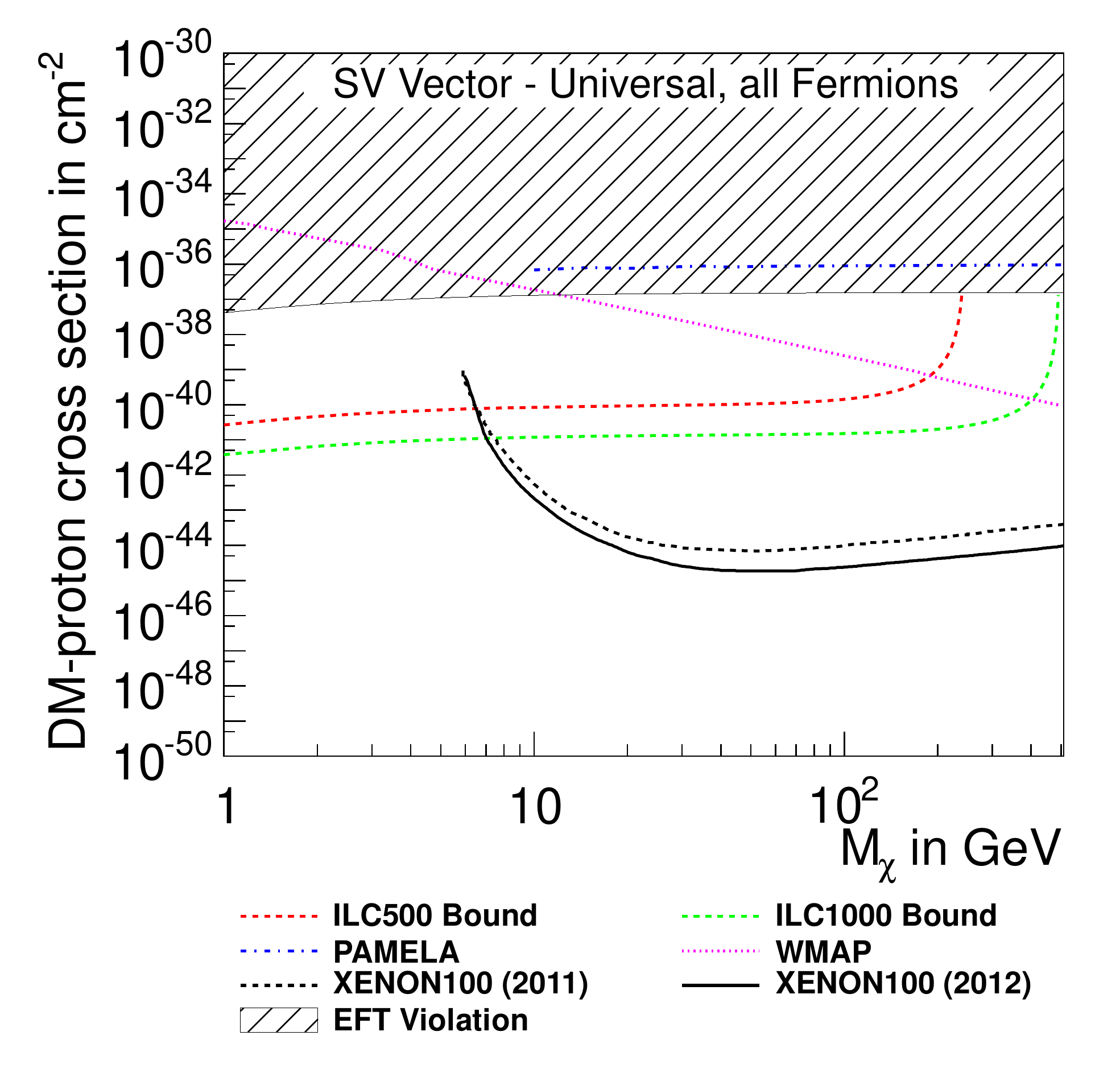} \hfill
\includegraphics[width=0.45\textwidth]{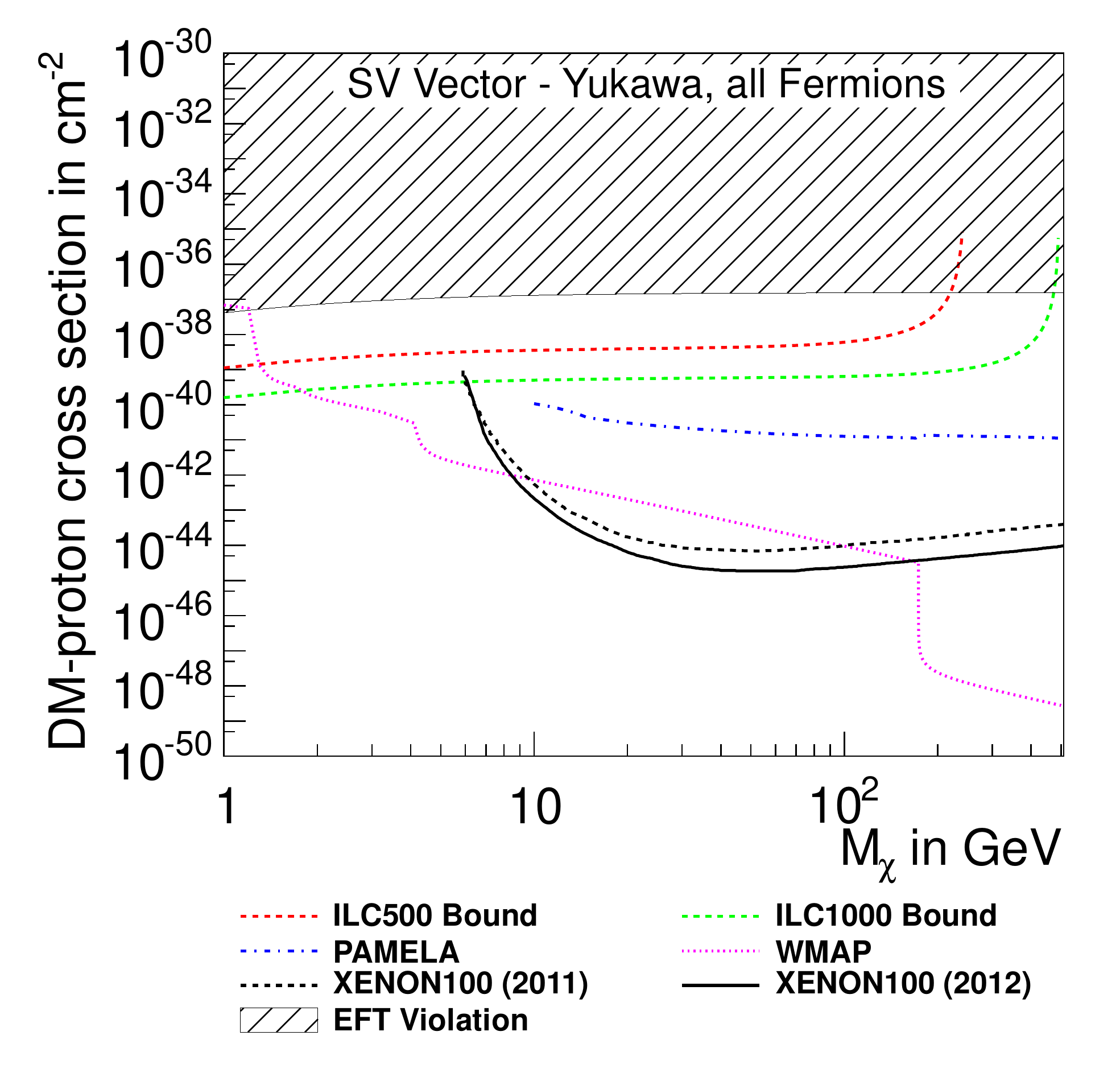}
\caption{Combined \unit{90}{\%} exclusion limits on the spin independent dark matter proton cross
  section from \textsc{Ilc}, \textsc{Pamela} and \textsc{Wmap} for a selection of scalar dark matter models.}
\label{img:totalbounds1}
\end{figure*}

\begin{figure*}
\centering
\includegraphics[width=0.45\textwidth]{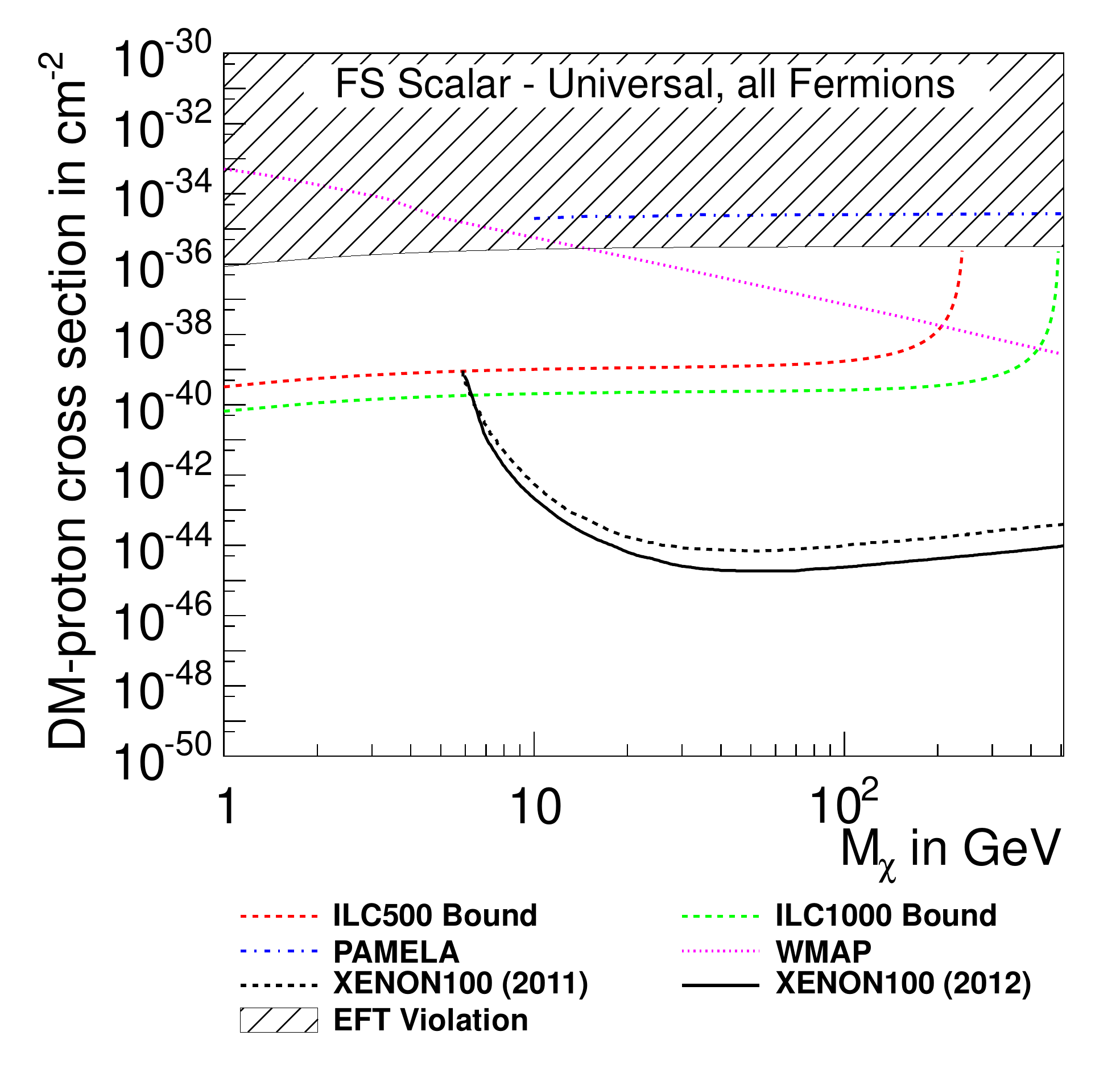} \hfill
\includegraphics[width=0.45\textwidth]{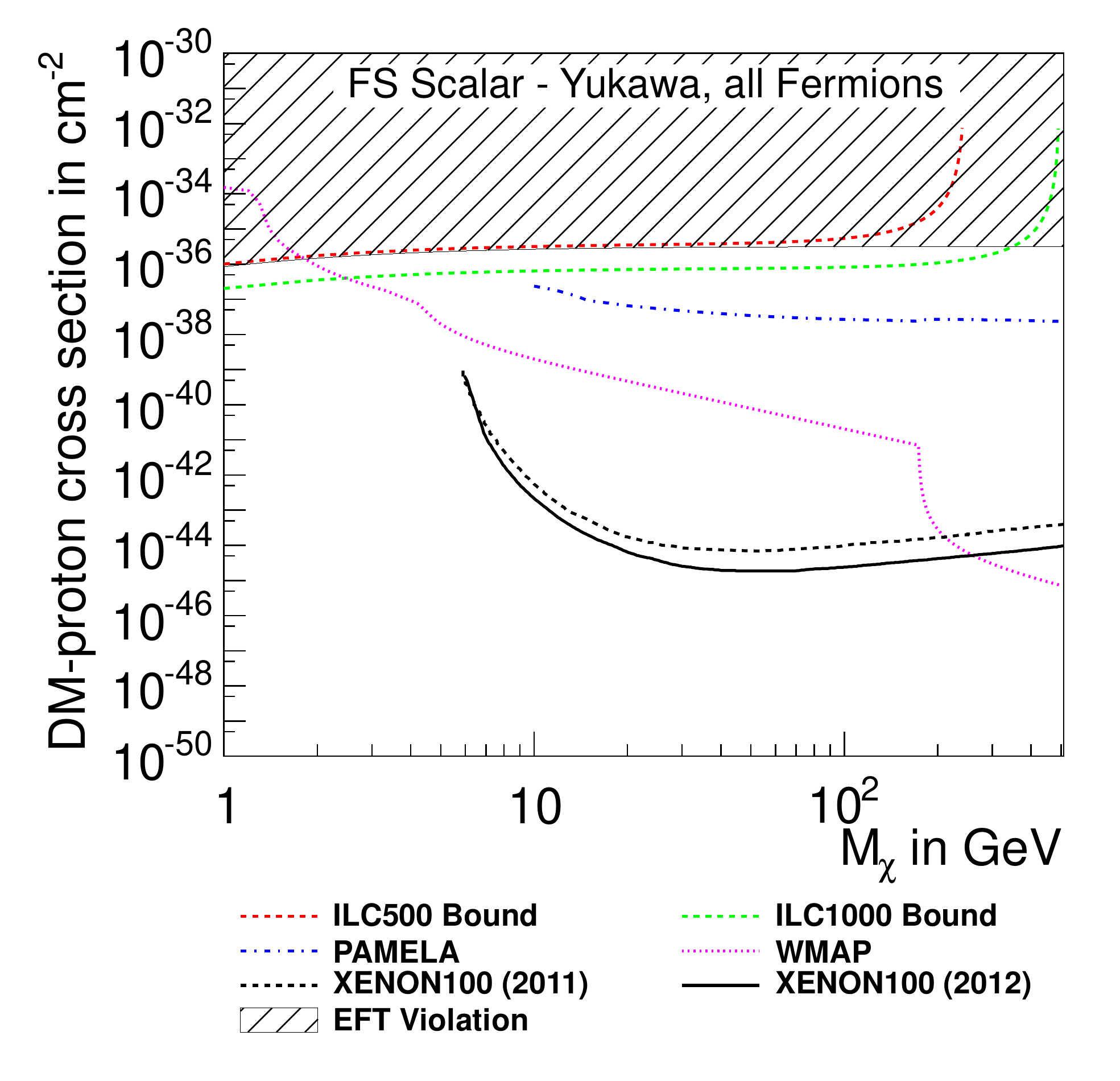}\\
\includegraphics[width=0.45\textwidth]{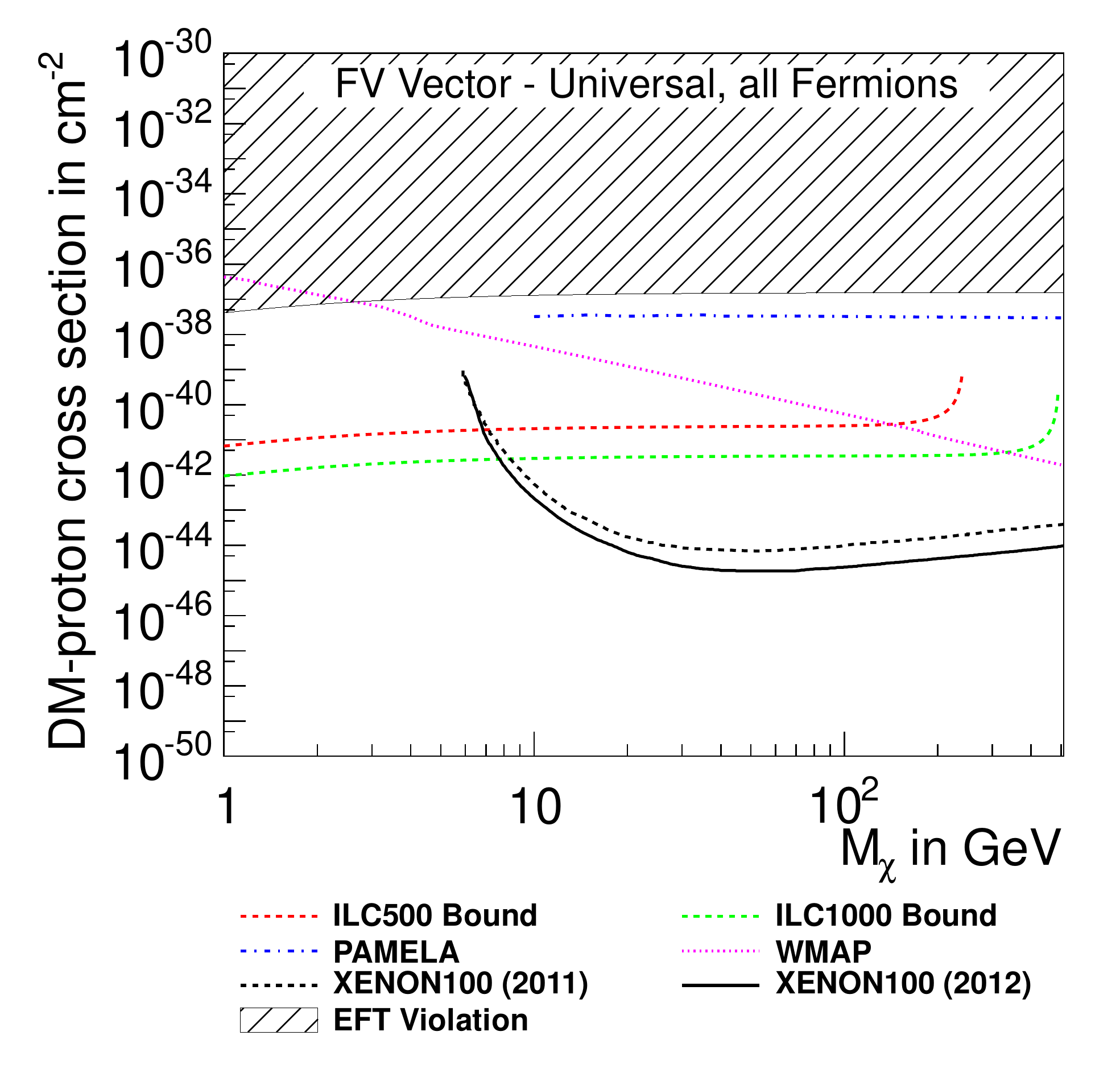} \hfill
\includegraphics[width=0.45\textwidth]{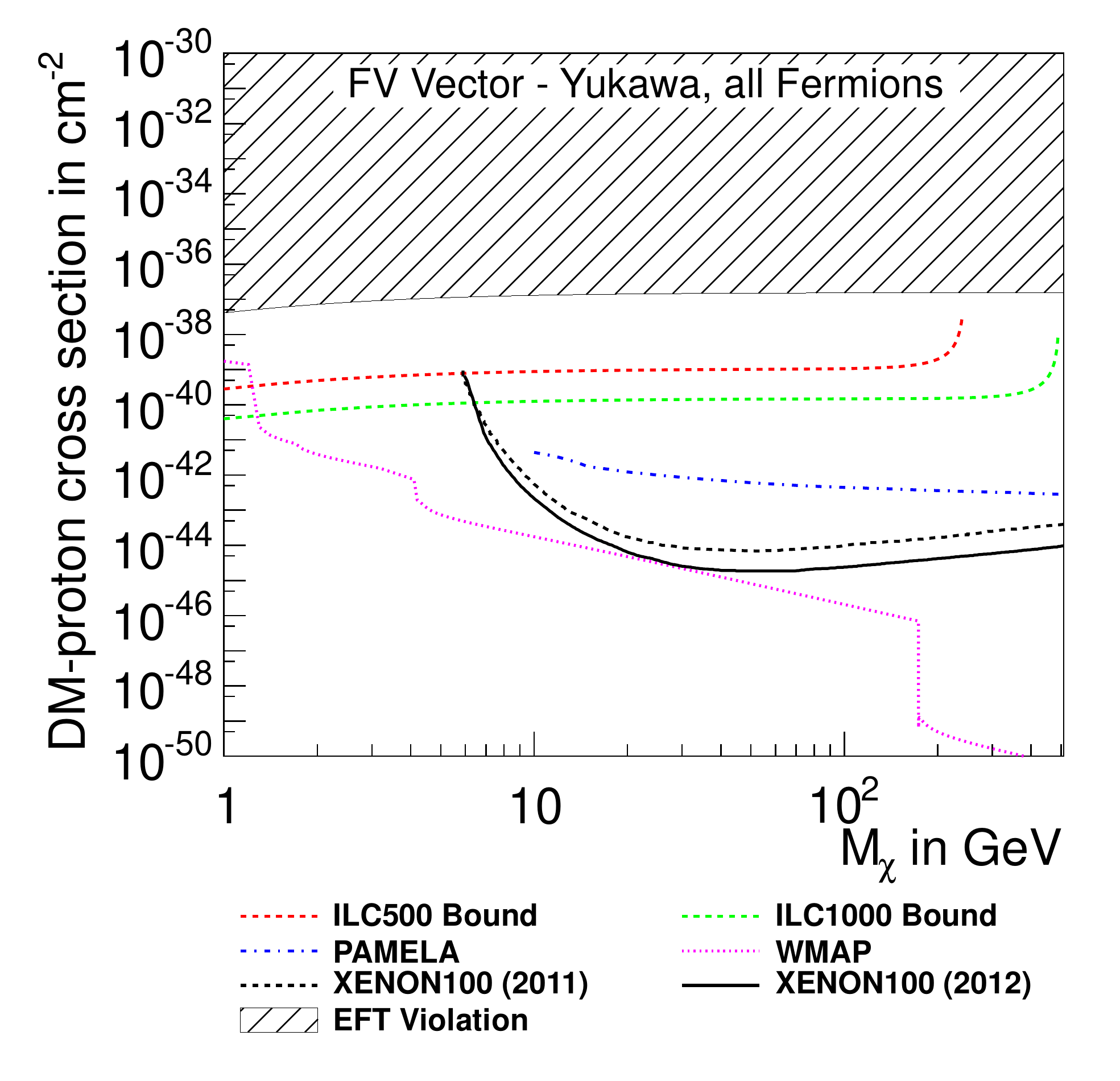}\\
\includegraphics[width=0.45\textwidth]{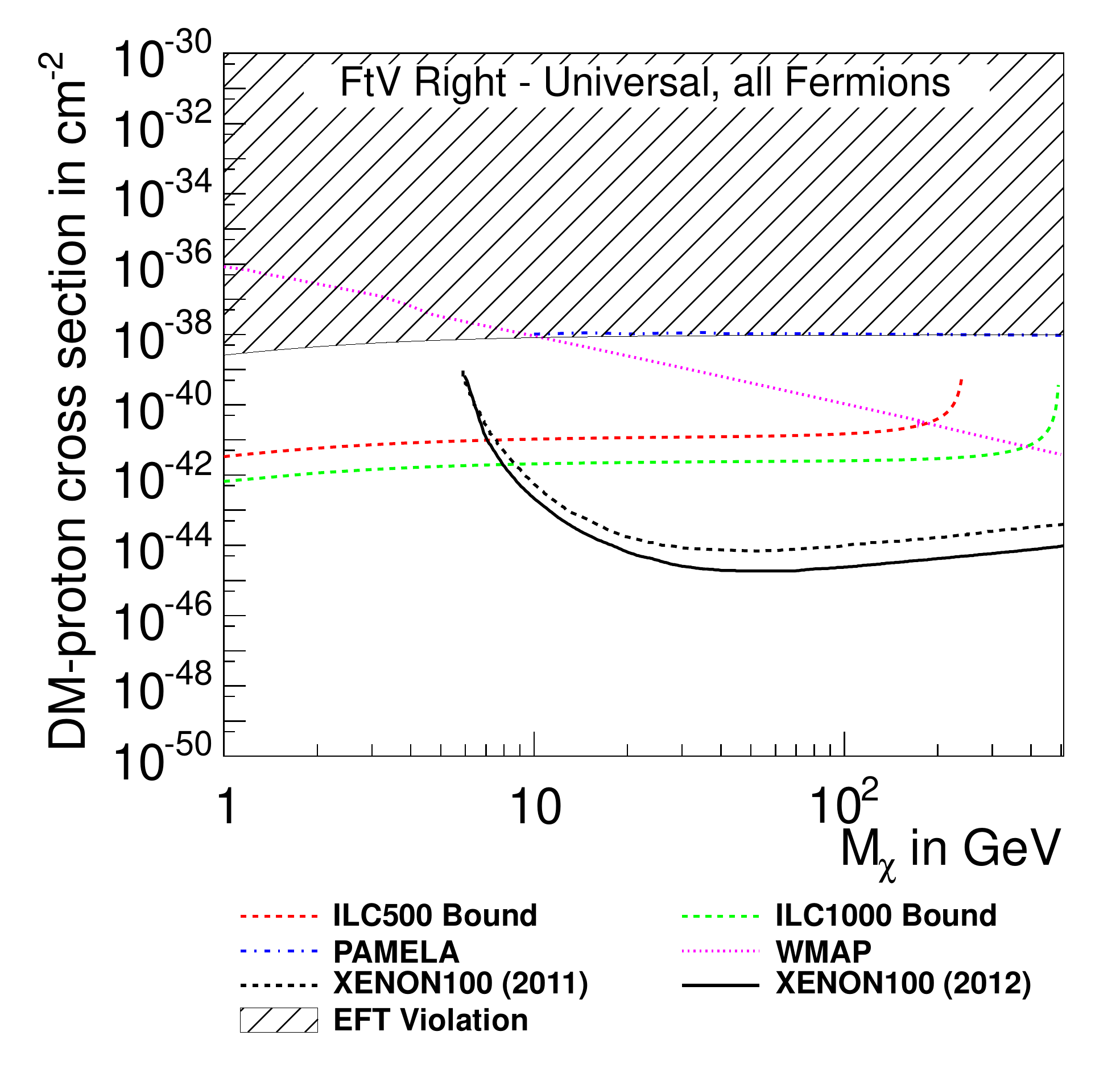} \hfill
\includegraphics[width=0.45\textwidth]{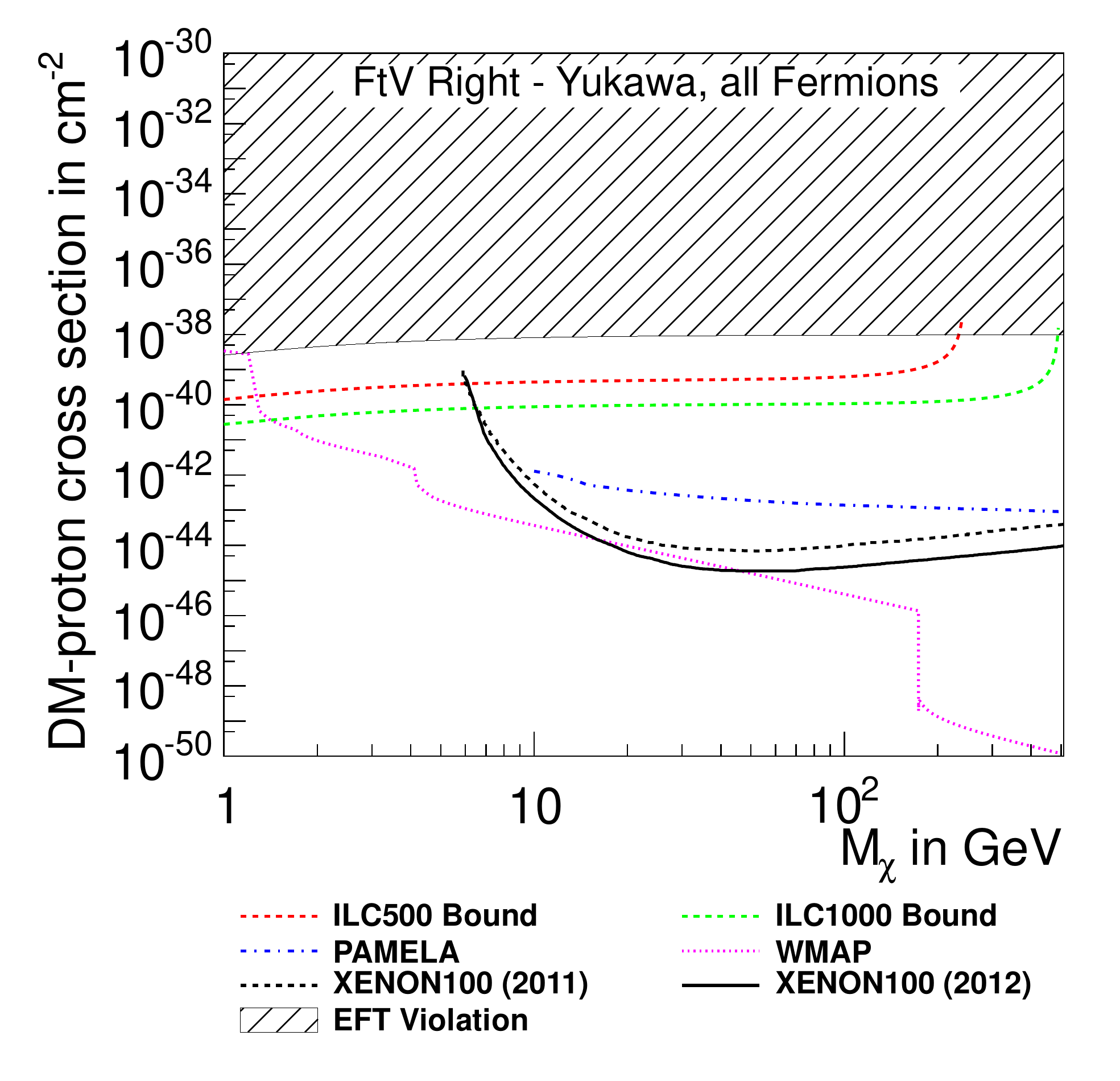}
\caption{Combined \unit{90}{\%} exclusion limits on the spin independent dark matter proton cross
  section from \textsc{Ilc}, \textsc{Pamela} and \textsc{Wmap} for a selection of fermionic dark matter models.}
\label{img:totalbounds2}
\end{figure*}

\begin{figure*}
\centering
\includegraphics[width=0.45\textwidth]{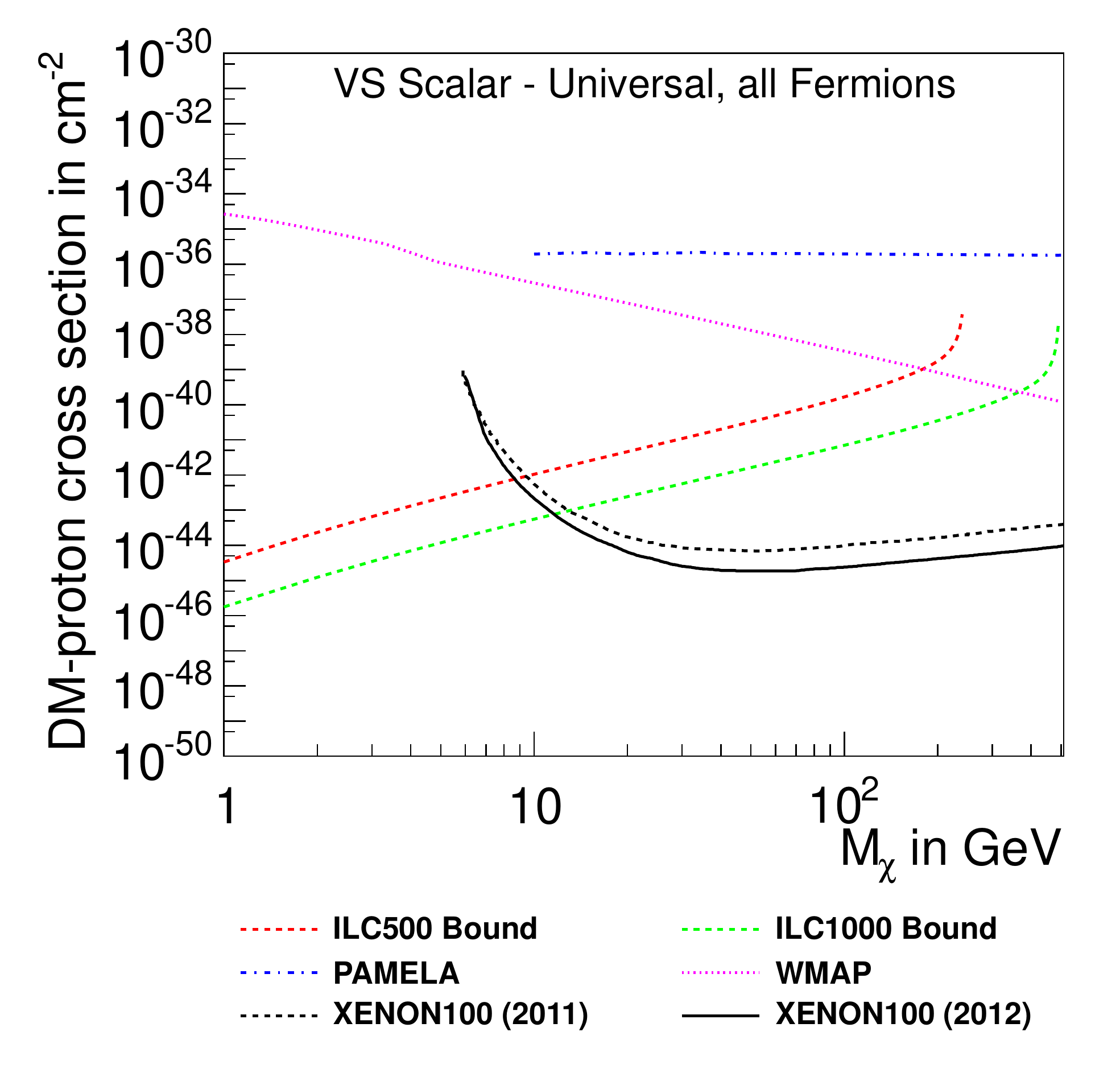} \hfill
\includegraphics[width=0.45\textwidth]{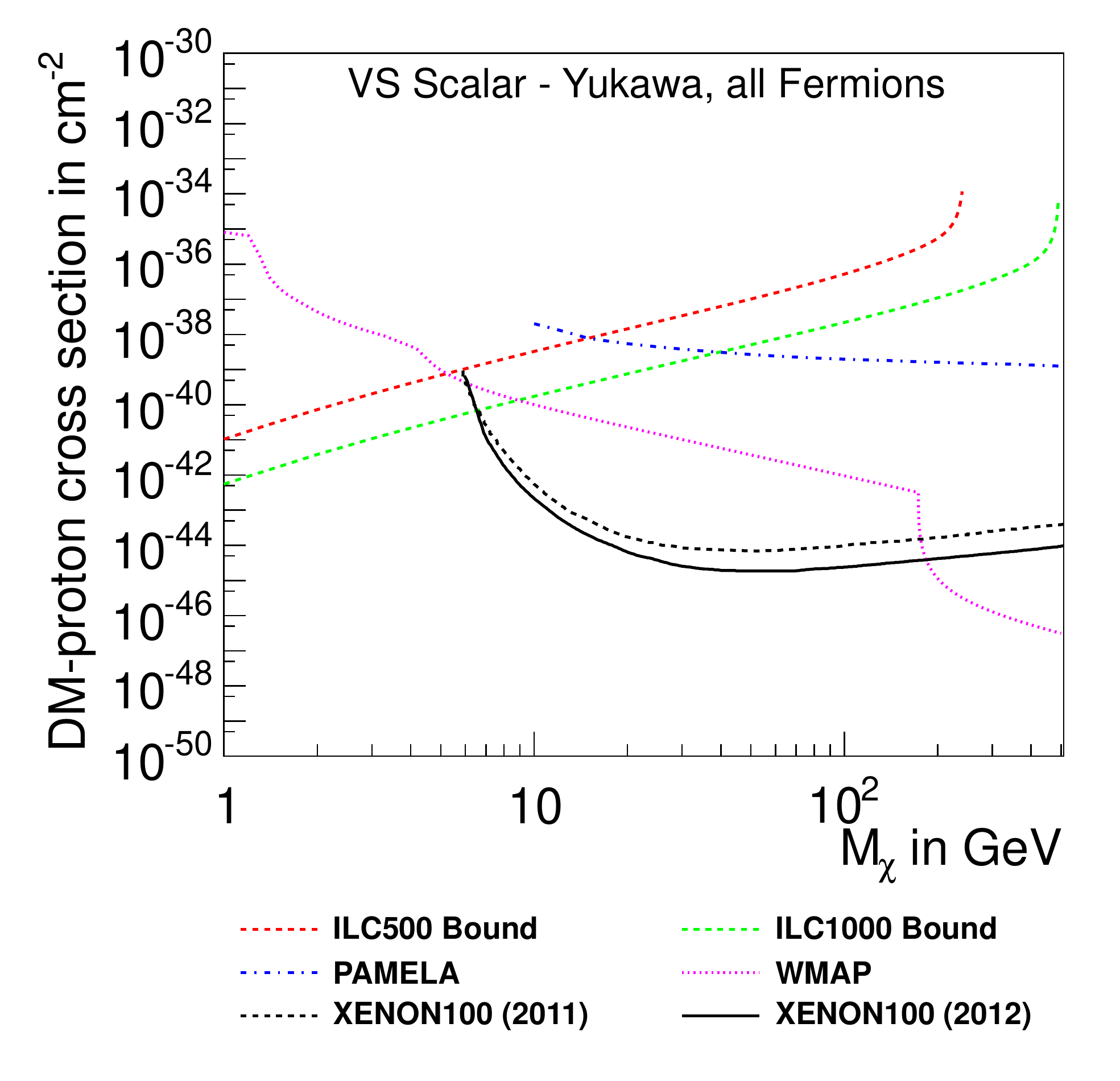}\\
\includegraphics[width=0.45\textwidth]{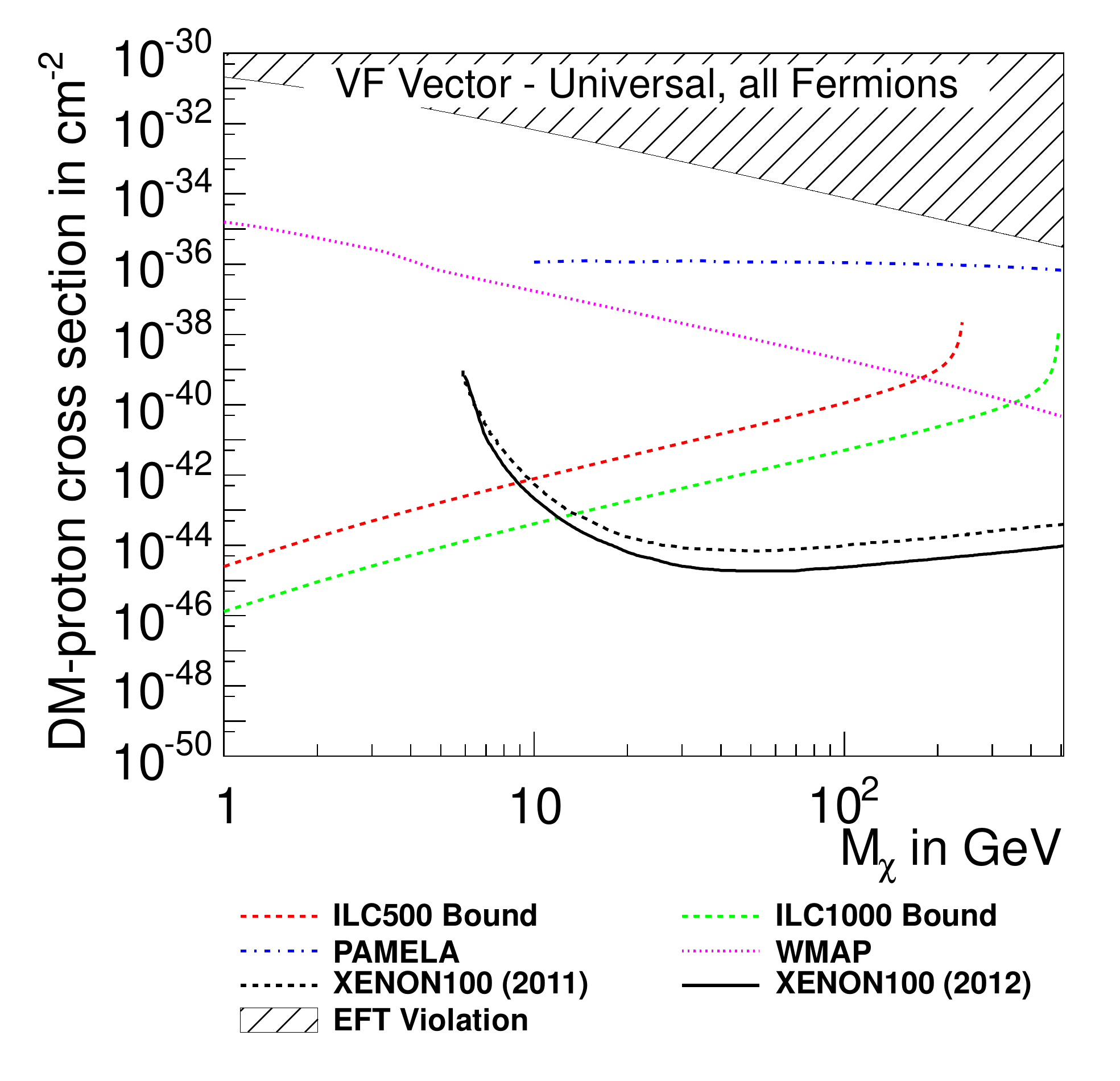} \hfill
\includegraphics[width=0.45\textwidth]{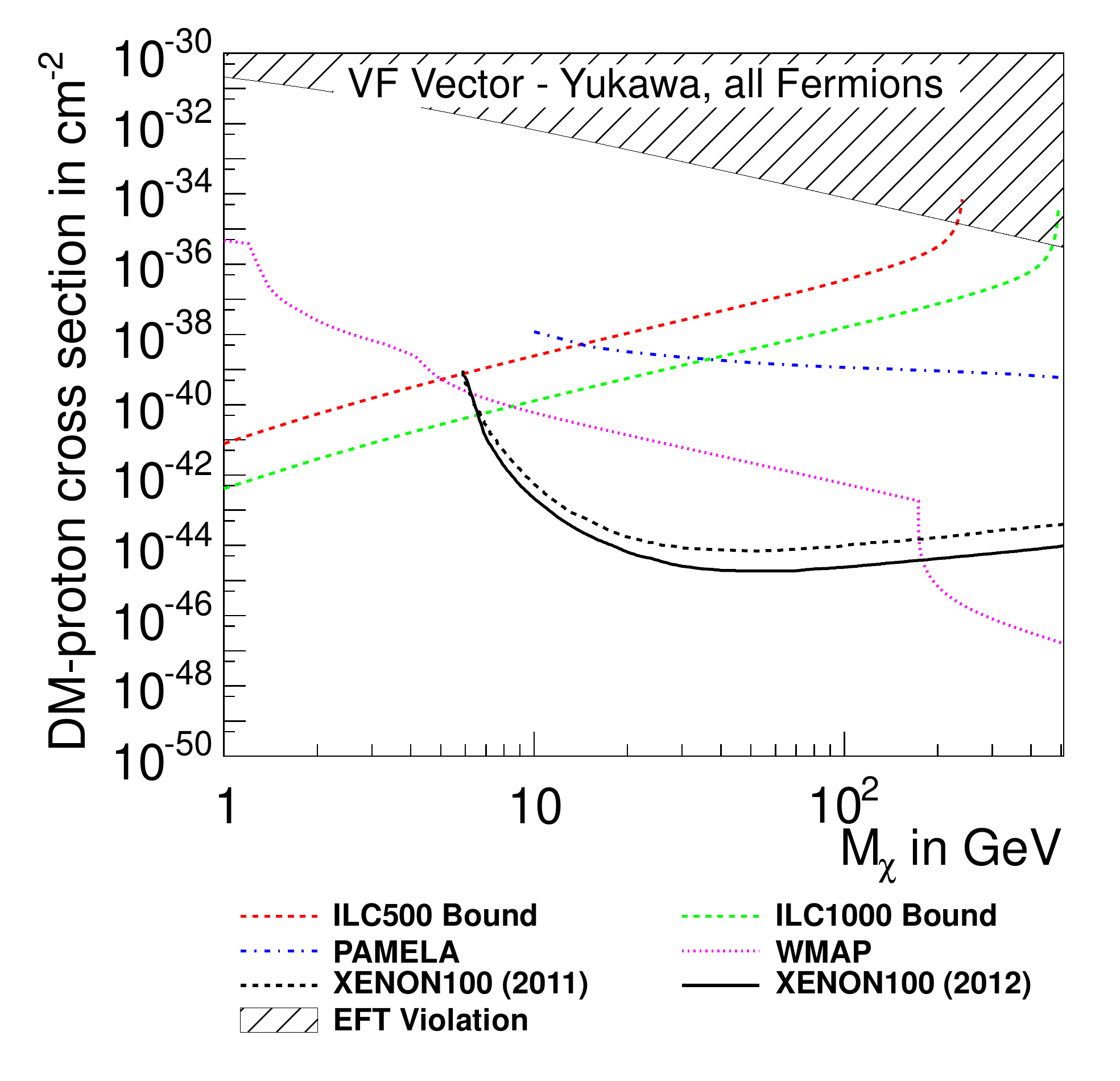}\\
\includegraphics[width=0.45\textwidth]{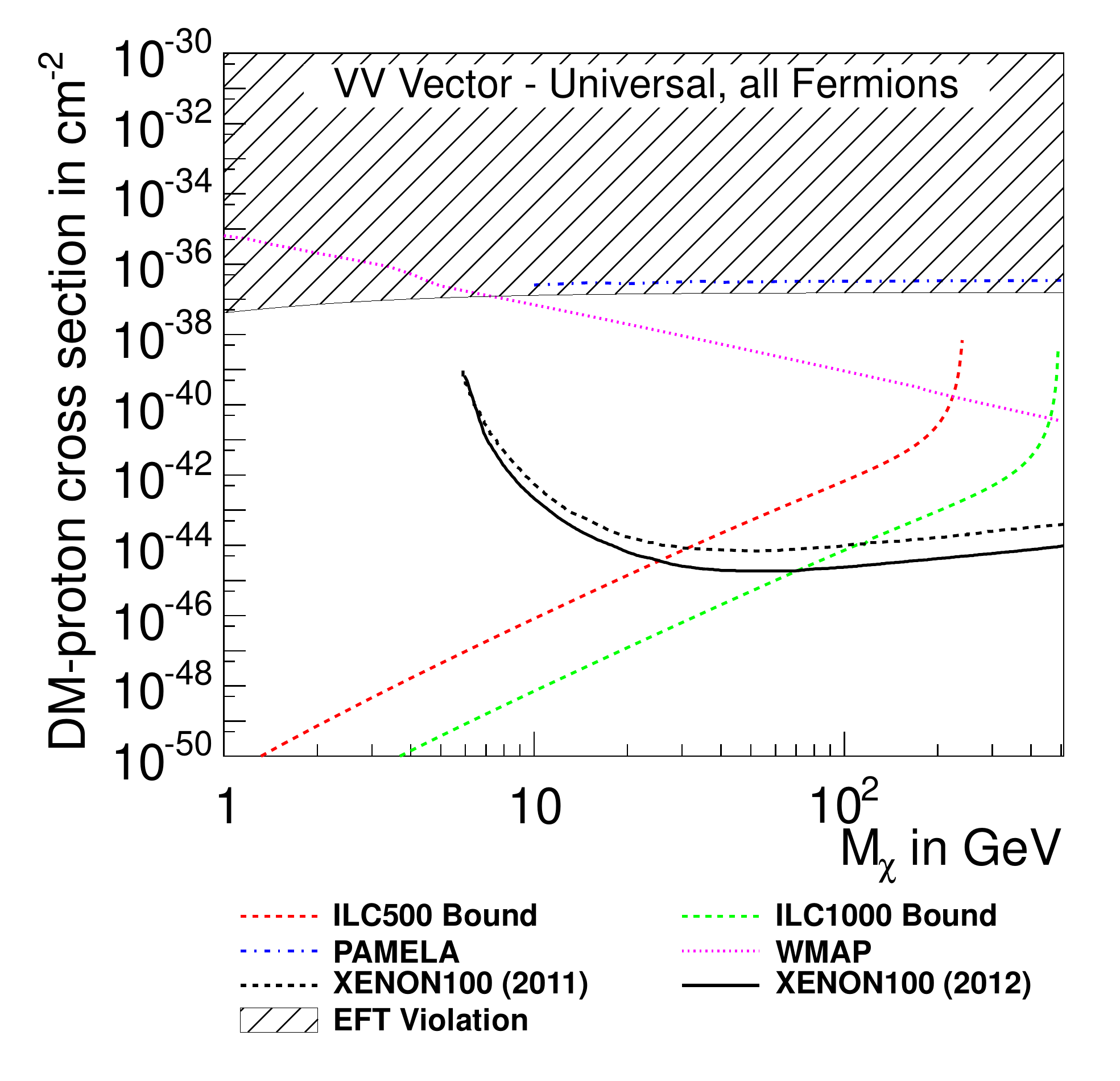} \hfill
\includegraphics[width=0.45\textwidth]{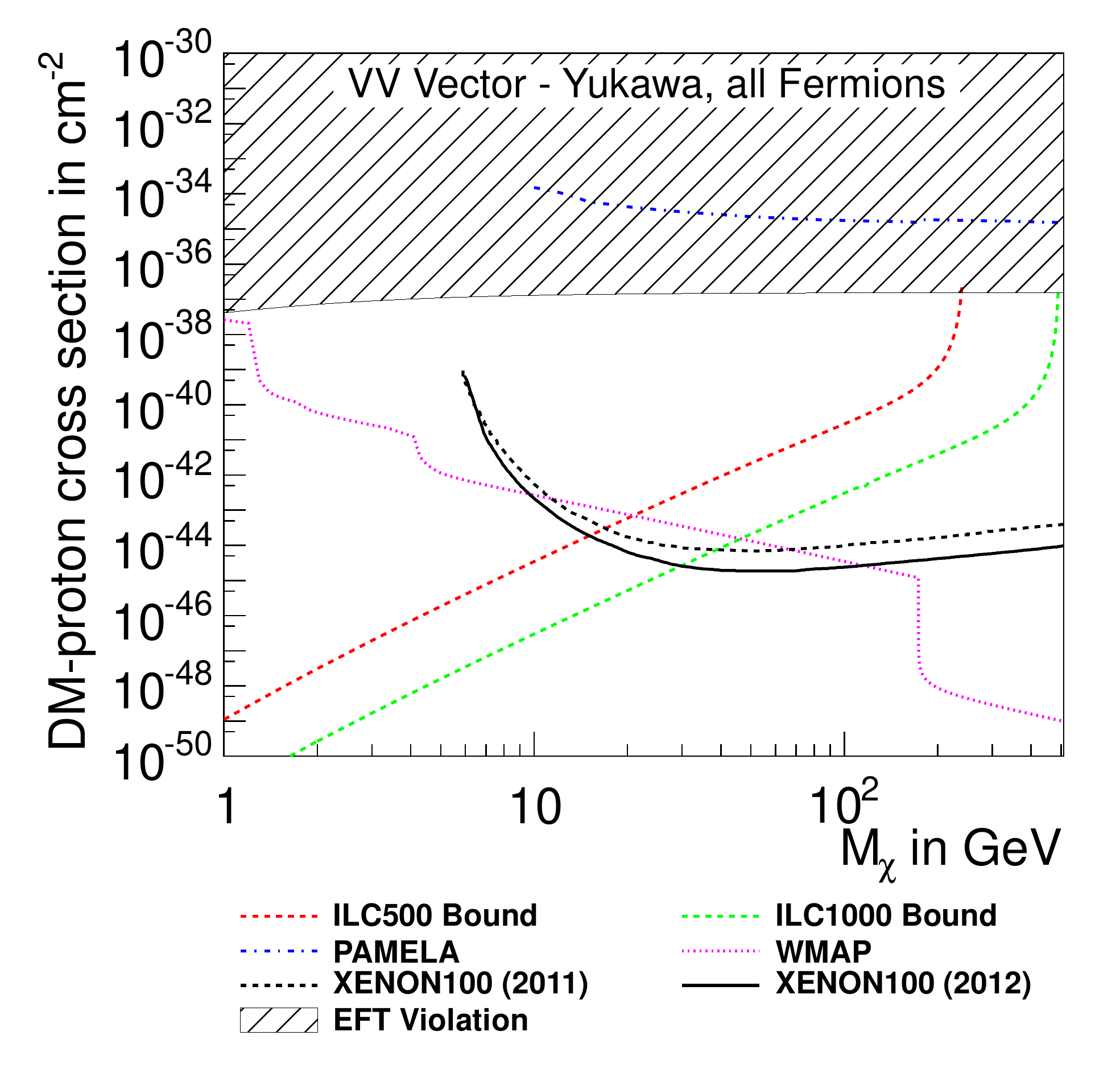}
\caption{Combined limits on the spin independent dark matter proton cross
  section from \textsc{Ilc}, \textsc{Pamela} and \textsc{Wmap} for a selection of vector dark matter models.}
\label{img:totalbounds3}
\end{figure*}

\begin{figure*}
\centering
\includegraphics[width=0.45\textwidth]{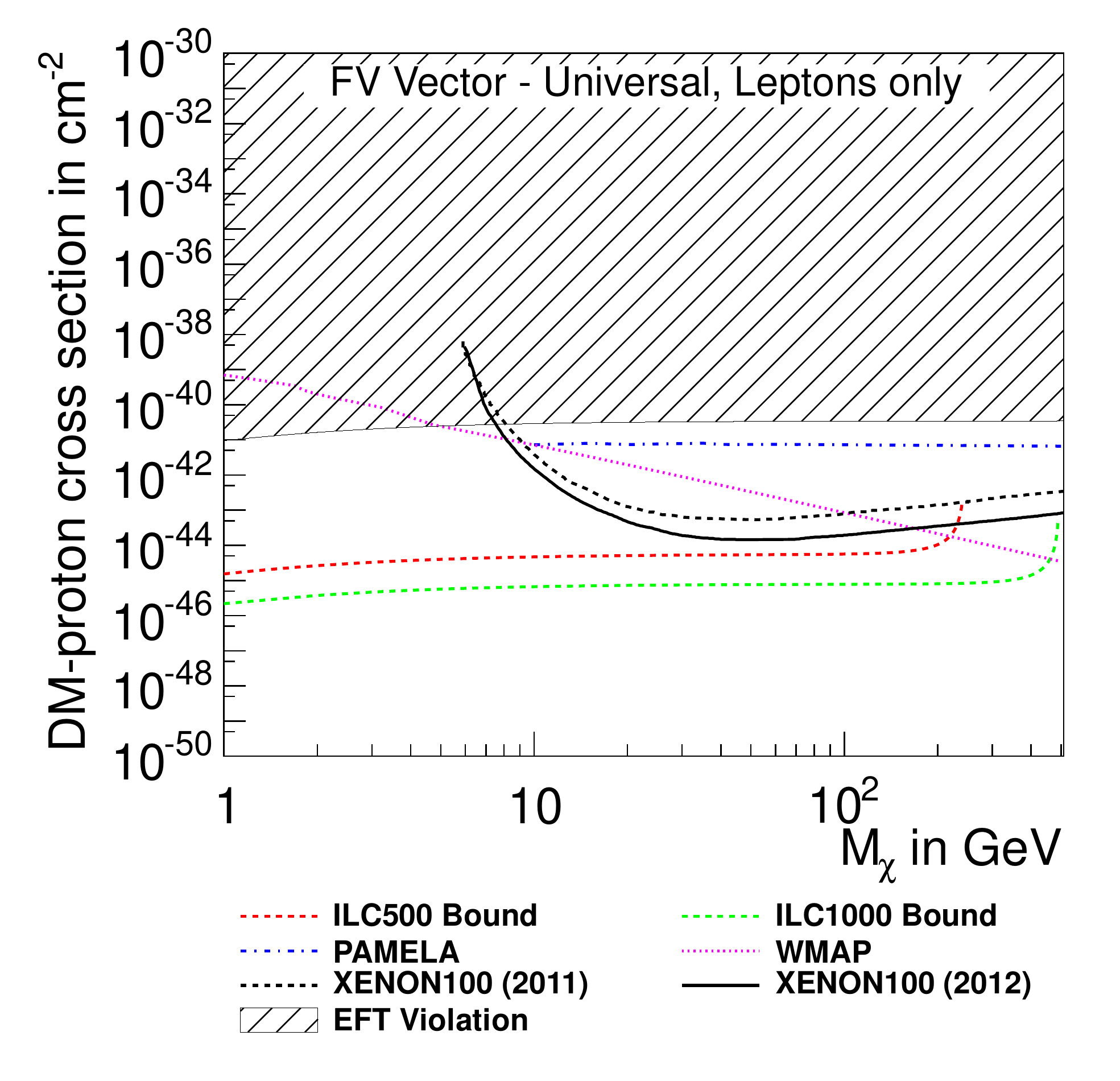} \hfill
\includegraphics[width=0.45\textwidth]{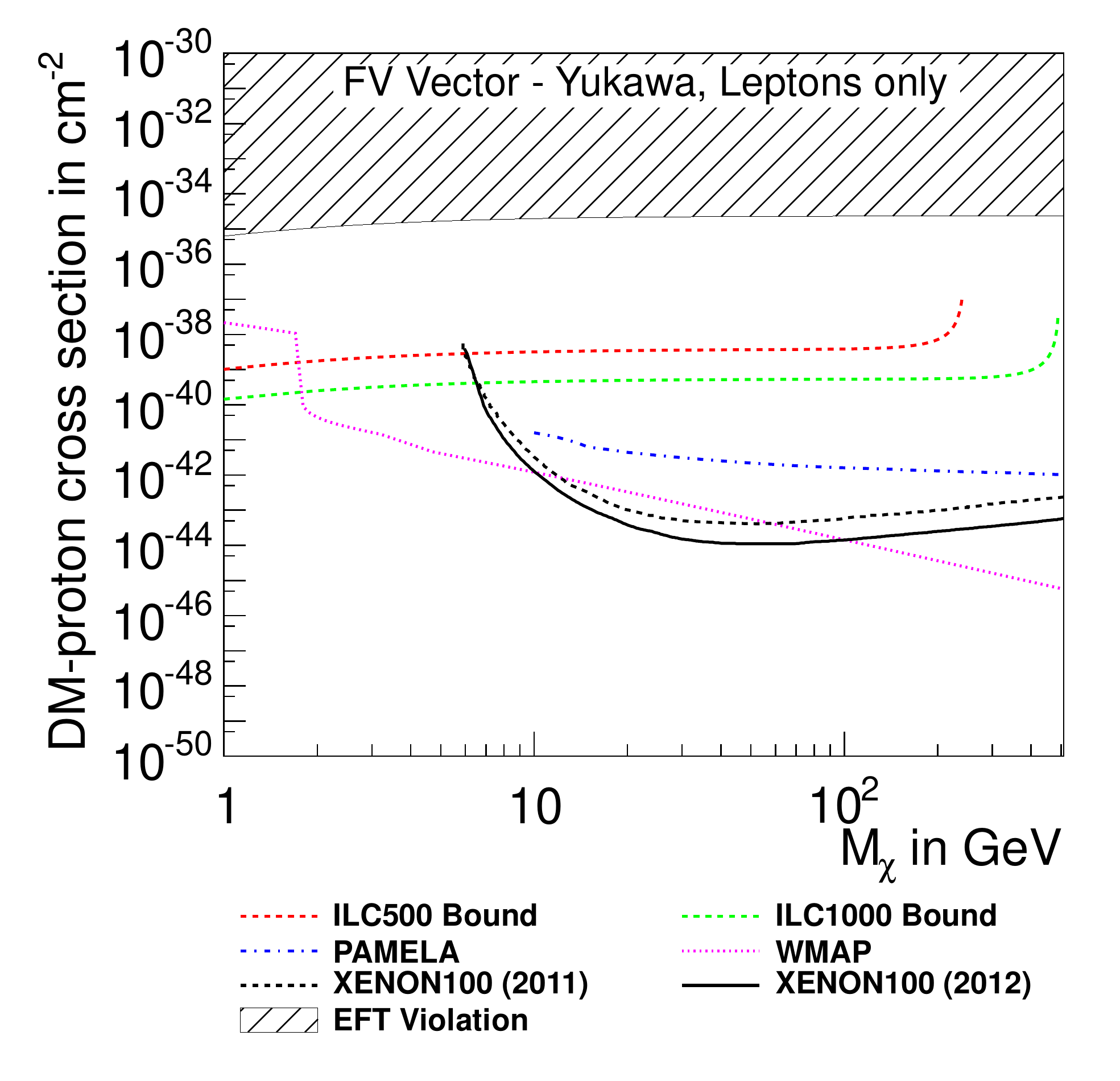}\\
\includegraphics[width=0.45\textwidth]{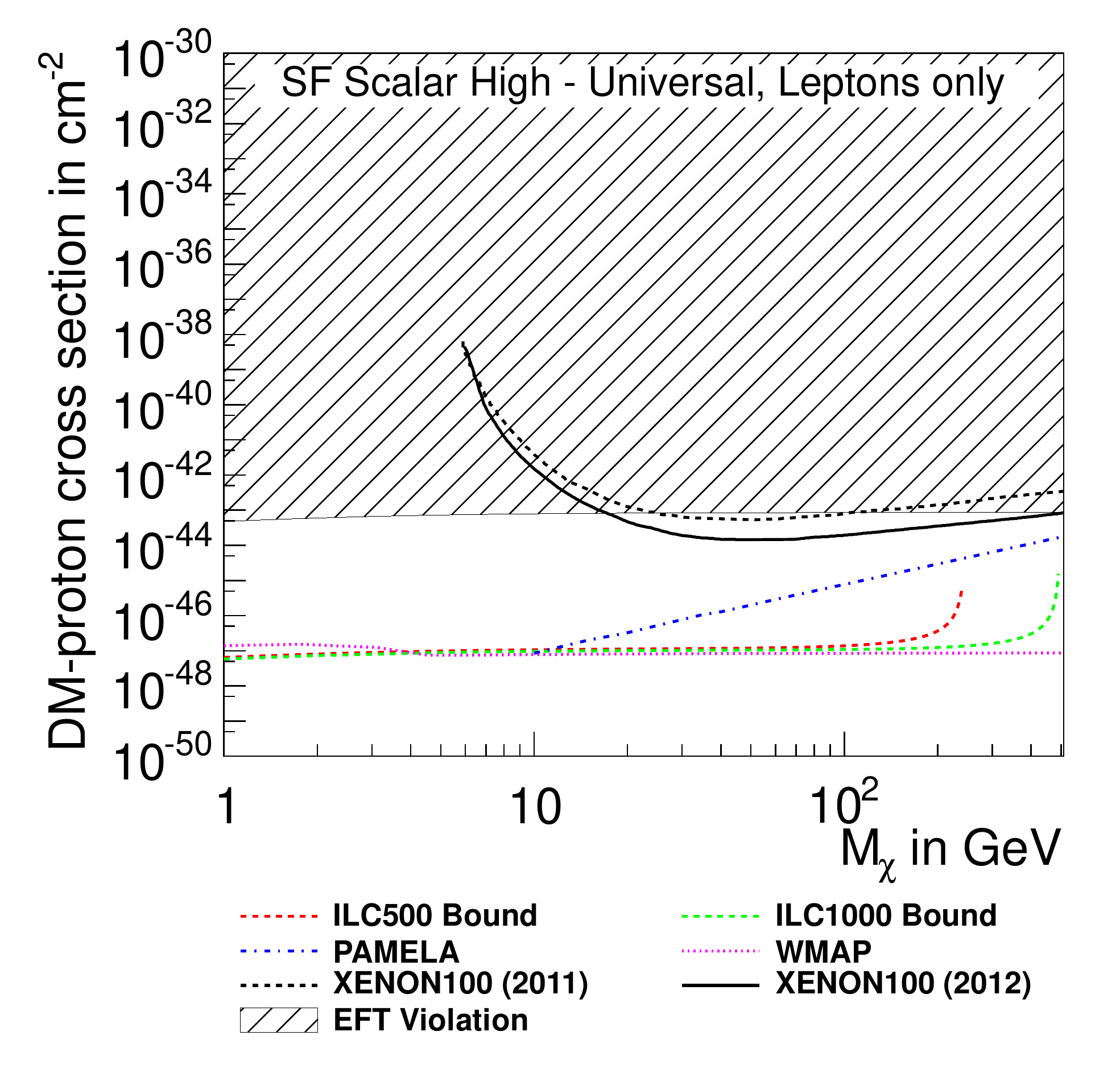} \hfill
\includegraphics[width=0.45\textwidth]{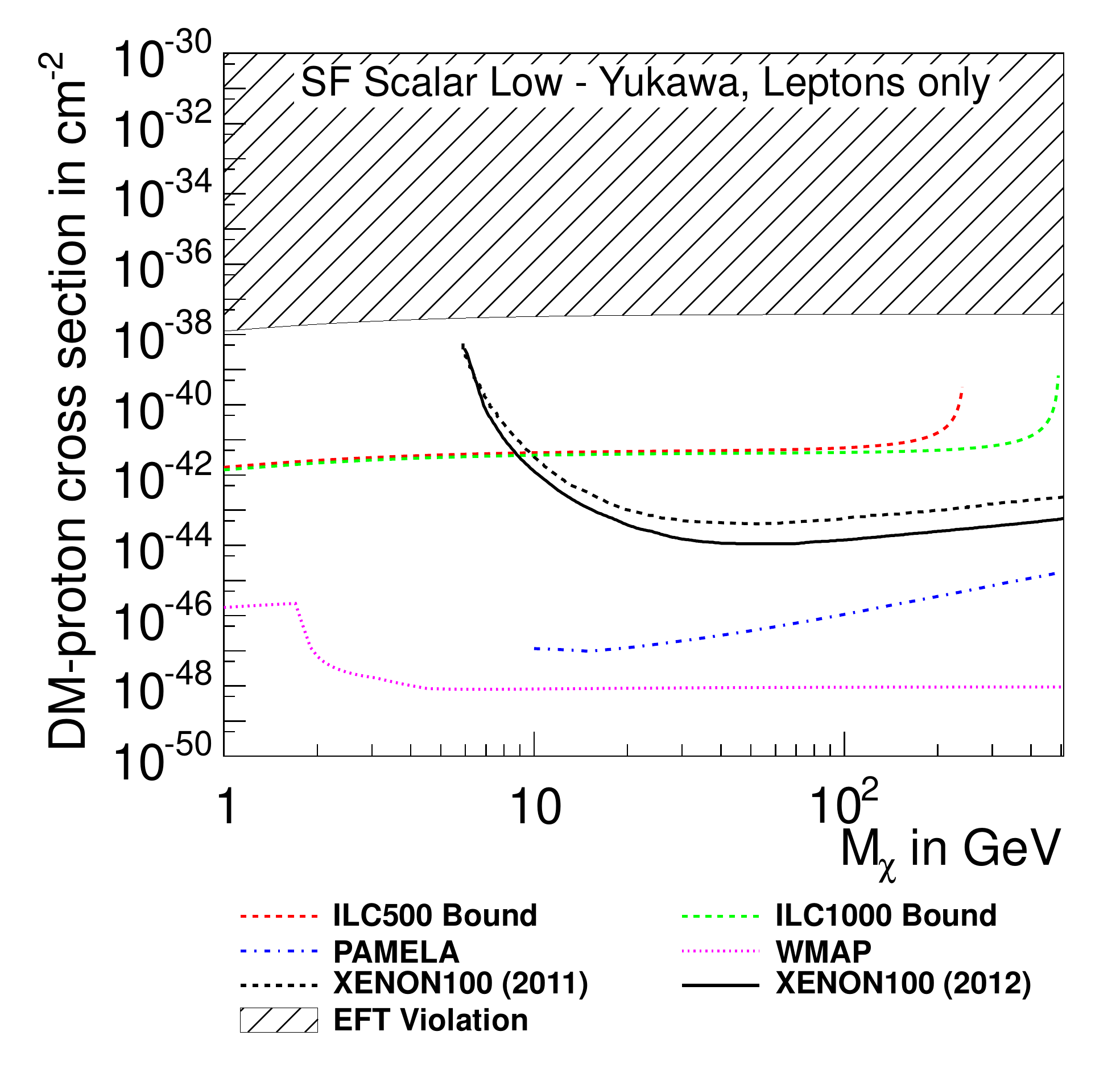}
\vspace{-0.5cm}
\caption{Combined limits for a selection of models
  with loop--coupling to leptons only. `Low' corresponds to $M_{\Omega} =$\unit{1}{\TeV} and `High' to $M_{\Omega} =$\unit{10}{\TeV}, Table~\ref{tbl:constraints}}
\label{img:totalbounds4}
\vspace{0.5cm}
\includegraphics[width=0.45\textwidth]{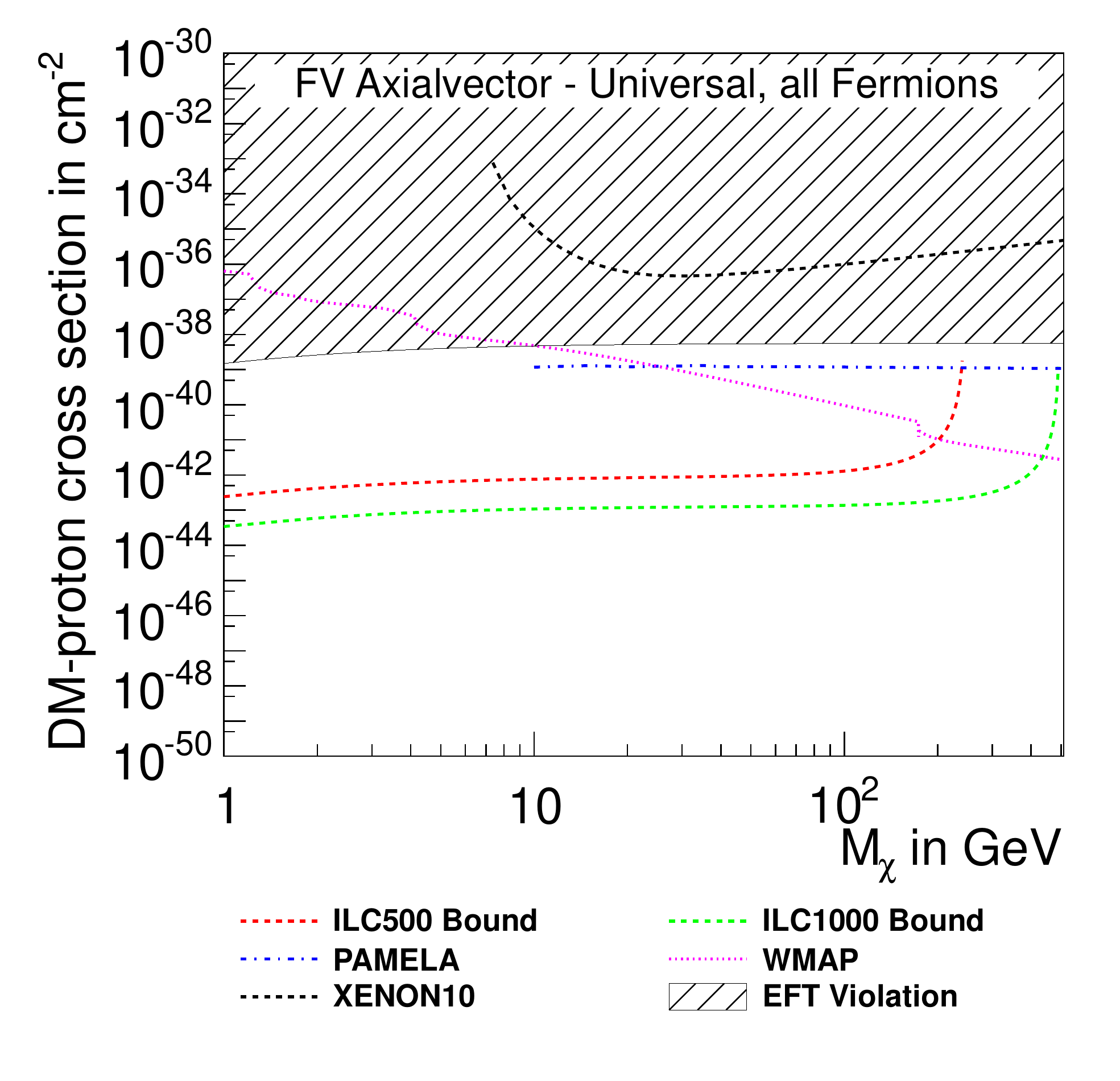} \hfill
\includegraphics[width=0.45\textwidth]{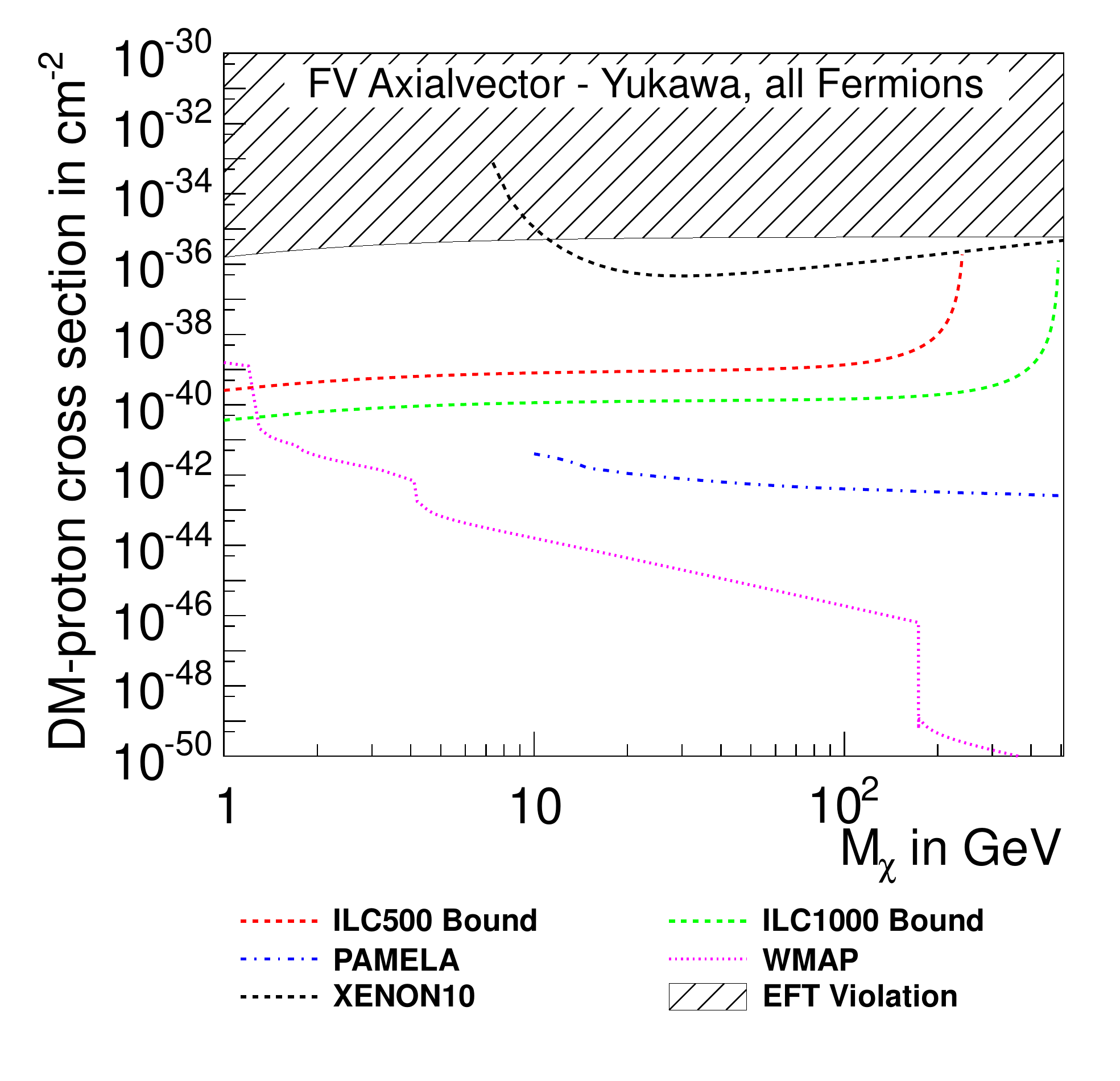}
\vspace{-0.5cm}
\caption{Combined limits on the spin dependent dark matter proton cross
  section.}
\label{img:totalbounds5}
\end{figure*}

\clearpage
\section{Conclusions}
\label{sec:conclusions}
In this paper we considered a broad range of effective models for dark
matter and investigated the possibility that these models could be
explored at the \textsc{Ilc}. The models considered the possibility
that dark matter was a new scalar, fermion or vector particle and
would be produced at the \textsc{Ilc} via a new, heavy intermediate
state, the mediator particle. For the mediator we also
 considered spins 0, 1/2 and 1. We obtained the corresponding
 effective theories by integrating out the mediator field.

 To be able to compare the reach of the \textsc{Ilc} with the other
 experimental searches, certain assumptions have to be made on how the
 mediator and dark matter couples to the Standard Model particles. We
 assume in all models that interactions only occur with the Standard
 Model fermions but the relative strength to different particles is
 varied. In the simplest variant we choose that the coupling is equal
 between all the Standard Model states. Another choice is that the
 interaction scales with the mass of the interacting Standard Model
 fermion, a `Yukawa-like' interaction. The last choice we make is the
 most optimistic for \textsc{Ilc} phenomenology with only the Standard Model
 leptons interacting with the heavy mediator.
\balance
 Since the produced dark matter particles will be invisible to the
 \textsc{Ilc} detectors, we require a radiated photon to be emitted
 from the initial state that will recoil against missing
 momentum. This topology provides a distinctive signal with which to
 discover  dark matter. For the \textsc{Ilc} study, we included the
 dominant backgrounds and most important detector effects. In addition
 we considered the possibility of using polarised initial states to
 reduce backgrounds and improve the signal strength.

 The effective theories that we consider provide an efficient way to
 compare the reach of the \textsc{Ilc} with other methods to discover
 dark matter. Firstly, we consider the dark matter annihilation cross
 section required for the relic density observed by \textsc{Wmap}. We
 also look at the direct detection bounds at \textsc{Xenon} by
 calculating the dark matter-nucleon scattering cross section. In
 addition, we include bounds from dark matter annihilation to
 positrons from the \textsc{Pamela} experiment.

In terms of the effective dark matter model, we found that the \textsc{Ilc}
should be able to probe couplings \unit{\power{10}{-7}}{\GeV \rpsquared}, or
\unit{\power{10}{-4}}{\reciprocal\GeV} depending on the mass dimension of the theory. In models that contain vector dark matter, the \textsc{Ilc} may be able to probe even weaker couplings in the case of low dark matter mass.

To compare with astrophysical bounds, we found that the
\textsc{Ilc} reach is strongly dependent on the exact dark matter
model. If we assume that dark matter is relatively heavy ($>$
\unit{100}{\GeV}) and interacts with a Standard Model particle in
proportion to its mass, then the \textsc{Ilc} is
uncompetitive. However, in the case that dark matter is relatively
light ($<$ \unit{10}{\GeV}) then the bounds from the \textsc{Ilc} are
competitive with astrophysical bounds in many models. In addition, if
dark matter happens to only interact with the Standard Model
leptons then the \textsc{Ilc} offers a unique possibility to discover
dark matter. For this reason, an \textsc{Ilc} search is complementary
to those done at the \textsc{Lhc} thanks to the different initial
state.

%% file: astrophysics_appendix.tex
\clearpage
\section{Cross Sections for Annihilation} \label{sec:sigmarelic}
\allowdisplaybreaks We give the full cross sections for annihilation
of a pair of dark matter particles with mass $M_\chi$ into a
pair of Standard Model fermions with mass $m_f$. To find the
expansion coefficients in $\sigma v \approx a + b v^2$, we perform the
non--relativistic approximation $s \approx 4 M_{\chi}^2+ M_{\chi}^2
v^2 + \frac{3}{4} M_{\chi}^2 v^4$ \cite{Beltran:2008xg}. Note that in order to
find the correct result for the $v^2$ term in $\sigma v$, it is
necessary to expand up to order $v^4$ because of the appearance of
$\sqrt{s}$ in the cross section formul{\ae}.

The total cross section is then given as the sum of the cross
  sections over all allowed final state fermions. This set is
restricted both by kinematics ($m_f \leq M_\chi$) and by the assumed
model. The latter also determines whether the coupling $G_f$ is
universal or particle--dependent.

We define the mass ratio $\xi\equiv m_{f}/M_{\chi}$ and the velocities of both
particles $\beta_X \equiv \sqrt{1-4m_X^2/s}$ to compactify the following expressions.

Some of our effective operators have been analysed before, for example \cite{Zheng:2010js,
    Yu:2011by}, and we agree with the respective results for the annihilation
  cross sections.
\subsection{Scalar \textsc{Wimp}}
\vspace{-1cm}
\begin{flalign}
\sigma^{\text{SS}}_{\text{Sc}} &= \frac{G_{f}^{2}}{8 \pi s} \frac{\beta_f}{\beta_\chi}(s-4 m_{f}^{2}),   &\\*
\sigma v &\approx \frac{G_{f}^{2}}{4 \pi} \sqrt{1-\xi^{2}} \Big[ (1-\xi^{2}) + \dfrac{v^{2}}{8}(5 \xi^{2}-2)\Big].&\\
\sigma^{\text{SS}}_{\text{Ps}} &= \frac{G_{f}^{2}}{8 \pi } \frac{\beta_f}{\beta_\chi},  &\\*
\sigma v &\approx \frac{G_{f}^{2}}{4 \pi} \Big[ \sqrt{1-\xi^{2}} + \dfrac{v^{2}}{8}\dfrac{3 \xi^2 -2}{\sqrt{1-\xi^{2}}}\Big].&\\
\sigma^{\text{SV}}_{\text{Vec}} &= \frac{G_{f}^{2}}{12 \pi}  \beta_f \beta_\chi (s+2 m_{f}^{2}),  &\\*
\sigma v &\approx \frac{G_{f}^{2}}{12 \pi} \Big[ M_{\chi}^{2} v^{2}\sqrt{1-\xi^{2}}(\xi^{2}+2)\Big].&\\
\sigma^{\text{SV}}_{\text{Ax}} &= \frac{G_{f}^{2}}{12 \pi }  \beta_f \beta_\chi (s-4 m_{f}^{2}),  &\\*
\sigma v &\approx \frac{G_{f}^{2}}{6 \pi} \Big[ M_{\chi}^{2}
v^{2}(1-\xi^{2})^{3/2}\Big]. &\\
\sigma^{\text{SV}}_{\text{Ch}} &= \frac{G_{f}^{2}}{24 \pi }  \beta_f\beta_\chi(s-m_f^2), &\\ 
\sigma v &\approx \frac{G_{f}^{2}M_{\chi}^2}{48 \pi} v^2\sqrt{1-\xi^{2}} (4-\xi^2). &\\
\sigma^{\text{SF}}_{\text{Sc/Ps}} &= \frac{G_{f}^{2}}{48 \pi s}
\frac{ \beta_f}{ \beta_\chi} \Big[ 2 s(4 m_{f}^{2}-2
M_{\chi}^{2}+3 M_{\Omega}^{2} \mp 6 m_{f} M_{\Omega}) \nonumber &\\ &\qquad -8
m_{f}^{2} \Big(3( M_{\Omega} \mp m_{f})^{2}+M_{\chi}^{2}\Big) +s^{2} \Big], &\\
\sigma v &\approx \frac{G_{f}^{2}}{4 \pi} \sqrt{1-\xi^{2}} \Big[ (1-\xi^{2})(\xi M_{\chi} \mp M_{\Omega})^{2}. \nonumber &\\*
&\qquad + \dfrac{v^{2}}{24}  \Big((15 \xi^{2}
-6) M_{\Omega}^{2}  \mp 6 \xi(5 \xi^{2}-2) M_{\chi} M_{\Omega}\nonumber &\\
&\qquad +(15 \xi^{4} -4 \xi^{2}+4) M_{\chi}^{2}\Big)\Big],&\\
\sigma^{\text{SFr}}_{\text{Sc/Ps}} &= \frac{G_{f}^{2}}{2\pi s} \frac{ \beta_f^3}{ \beta_\chi} (m_{f} \mp M_{\Omega})^{2}   &\\*
\sigma v &\approx \frac{G_{f}^{2}}{\pi} \sqrt{1-\xi^{2}}^3 (\xi M_{\chi} \mp
M_{\Omega})^{2}  \nonumber &\\ & \qquad \times \Big[ 1
+ \dfrac{v^{2}}{8} (5 \xi^{2}-2)\Big].
\end{flalign}
\subsection{Fermion \textsc{Wimp}}
\vspace{-1cm}
\begin{flalign}
\sigma^{\text{FS}}_{\text{Sc}} &= \frac{G_{f}^{2}}{16 \pi }  \beta_f \beta_\chi(s-4 m_{f}^{2}),   &\\
\sigma v &\approx \frac{G_{f}^{2}}{8 \pi}  v^{2} M_{\chi}^{2} (1-\xi^{2})^{3/2}.&\\
\sigma^{\text{FS}}_{\text{Ps}} &= \frac{G_{f}^{2}}{16 \pi } \frac{ \beta_f}{ \beta_\chi}s,  &\\
\sigma v &\approx \frac{G_{f}^{2} M_{\chi}^{2}}{2 \pi} \Big[ \sqrt{1-\xi^{2}} + \dfrac{v^{2}}{8}\dfrac{\xi^2 }{\sqrt{1-\xi^{2}}}\Big].&\\
\sigma^{\text{FV}}_{\text{Vec}} &= \frac{G_{f}^{2}}{12 \pi s} \dfrac{ \beta_f}{ \beta_\chi}(s+2 M_{\chi}^{2}) (s+2 m_{f}^{2}),  &\\
\sigma v &\approx \frac{G_{f}^{2} M_{\chi}^{2} }{2 \pi} \Big[
\sqrt{1-\xi^{2}}(2+\xi^{2}) \nonumber &\\ &\qquad +v^{2}\dfrac{-4+2 \xi^{2}+11 \xi^{4}}{24 \sqrt{1-\xi^{2}}}\Big].&\\
\sigma^{\text{FV}}_{\text{Ax}} &= \frac{G_{f}^{2}}{12 \pi s} \dfrac{
  \beta_f}{ \beta_\chi}\Big[s\Big(s-4(m_{f}^{2}+M_{\chi}^{2})\Big) \nonumber &\\
& \qquad \qquad  +28 m_{f}^{2} M_{\chi}^{2}\Big],   &\\
\sigma v &\approx \frac{G_{f}^{2} M_{\chi}^{2} }{2 \pi} \Big[
\sqrt{1-\xi^{2}}\xi^{2} \nonumber &\\
& \qquad \qquad  + v^{2}\dfrac{8-28 \xi^{2}+23 \xi^{4}}{24 \sqrt{1-\xi^{2}}}\Big].&\\
\sigma^{\text{FV}}_{\text{Ch}} &= \frac{G_{f}^{2}}{48 \pi s} \dfrac{
  \beta_f}{ \beta_\chi}\Big(s(s-m_f^2+M_{\chi}^2) \nonumber &\\ & \qquad \qquad +4 m_f^2 M_{\chi}^2\Big), &\\
\sigma v &\approx \frac{G_{f}^{2}  M_{\chi}^{2} }{8 \pi} \Big[
\sqrt{1-\xi^{2}} \nonumber &\\
& \qquad \qquad +  v^{2}\dfrac{(2 -\xi^2 +2 \xi^4)}{24
  \sqrt{1-\xi^{2}}}\Big]. &\\
\sigma^{\text{FVr}}_{\text{Ch}} &= \frac{G_{f}^{2}}{24 \pi s} \frac{ \beta_f}{
  \beta_\chi}\Big((s-4 M_{\chi}^{2})(s-m_f^2) \nonumber &\\ & \qquad + 6 m_f^2 M_{\chi}^2 \Big), &\\
\sigma v &\approx \frac{G_{f}^{2} M_{\chi}^{2} }{4\pi} \Big[ \xi^2
\sqrt{1-\xi^2} \nonumber &\\ & \qquad + v^{2} \frac{16-32 \xi^2 +19 \xi^4}{24 \sqrt{1-\xi^{2}}} \Big]. &\\
\sigma^{\text{FtS}}_{\text{Sc/Ps}} &= \frac{G_{f}^{2}}{48 \pi s} \dfrac{
  \beta_f}{ \beta_\chi}\Big(s(s-M_{\chi}^{2}) \mp 6 m_{f} M_{\chi}s \nonumber &\\ &
\qquad \qquad + m_{f}^{2}(16 M_{\chi}^{2}-s)\Big),  &\\*
\sigma v &\approx \frac{G_{f}^{2} M_{\chi}^{2}}{8 \pi} (1 \mp \xi)^{2}
\Big[ \sqrt{1-\xi^{2}} \nonumber &\\ & \qquad + v^{2} \dfrac{2 \pm 16 \xi+17
  \xi^{2}}{24 \sqrt{1-\xi^{2}}}\Big].&\\
\sigma^{\text{FtSr}}_{\text{Sc/Ps}} &= \frac{G_{f}^{2}}{96 \pi s} \dfrac{
  \beta_f}{ \beta_\chi}\Big(5 s^{2}+80 m_{f}^{2} M_{\chi}^{2} \nonumber &\\ &
\qquad -2s(7 m_{f}^{2}+7 M_{\chi}^{2}\mp 6 m_{f} M_{\chi})\Big),   &\\
\sigma v &\approx \frac{G_{f}^{2} M_{\chi}^{2}}{8 \pi}(1 \pm \xi)^{2}  \Big[
\sqrt{1-\xi^{2}} \nonumber &\\ &\qquad + v^{2} \dfrac{14\mp 40 \xi+29
  \xi^{2}}{24 \sqrt{1-\xi^{2}}}\Big].&\\
\sigma^{\text{FtV}}_{\text{Vec/Ax}} &= \frac{G_{f}^{2}}{24 \pi s} \dfrac{
  \beta_f}{ \beta_\chi}\Big(s(4s-7M_{\chi}^{2}) \pm 6 m_{f} M_{\chi}s \nonumber &\\
& \qquad \qquad - m_{f}^{2}(7s-40 M_{\chi}^{2})\Big),  &\\*
\sigma v &\approx \frac{G_{f}^{2} M_{\chi}^{2}}{4 \pi} \Big[(3 \pm 2
\xi+\xi^{2}) \sqrt{1-\xi^{2}} \nonumber &\\ & \qquad + v^{2} \dfrac{14 \mp 12
  \xi -31 \xi^{2} \pm 18 \xi^3+29 \xi^{4}}{24 \sqrt{1-\xi^{2}}}\Big]. \hspace{-2cm} &\\
\sigma^{\text{FtVr}}_{\text{Vec/Ax}} &= \frac{G_{f}^{2}}{12 \pi s} \dfrac{\beta_f}{\beta_\chi}\Big(7 s^{2}+76 m_{f}^{2} M_{\chi}^{2} \nonumber &\\
& \qquad -4s(4 m_{f}^{2}+4 M_{\chi}^{2} \pm 3 m_{f} M_{\chi})\Big),  &\\
\sigma v &\approx \frac{G_{f}^{2} M_{\chi}^{2}}{2 \pi} \Big[(2 \mp \xi)^{2}
\sqrt{1-\xi^{2}} \nonumber &\\ & \qquad + v^{2} \dfrac{32 \pm 24 \xi-64 \xi^{2} \mp 36 \xi^3 +47\xi^{4}}{24 \sqrt{1-\xi^{2}}}\Big]. \hspace{-2cm} &\\
\sigma^{\text{FtV}}_{\text{Ch}} &= \frac{G_{f}^{2}}{48 \pi s} \dfrac{
  \beta_f}{ \beta_\chi}\Big(4 m_f^2 M_{\chi}^2 + s(s-m_f^2-M_\chi^2)\Big), &\\
\sigma v &\approx \frac{G_{f}^{2} M_{\chi}^{2} }{8 \pi} \Big[
\sqrt{1-\xi^{2}}  +  v^{2}\dfrac{(2 -\xi^2 +2 \xi^4)}{24
  \sqrt{1-\xi^{2}}}\Big].  \hspace{-2cm}  &\\
\sigma^{\text{FtVr}}_{\text{Ch}} &= \frac{G_{f}^{2}}{24 \pi s} \frac{\beta_f}{\beta_\chi}\Big((s-4 M_{\chi}^{2})(s-m_f^2) \nonumber &\\ & \qquad \qquad + 6 m_f^2 M_{\chi}^2\Big),  &\\
\sigma v &\approx \frac{G_{f}^{2} M_{\chi}^{2} }{4\pi} \Big[ \xi^2
\sqrt{1-\xi^2}  \nonumber &\\ & \qquad \qquad +v^{2} \frac{16-32 \xi^2 +19 \xi^4}{24 \sqrt{1-\xi^{2}}}
\Big] .
\end{flalign}
\subsection{Vector \textsc{Wimp}}
\vspace{-1cm}
\begin{flalign}
\sigma^{\text{VS}}_{\text{Sc}} &= \frac{G_{f}^{2}}{288 M_{\chi}^{4} \pi s}\frac{ \beta_f}{ \beta_\chi}(s-4 m_{f}^{2}) \nonumber \\
& \qquad \times (12 M_{\chi}^{4} +s^{2}-4 M_{\chi}^{2} s),   \\
\sigma v &\approx \frac{G_{f}^{2}}{12 \pi}\sqrt{1-\xi^{2}} \Big[ (1-\xi^{2})
\nonumber \\ & \qquad \qquad + \dfrac{v^{2}}{24}(2+7 \xi^2)\Big].\\
\sigma^{\text{VS}}_{\text{Ps}} &= \frac{G_{f}^{2}}{288 M_{\chi}^{4} \pi } \frac{ \beta_f}{ \beta_\chi}(12 M_{\chi}^{4} +s^{2}-4 M_{\chi}^{2} s),   \\
\sigma v &\approx \frac{G_{f}^{2}}{12 \pi}\sqrt{1-\xi^{2}} \Big[ 1 + \dfrac{v^{2}}{24}\dfrac{2+\xi^{2}}{1-\xi^{2}}\Big].\\
\sigma^{\text{VV}}_{\text{Vec}} &= \frac{G_{f}^{2}}{432  \pi M_{\chi}^{4} }
\beta_f \beta_\chi (s+2 m_{f}^{2}) \nonumber \\ & \qquad \times (s^{2}+20 M_{\chi}^{2} s +12 M_{\chi}^{4}),  \\
\sigma v &\approx \frac{G_{f}^{2}}{4 \pi} M_{\chi}^{2} v^{2}\sqrt{1-\xi^{2}}(\xi^{2}+2).\\
\sigma^{\text{VV}}_{\text{Ax}} &= \frac{G_{f}^{2}}{432  \pi M_{\chi}^{4} }
\beta_f \beta_\chi (s-4 m_{f}^{2}) \nonumber \\ 
& \qquad \times (s^{2}+20 M_{\chi}^{2} s +12 M_{\chi}^{4}),  \\
\sigma v &\approx \frac{G_{f}^{2}}{2 \pi} M_{\chi}^{2} v^{2} (1-\xi^{2})^{3/2}.\\
\sigma^{\text{VV Ch}} &= \frac{G_{f}^{2}}{864 \pi M_{\chi}^{4}} \beta_f
 \beta_\chi  (s-m_{f}^{2})\nonumber \\ & \qquad \times (s^{2}+20 M_{\chi}^{2} s +12 M_{\chi}^{4}), \\
\sigma v &\approx \frac{G_{f}^{2}}{16 \pi} M_{\chi}^{2}
v^{2}\sqrt{1-\xi^{2}}(4-\xi^4). \\
\sigma^{\text{VF}}_{\text{Vec/Ax}} &= \frac{G^2}{4320 \pi  M_{\chi}^4 s } \frac{
  \beta_f}{\beta_\chi} \Big[ 8 m_{f}^{4}(-174 M_{\chi}^{4}+2M_{\chi}^{2}s+s^{2}) \nonumber \\
& \qquad+4 M_{\chi}^{2}s (s-20 M_{\Omega}^{2})+s^{2}(10 M_{\Omega}^{2} + 7s)\Big) \nonumber \\ 
&\qquad -2 m_{f}^{2}\Big(680 M_{\chi}^{6}+152 M_{\chi}^{4} (5 M_{\Omega}^{2}+s) \nonumber \\ 
& \qquad+s\Big(40 M_{\chi}^{6}+2 M_{\chi}^{4}(70 M_{\Omega}^{2}-31s) \nonumber \\
& \qquad\pm 240 m_{f}^{3} M_{\Omega} M_{\chi}^2 (10 M_{\chi}^{2}- s^{2}) \nonumber \\
&\qquad \pm 120 m_{f} M_{\chi}^2 M_{\Omega} s (M_{\chi}^{2}-s) \nonumber \\ 
&\qquad +M_{\chi}^{2}(76 s^{2}-40 M_{\Omega}^{2} s )\nonumber \\ & \qquad+s^{2}(20 M_{\Omega}^{2}+3s)\Big) \Big], \\
\sigma v &\approx \frac{G^2}{36 \pi }\sqrt{1-\xi^2} \Big[ (1-\xi^2)\Big((5
\xi^2+4) M_{\chi}^2 \nonumber \\ & \qquad \mp 6 \xi M_{\chi} M_{\Omega}+5 M_{\Omega}^2\Big) \nonumber \\
& \qquad +\frac{v^2}{24} \Big(\mp 6 \xi (19 \xi^2+2) M_{\chi} M_{\Omega}\nonumber
\\ & \qquad +3 (25 \xi^2+6) M_{\Omega}^2\nonumber \\ & \qquad+(83 \xi^4+ 136 \xi^2+156) M_{\chi}^2\Big) \Big].\\
\sigma^{\text{VFr}}_{\text{Vec/Ax}} &= \frac{ G_{f}^{2}}{2160 \pi M_{\chi}^{4} s} \frac{ \beta_f}{ \beta_\chi} \Big[s^4 + 22 m_f^2 M_{\chi}^2 + 13 M_{\chi}^4 \nonumber \\ 
& \qquad -8 s^2\Big(8 m_f^4 +15 M_{\Omega}^2 (m_f^2+M_{\chi}^2) \nonumber \\ 
& \qquad \pm 5 m_f M_{\Omega} (4 m_f^2 + 5 M_{\chi}^2)\Big)\nonumber \\
&\qquad-32 m_f^2 M_{\chi}^4 (37 m_f^2 \pm 50 m_f M_{\Omega}\nonumber \\
&\qquad  +70 M_{\chi}^2+45 M_{\Omega}^2) \nonumber \\ 
&\qquad+ 2 s^3 (6 m_f^2 \pm 20 m_f M_{\Omega}+16 M_{\chi}^2+15 M_{\Omega}^2) \nonumber \\*  
&\qquad+8 M_{\chi}^2 s \Big(24 m_f^4 +15 M_{\Omega}^2 (4 m_f^2+3M_{\chi}^2)
\nonumber \\
&\qquad \pm 50 m_f M_{\Omega} (2 m_f^2 + M_{\chi}^2)\nonumber \\
&\qquad+119 m_f^2 M_{\chi}^2+40 M_{\chi}^4\Big), \\ 
\sigma v &\approx \frac{ G_{f}^{2}}{ 9 \pi}  \sqrt{1-\xi^{2}} \Big[
(1-\xi^{2})\Big(3 M_{\Omega}^{2} \pm 2 \xi M_{\chi} M_{\Omega} + \nonumber \\
&\qquad (3 \xi^{2}+4) M_{\chi}^{2}\Big) + \dfrac{v^{2}}{24} \Big(3(2+7 \xi^{2}) M_{\Omega}^{2} \nonumber \\*
& \qquad \pm 6 \xi (2+\xi^{2}) M_{\chi} M_{\Omega} \nonumber\\
&\qquad +(16+30 \xi^{2}+29 \xi^{4}) M_{\chi}^{2}\Big)\Big].
\end{flalign}
\section{Cross Sections for Direct Detection} \label{sec:directdetect}
We now give results for the dark matter--nucleon scattering cross
section at zero momentum transfer, $\sigma^0$, for all defined
benchmark models. In a universal scenario, the effective coupling is
independent of the quark ($G_q = G$), whereas it grows proportionally
to the quark mass in a Yukawa-like model ($G_q = G\ m_q / m_e$).  We
use the following definitions:
\begin{align}
\frac{f_{p}}{M_P} &\equiv \sum_{q=u,d,s} \hspace{-0.2cm}f_{q}^{p}\frac{G_q}{m_{q}} + \frac{2}{27}(1-\sum_{q=u,d,s}\hspace{-0.2cm}f_{q}^{p})\sum_{q=c,b,t}\hspace{-0.05cm}\frac{G_{q}}{m_{q}}, \\
d_p &\equiv \sum_{q=u,d,s} G_q \Delta_q^p, \\
b_{p} &\equiv 2 G_{u} +G_{d}, \\
\tilde{b}_p &\equiv b_p M_\chi + 2 G_u m_u + G_d m_d. \\
\intertext{with the numerical values for $f_q^p$ and  $\Delta_q^p$ listed in \cite{fnumbers, deltanumbers}:}
f_{u}^{p} &=  0.020 \pm 0.004, \\
f_{d}^{p} &= 0.026 \pm 0.005, \\
f_{s}^{p} &= 0.118 \pm 0.062, \\ 
\Delta_{u}^{p} &= -0.427 \pm 0.013, \\
\Delta_{d}^{p} &= 0.842\pm 0.012, \\
\Delta_{s}^{p} &= -0.085 \pm 0.018. \\
\intertext{Furthermore we define the reduced mass of the \textsc{Wimp} proton system,}
\mu  &\equiv \frac{M_{\chi} M_{p}}{M_{\chi}+M_{p}}.
\end{align}
The cross sections can be evaluated in a nonrelativistic approximation for the
\textsc{Wimp} and by using the quark proton form factors listed above. See
e.g.\ \cite{Agrawal:2010fh}. If a model is not listed, its
scattering cross section
equals zero, e.g.\ for pseudoscalar interactions that always vanish in a
nonrelativistic model. Again, we agree with the respective results in  \cite{Zheng:2010js,
    Yu:2011by} for comparable operators. 

Cross sections for real final state particles can
easily be derived from the following list by setting the vector form factors
$b_p$ and $\tilde{b}_p$ to zero and rescaling $f_p$ and $d_p$ by a
factor of 2.
\subsection{Scalar \textsc{Wimp}}
\vspace{-1cm}
\begin{flalign}
\sigma^0_{\text{SS Sc.}} &= \frac{\mu^{2}}{ 4 \pi M_{\chi}^{2}}f_{p}^{2}, &\\
\sigma^0_{\text{SV Vec.}} &= \frac{\mu^{2}}{\pi}b_{p}^{2}, &\\
\sigma^0_{\text{SF Sc.}} &= \frac{\mu^{2}}{4 \pi}(+f_{p} + \frac{\tilde{b}_{p}}{M_\Omega})^{2}, &\\
\sigma^0_{\text{SF Ps.}} &= \frac{\mu^{2}}{4 \pi}(-f_{p} + \frac{\tilde{b}_{p}}{M_\Omega})^{2},  &\\
\sigma^0_{\text{SV Chi.}} &= \frac{\mu^{2}}{ 4 \pi}b_{p}^{2}. &
\end{flalign}
\subsection{Fermion \textsc{Wimp}}
\vspace{-1cm}
\begin{flalign}
\sigma^0_{\text{FS Sc.}} &= \frac{\mu^{2}}{\pi}f_{p}^{2},&\\ 
\sigma^0_{\text{FV Vec.}} &= \frac{\mu^{2}}{\pi}b_{p}^{2}, &\\ 
\sigma^0_{\text{FV Ax.}} &= 3  \frac{\mu^{2}}{ \pi} d_p^{2},&\\ 
\sigma^0_{\text{FV Chi.}} &= \frac{\mu^{2}}{ 16 \pi}b_{p}^{2}, &\\
\sigma^0_{\text{FVr Chi.}} &= 3  \frac{\mu^{2}}{\pi} d_p^{2}, &\\
\sigma^0_{\text{FtS Sc.}} &= \frac{\mu^{2}}{16 \pi}(b_{p} + f_{p} )^{2},&\\ 
\sigma^0_{\text{FtS Ps.}} &= \frac{\mu^{2}}{16 \pi}(b_{p} - f_{p} )^{2},&\\
\sigma^0_{\text{FtV Vec.}} &= \frac{\mu^{2}}{\pi}(1/2 \cdot b_{p} -f_{p})^{2},&\\ 
\sigma^0_{\text{FtV Ax.}} &= \frac{\mu^{2}}{\pi}(1/2 \cdot b_{p} + f_{p})^{2},&\\ 
\sigma^0_{\text{FtV Chi.}} &= \frac{\mu^{2}}{ 16 \pi}b_{p}^{2}. &
\end{flalign}
\subsection{Vector \textsc{Wimp}}
\vspace{-1cm}
\begin{flalign}
\sigma^0_{\text{VS Sc.}} &= \frac{\mu^{2}}{4 \pi M_{\chi}^{2}}f_{p}^{2},&\\ 
\sigma^0_{\text{VF Vec.}} &= \frac{\mu^{2}}{4 \pi}(-f_{p}+\frac{\tilde{b}_{p}}{M_\Omega})^{2},&\\ 
\sigma^0_{\text{VF Ax.}} &= \frac{\mu^{2}}{4 \pi}(+f_{p}+\frac{\tilde{b}_{p}}{M_\Omega})^{2}, &\\
\sigma^0_{\text{VF Chi.}} &= \frac{\mu^{2}}{4 \pi}b_{p}^{2}, &\\
\sigma^0_{\text{VV Vec.}} &= \frac{\mu^{2}}{\pi}b_{p}^{2}. & 
\end{flalign}

\subsection{Photon Loop} 
\label{app:PhotonLoop} 

If the \textsc{Wimp} only couples to leptons, the \textsc{Wimp}--proton
interaction can only happen at the loop level. In that case, a low
energy photon that couples to a virtual lepton pair interacts with the
whole proton. This only happens for models with s--channel vector
bilinears $\bar{\psi} \gamma^\mu \psi$, i.e.\ models which
include either a, $b_p$, or a, $\tilde{b}_p$, term in the low energy
tree level cross section. Results can therefore be derived as follows,
\begin{align}
\sigma_0^\text{Loop} &= \frac{\alpha^2_{\text{em}}}{81 \pi^2} F^2(q^2)
\left. \sigma_0^\text{Tree} \right|_{\text{reduced}}, \\
\intertext{where the reduced cross section has to be understood as the tree level
  cross section given above after setting $b_p,\tilde{b}_p = 1$ and $f_p, d_p
  = 0$. This ensures
  that we only take the vector interaction parts. If the tree level cross section includes
  a $b_p$ term, the loop factor is given as,}
F(q^2) &\equiv \sum_lG_l\ f(q^2, m_l). \\
\intertext{For $\tilde{b}_p$ terms, it reads,}
F(q^2) &\equiv \sum_l \left(m_l + M_\chi\right) G_l\ f(q^2, m_l). \\
\intertext{In both cases, the loop function can be evaluated as,}
f(q^{2},m)&\equiv\frac{1}{q^{2}} \left[ 5 q^{2}+12 m^{2}-6(q^{2}+2
m^{2})\beta_q \text{arcoth}~ \beta_q \right. \nonumber \\
& \qquad \qquad \left. - 3 q^2 \ln m^2 / \Lambda^2 \right], \\
\beta_q &\equiv \sqrt{1-4 m^{2} / q^{2}}.
\end{align}
We follow the conservative assumption of a maximum scattering angle to find $q^2 = - 4
\mu^2 v^2$ with $\mu$ describing the reduced mass of the \textsc{Wimp} nucleus system
and $v = \unit{500}{\kilo\meter\per\second}$ being the typical
escape velocity of a \textsc{Wimp} in a dark matter halo. Because of the new
$q$--dependence of the cross section and the fact that the photon only couples
to the protons inside the nucleus, the official \textsc{Xenon} results have to
be rescaled according to,
\begin{align}
\sigma^\text{Loop} = \sigma^\text{Tree} \left[ \frac{F(\tilde{q}^2)}{F(q^2)} \cdot \frac{A}{Z}\right]^2,
\end{align}
where $\tilde{q} = q(M_N = M_P)$ uses the reduced mass $\mu$ of the \textsc{Wimp} proton
system instead. This weakens the cross section limits by about a factor
  of 10.

%% file: appendix_ilc.tex
\vspace{1.0cm}
\section{Differential Cross Section for $ \Ppositron \Pelectron \rightarrow \boldsymbol{
  \chi \chi} \Pphoton$}

\subsection{Abbreviations}
\label{app:xsectterms}
We use the following abbreviations for the final cross section list in
Table~\ref{tbl:2t3crosssections}:
\begin{align}
\intertext{Polarisation prefactors:}
C_S \equiv 1+P^+ P^-&, \quad C_V \equiv 1-P^+ P^-, \\
C_L \equiv (1- P^-) ( 1+P^+)&,\quad  C_R \equiv (1+ P^-) ( 1-P^+). \nonumber \\
\intertext{Terms with combined couplings:}
G_{X \pm Y} \equiv g_{X}^2 \pm g_{Y}^2&,\quad G_{XY} \equiv g_{X} g_{Y}. \\
\intertext{Relativistic velocities:}
\beta \equiv \sqrt{\displaystyle 1-\frac{4 M_\chi^2}{s}}&,\quad \hat{\beta} \equiv
\sqrt{\displaystyle 1-\frac{4 M_\chi^2}{s(1-x)}}\;.
\end{align}
Kinematical functions:
\begin{align}
F_{x \theta} &\equiv \frac{\alpha}{\pi} \frac{(x-1)^2+1}{x \sin^2 \theta}\;, \\
V_{x \theta}  &\equiv  \frac{x^2\cos(2 \theta) + (3x -8)x   + 8 }{4 \left((x-1)^2 +
    1\right)}\;.
\end{align}

We show terms that arise in the analytical evaluation of the differential photon cross
section in $\Ppositron \Pelectron \rightarrow \chi \chi \Pphoton$ but not in
the Weizs\"acker--Williams approximation in (\ref{eq:analyticfirst})-(\ref{eq:analyticlast}). They all vanish in the soft--photon
limit $x \rightarrow 0$.
\begin{widetext}
\begin{align}
A_{SF} &= \frac{ \left(1 - V_{x \theta} \right)}{4 M^2_\Omega}
\frac{\hat{s}}{1-x}  \left[(g_s+g_p)^4 C_R + (g_s-g_p)^4 C_L \right] \label{eq:analyticfirst}\\
A_{SFr} &= \frac{\alpha}{8 \pi}
\frac{\hat{s}}{M_\Omega^2} \frac{x}{1-x}  \left[(g_s+g_p)^4 C_R + (g_s-g_p)^4 C_L \right] \\
A_{FtS} &= \frac{(1-V_{x \theta})}{4} \left[ C_S (\hat{s} - 4 M_\chi^2) + \frac{1}{1 - x} C_S (2 M_\chi^2 + \hat{s}) \right] \\
A_{VF} & = 20 G^2_{lr} C_S (1-V_{x \theta}) \frac{x }{1-x} (\hat{s}^2 + 4
M_\chi^2 \hat{s}-8  M_\chi^4  )+  \frac{(g_l^4 C_L + g_t^4 C_R)}{M_\Omega^2 }
\Big[   -\frac{1}{32}\frac{x^4 \sin^2(2 \theta) \hat{s} ( 3 \hat{s}^2 + 26 M_\chi^2 \hat{s} - 32 M_\chi^4)}{(x-1)^2
    ((x-1)^2+1)} \nonumber \\
& \ + 6 \frac{x}{((x-1)^2+1)} \hat{s} ( \hat{s}^2  + 7 M_\chi^2 \hat{s} - 24
M_\chi^4 ) - \frac{1}{4} (1-V_{x \theta}) (21 \hat{s}^3 + 282 M_\chi^2
\hat{s}^2 - 1144 M_\chi^4 \hat{s}  + 160 M_\chi^6) \nonumber   \\
& \ +\frac{3}{2}\frac{(1-V_{x \theta})}{(1-x)}  \hat{s} (\hat{s}^2-28 M_\chi^2
\hat{s}+ 16 M_\chi^4) +\frac{1}{4} \frac{(1-V_{x \theta})}{(1-x)^2} \hat{s} (7
\hat{s}^2-126 M_\chi^2 \hat{s}+ 32 M_\chi^4) \left. +\frac{(1-V_{x
      \theta})}{(1-x)^3} \hat{s} (\hat{s}^2 +2 M_\chi^2  \hat{s}+6 M_\chi^4)
\right] \\ 
A_{VFr} & = \frac{(g_l^4 C_L + g_r^4 C_R)}{M_\Omega^2 } \Big[ - \frac{1}{32} \frac{x^4 \sin^2(2 \theta)}{(x-1)^2
    ((x-1)^2+1)} \hat{s}(  \hat{s}^2 + 32 M_\chi^2 \hat{s} -24 M_\chi^4    ) + 2 \frac{x}{((x-1)^2+1)} \hat{s} (  \hat{s}^2 + 12 M_\chi^2 \hat{s} +  56 M_\chi^4  )  \nonumber \\
&\ - \frac{1}{4}(1-V_{x \theta}) (7\hat{s}^3 + 144 M_\chi^2 \hat{s}^2 -168
M_\chi^4 \hat{s}+1280 M_\chi^6) +\frac{1}{2}\frac{(1-V_{x \theta})}{(1-x)}  \hat{s} (\hat{s}^2 -48 M_\chi^2 \hat{s}+ 56 M_\chi^4)\nonumber \\
&\ +\frac{1}{4}\frac{(1-V_{x \theta})}{(1-x)^2} \hat{s} (9 \hat{s}^2 -272
M_\chi^2 \hat{s}+104 M_\chi^4) \left. +2 \frac{(1-V_{x \theta})}{(1-x)^3}
  \hat{s} (\hat{s}^2+2 M_\chi^2 \hat{s}+ 6 M_\chi^4) \right]  \,.\label{eq:analyticlast}
\end{align}
\end{widetext}

%% file: DM.bbl
\providecommand{\href}[2]{#2}\begingroup\raggedright\begin{thebibliography}{10}

\bibitem{Bertone:2004pz}
G.~Bertone, D.~Hooper, and J.~Silk, ``{Particle dark matter: Evidence,
  candidates and constraints},''
  \href{http://dx.doi.org/10.1016/j.physrep.2004.08.031}{{\em Phys.Rept.} {\bf
  405} (2005)  279--390},
\href{http://arxiv.org/abs/hep-ph/0404175}{{\tt arXiv:hep-ph/0404175
  [hep-ph]}}.

\bibitem{Martin:1997ns}
S.~P. Martin, ``{A Supersymmetry primer},'' {\em arXiv:hep-ph} (1997)  9709356,
\href{http://arxiv.org/abs/hep-ph/9709356}{{\tt arXiv:hep-ph/9709356
  [hep-ph]}}.

\bibitem{Drees:2004jm}
M.~Drees, R.~Godbole, and P.~Roy,
``{Theory and phenomenology of sparticles: An account of four-dimensional N=1
  supersymmetry in high energy physics},''.

\bibitem{Appelquist:2000nn}
T.~Appelquist, H.-C. Cheng, and B.~A. Dobrescu, ``{Bounds on universal extra
  dimensions},'' \href{http://dx.doi.org/10.1103/PhysRevD.64.035002}{{\em
  Phys.Rev.} {\bf D64} (2001)  035002},
\href{http://arxiv.org/abs/hep-ph/0012100}{{\tt arXiv:hep-ph/0012100
  [hep-ph]}}.

\bibitem{ArkaniHamed:2001nc}
N.~Arkani-Hamed, A.~G. Cohen, and H.~Georgi, ``{Electroweak symmetry breaking
  from dimensional deconstruction},''
  \href{http://dx.doi.org/10.1016/S0370-2693(01)00741-9}{{\em Phys.Lett.} {\bf
  B513} (2001)  232--240},
\href{http://arxiv.org/abs/hep-ph/0105239}{{\tt arXiv:hep-ph/0105239
  [hep-ph]}}.

\bibitem{Goodman:1984dc}
M.~W. Goodman and E.~Witten, ``{Detectability of Certain Dark Matter
  Candidates},''
\href{http://dx.doi.org/10.1103/PhysRevD.31.3059}{{\em Phys.Rev.} {\bf D31}
  (1985)  3059}.

\bibitem{Bouquet:1989sr}
A.~Bouquet, P.~Salati, and J.~Silk, ``{$\gamma^-$ ray LINES AS A PROBE FOR A
  COLD DARK MATTER HALO},''
\href{http://dx.doi.org/10.1103/PhysRevD.40.3168}{{\em Phys.Rev.} {\bf D40}
  (1989)  3168}.

\bibitem{Birkedal:2004xn}
A.~Birkedal, K.~Matchev, and M.~Perelstein, ``{Dark matter at colliders: A
  Model independent approach},''
  \href{http://dx.doi.org/10.1103/PhysRevD.70.077701}{{\em Phys.Rev.} {\bf D70}
  (2004)  077701},
\href{http://arxiv.org/abs/hep-ph/0403004}{{\tt arXiv:hep-ph/0403004
  [hep-ph]}}.

\bibitem{Konar:2009ae}
P.~Konar, K.~Kong, K.~T. Matchev, and M.~Perelstein, ``{Shedding Light on the
  Dark Sector with Direct WIMP Production},''
  \href{http://dx.doi.org/10.1088/1367-2630/11/10/105004}{{\em New J.Phys.}
  {\bf 11} (2009)  105004},
\href{http://arxiv.org/abs/0902.2000}{{\tt arXiv:0902.2000 [hep-ph]}}.

\bibitem{Bartels:2007cv}
C.~Bartels and J.~List, ``{Model-independent WIMP searches at the ILC},'' {\em
  eConf} {\bf C0705302} (2007)  COS02,
\href{http://arxiv.org/abs/0709.2629}{{\tt arXiv:0709.2629 [hep-ex]}}.

\bibitem{Bartels:2009fa}
C.~Bartels and J.~List, ``{WIMP Searches at the ILC using a model-independent
  Approach},''
\href{http://arxiv.org/abs/0901.4890}{{\tt arXiv:0901.4890 [hep-ex]}}.

\bibitem{Bartels:2010qv}
C.~Bartels and J.~List, ``{Model independent WIMP Searches in full Simulation
  of the ILD Detector},''
\href{http://arxiv.org/abs/1007.2748}{{\tt arXiv:1007.2748 [hep-ex]}}.

\bibitem{Bartels:2012ex}
C.~Bartels, M.~Berggren, and J.~List, ``{Characterising WIMPs at a future
  $e^+e^-$ Linear Collider},''
\href{http://arxiv.org/abs/1206.6639}{{\tt arXiv:1206.6639 [hep-ex]}}.

\bibitem{Bernal:2008zk}
N.~Bernal, A.~Goudelis, Y.~Mambrini, and C.~Munoz, ``{Determining the WIMP mass
  using the complementarity between direct and indirect searches and the
  ILC},'' \href{http://dx.doi.org/10.1088/1475-7516/2009/01/046}{{\em JCAP}
  {\bf 0901} (2009)  046},
\href{http://arxiv.org/abs/0804.1976}{{\tt arXiv:0804.1976 [hep-ph]}}.

\bibitem{Dreiner:2006sb}
H.~K. Dreiner, O.~Kittel, and U.~Langenfeld, ``{Discovery potential of
  radiative neutralino production at the ILC},''
  \href{http://dx.doi.org/10.1103/PhysRevD.74.115010}{{\em Phys.Rev.} {\bf D74}
  (2006)  115010},
\href{http://arxiv.org/abs/hep-ph/0610020}{{\tt arXiv:hep-ph/0610020
  [hep-ph]}}.

\bibitem{Dreiner:2007vm}
H.~K. Dreiner, O.~Kittel, and U.~Langenfeld, ``{The Role of Beam polarization
  for Radiative Neutralino Production at the ILC},''
  \href{http://dx.doi.org/10.1140/epjc/s10052-007-0520-3}{{\em Eur.Phys.J.}
  {\bf C54} (2008)  277--284},
\href{http://arxiv.org/abs/hep-ph/0703009}{{\tt arXiv:hep-ph/0703009
  [HEP-PH]}}.

\bibitem{Cao:2009uw}
Q.-H. Cao, C.-R. Chen, C.~S. Li, and H.~Zhang, ``{Effective Dark Matter Model:
  Relic density, CDMS II, Fermi LAT and LHC},''
  \href{http://dx.doi.org/10.1007/JHEP08(2011)018}{{\em JHEP} {\bf 1108} (2011)
   018},
\href{http://arxiv.org/abs/0912.4511}{{\tt arXiv:0912.4511 [hep-ph]}}.

\bibitem{Bai:2010hh}
Y.~Bai, P.~J. Fox, and R.~Harnik, ``{The Tevatron at the Frontier of Dark
  Matter Direct Detection},''
  \href{http://dx.doi.org/10.1007/JHEP12(2010)048}{{\em JHEP} {\bf 1012} (2010)
   048},
\href{http://arxiv.org/abs/1005.3797}{{\tt arXiv:1005.3797 [hep-ph]}}.

\bibitem{Fox:2011pm}
P.~J. Fox, R.~Harnik, J.~Kopp, and Y.~Tsai, ``{Missing Energy Signatures of
  Dark Matter at the LHC},''
  \href{http://dx.doi.org/10.1103/PhysRevD.85.056011}{{\em Phys.Rev.} {\bf D85}
  (2012)  056011},
\href{http://arxiv.org/abs/1109.4398}{{\tt arXiv:1109.4398 [hep-ph]}}.

\bibitem{Goodman:2010ku}
J.~Goodman, M.~Ibe, A.~Rajaraman, W.~Shepherd, T.~M. Tait, {\em et al.},
  ``{Constraints on Dark Matter from Colliders},''
  \href{http://dx.doi.org/10.1103/PhysRevD.82.116010}{{\em Phys.Rev.} {\bf D82}
  (2010)  116010},
\href{http://arxiv.org/abs/1008.1783}{{\tt arXiv:1008.1783 [hep-ph]}}.

\bibitem{Goodman:2010yf}
J.~Goodman, M.~Ibe, A.~Rajaraman, W.~Shepherd, T.~M. Tait, {\em et al.},
  ``{Constraints on Light Majorana dark Matter from Colliders},''
  \href{http://dx.doi.org/10.1016/j.physletb.2010.11.009}{{\em Phys.Lett.} {\bf
  B695} (2011)  185--188},
\href{http://arxiv.org/abs/1005.1286}{{\tt arXiv:1005.1286 [hep-ph]}}.

\bibitem{Beltran:2010ww}
M.~Beltran, D.~Hooper, E.~W. Kolb, Z.~A. Krusberg, and T.~M. Tait, ``{Maverick
  dark matter at colliders},''
  \href{http://dx.doi.org/10.1007/JHEP09(2010)037}{{\em JHEP} {\bf 1009} (2010)
   037},
\href{http://arxiv.org/abs/1002.4137}{{\tt arXiv:1002.4137 [hep-ph]}}.

\bibitem{Rajaraman:2011wf}
A.~Rajaraman, W.~Shepherd, T.~M. Tait, and A.~M. Wijangco, ``{LHC Bounds on
  Interactions of Dark Matter},''
  \href{http://dx.doi.org/10.1103/PhysRevD.84.095013}{{\em Phys.Rev.} {\bf D84}
  (2011)  095013},
\href{http://arxiv.org/abs/1108.1196}{{\tt arXiv:1108.1196 [hep-ph]}}.

\bibitem{Bai:2012he}
Y.~Bai and T.~M. Tait, ``{Searches with Mono-Leptons},''
\href{http://arxiv.org/abs/1208.4361}{{\tt arXiv:1208.4361 [hep-ph]}}.

\bibitem{Cheung:2012gi}
K.~Cheung, P.-Y. Tseng, Y.-L.~S. Tsai, and T.-C. Yuan, ``{Global Constraints on
  Effective Dark Matter Interactions: Relic Density, Direct Detection, Indirect
  Detection, and Collider},''
  \href{http://dx.doi.org/10.1088/1475-7516/2012/05/001}{{\em JCAP} {\bf 1205}
  (2012)  001},
\href{http://arxiv.org/abs/1201.3402}{{\tt arXiv:1201.3402 [hep-ph]}}.

\bibitem{Chatrchyan:2012pa}
{\bf CMS Collaboration} Collaboration, S.~Chatrchyan {\em et al.}, ``{Search
  for dark matter and large extra dimensions in monojet events in pp collisions
  at sqrt(s)= 7 TeV},''
\href{http://arxiv.org/abs/1206.5663}{{\tt arXiv:1206.5663 [hep-ex]}}.

\bibitem{ATLAS-CONF-2012-084}
``Search for dark matter candidates and large extra dimensions in events with a
  jet and missing transverse momentum with the atlas detector,'' Tech. Rep.
  ATLAS-CONF-2012-084, CERN, Geneva, Jul, 2012.

\bibitem{Fox:2011fx}
P.~J. Fox, R.~Harnik, J.~Kopp, and Y.~Tsai, ``{LEP Shines Light on Dark
  Matter},'' \href{http://dx.doi.org/10.1103/PhysRevD.84.014028}{{\em
  Phys.Rev.} {\bf D84} (2011)  014028},
\href{http://arxiv.org/abs/1103.0240}{{\tt arXiv:1103.0240 [hep-ph]}}.

\bibitem{Kurylov:2003ra}
A.~Kurylov and M.~Kamionkowski, ``{Generalized analysis of weakly interacting
  massive particle searches},''
  \href{http://dx.doi.org/10.1103/PhysRevD.69.063503}{{\em Phys.Rev.} {\bf D69}
  (2004)  063503},
\href{http://arxiv.org/abs/hep-ph/0307185}{{\tt arXiv:hep-ph/0307185
  [hep-ph]}}.

\bibitem{Beltran:2008xg}
M.~Beltran, D.~Hooper, E.~W. Kolb, and Z.~C. Krusberg, ``{Deducing the nature
  of dark matter from direct and indirect detection experiments in the absence
  of collider signatures of new physics},''
  \href{http://dx.doi.org/10.1103/PhysRevD.80.043509}{{\em Phys.Rev.} {\bf D80}
  (2009)  043509},
\href{http://arxiv.org/abs/0808.3384}{{\tt arXiv:0808.3384 [hep-ph]}}.

\bibitem{Agrawal:2010fh}
P.~Agrawal, Z.~Chacko, C.~Kilic, and R.~K. Mishra, ``{A Classification of Dark
  Matter Candidates with Primarily Spin-Dependent Interactions with Matter},''
\href{http://arxiv.org/abs/1003.1912}{{\tt arXiv:1003.1912 [hep-ph]}}.

\bibitem{Conley:2010jk}
J.~Conley, H.~Dreiner, and P.~Wienemann, ``{Measuring a Light Neutralino Mass
  at the ILC: Testing the MSSM Neutralino Cold Dark Matter Model},''
  \href{http://dx.doi.org/10.1103/PhysRevD.83.055018}{{\em Phys.Rev.} {\bf D83}
  (2011)  055018},
\href{http://arxiv.org/abs/1012.1035}{{\tt arXiv:1012.1035 [hep-ph]}}.

\bibitem{Zheng:2010js}
J.-M. Zheng, Z.-H. Yu, J.-W. Shao, X.-J. Bi, Z.~Li, {\em et al.},
  ``{Constraining the interaction strength between dark matter and visible
  matter: I. fermionic dark matter},''
  \href{http://dx.doi.org/10.1016/j.nuclphysb.2011.09.009}{{\em Nucl.Phys.}
  {\bf B854} (2012)  350--374},
\href{http://arxiv.org/abs/1012.2022}{{\tt arXiv:1012.2022 [hep-ph]}}.

\bibitem{Yu:2011by}
Z.-H. Yu, J.-M. Zheng, X.-J. Bi, Z.~Li, D.-X. Yao, {\em et al.},
  ``{Constraining the interaction strength between dark matter and visible
  matter: II. scalar, vector and spin-3/2 dark matter},''
  \href{http://dx.doi.org/10.1016/j.nuclphysb.2012.02.016}{{\em Nucl.Phys.}
  {\bf B860} (2012)  115--151},
\href{http://arxiv.org/abs/1112.6052}{{\tt arXiv:1112.6052 [hep-ph]}}.

\bibitem{Haba:2011vi}
N.~Haba, K.~Kaneta, S.~Matsumoto, and T.~Nabeshima, ``{A Simple Method of
  Calculating Effective Operators},''
  \href{http://dx.doi.org/10.5506/APhysPolB.43.405}{{\em Acta Phys.Polon.} {\bf
  B43} (2012)  405--444},
\href{http://arxiv.org/abs/1106.6106}{{\tt arXiv:1106.6106 [hep-ph]}}.

\bibitem{Denner:1992vza}
A.~Denner, H.~Eck, O.~Hahn, and J.~Kublbeck, ``{Feynman rules for fermion
  number violating interactions},''
\href{http://dx.doi.org/10.1016/0550-3213(92)90169-C}{{\em Nucl.Phys.} {\bf
  B387} (1992)  467--484}.

\bibitem{Komatsu:2010fb}
{\bf WMAP Collaboration} Collaboration, E.~Komatsu {\em et al.}, ``{Seven-Year
  Wilkinson Microwave Anisotropy Probe (WMAP) Observations: Cosmological
  Interpretation},'' \href{http://dx.doi.org/10.1088/0067-0049/192/2/18}{{\em
  Astrophys.J.Suppl.} {\bf 192} (2011)  18},
\href{http://arxiv.org/abs/1001.4538}{{\tt arXiv:1001.4538 [astro-ph.CO]}}.

\bibitem{Aprile:2012nq}
{\bf XENON100 Collaboration} Collaboration, E.~Aprile {\em et al.}, ``{Dark
  Matter Results from 225 Live Days of XENON100 Data},''
\href{http://arxiv.org/abs/1207.5988}{{\tt arXiv:1207.5988 [astro-ph.CO]}}.

\bibitem{Adriani:2008zr}
{\bf PAMELA Collaboration} Collaboration, O.~Adriani {\em et al.}, ``{An
  anomalous positron abundance in cosmic rays with energies 1.5-100 GeV},''
  \href{http://dx.doi.org/10.1038/nature07942}{{\em Nature} {\bf 458} (2009)
  607--609},
\href{http://arxiv.org/abs/0810.4995}{{\tt arXiv:0810.4995 [astro-ph]}}.

\bibitem{Coleman:2003hs}
T.~S. Coleman and M.~Roos, ``{Effective degrees of freedom during the radiation
  era},'' \href{http://dx.doi.org/10.1103/PhysRevD.68.027702}{{\em Phys.Rev.}
  {\bf D68} (2003)  027702},
\href{http://arxiv.org/abs/astro-ph/0304281}{{\tt arXiv:astro-ph/0304281
  [astro-ph]}}.

\bibitem{Angle:2008we}
J.~Angle, E.~Aprile, F.~Arneodo, L.~Baudis, A.~Bernstein, {\em et al.},
  ``{Limits on spin-dependent WIMP-nucleon cross-sections from the XENON10
  experiment},'' \href{http://dx.doi.org/10.1103/PhysRevLett.101.091301}{{\em
  Phys.Rev.Lett.} {\bf 101} (2008)  091301},
\href{http://arxiv.org/abs/0805.2939}{{\tt arXiv:0805.2939 [astro-ph]}}.

\bibitem{Baltz:1998xv}
E.~A. Baltz and J.~Edsjo, ``{Positron propagation and fluxes from neutralino
  annihilation in the halo},''
  \href{http://dx.doi.org/10.1103/PhysRevD.59.023511}{{\em Phys.Rev.} {\bf D59}
  (1998)  023511},
\href{http://arxiv.org/abs/astro-ph/9808243}{{\tt arXiv:astro-ph/9808243
  [astro-ph]}}.

\bibitem{Delahaye:2008ua}
T.~Delahaye, F.~Donato, N.~Fornengo, J.~Lavalle, R.~Lineros, {\em et al.},
  ``{Galactic secondary positron flux at the Earth},''
  \href{http://dx.doi.org/10.1051/0004-6361/200811130}{{\em Astron.Astrophys.}
  {\bf 501} (2009)  821--833},
\href{http://arxiv.org/abs/0809.5268}{{\tt arXiv:0809.5268 [astro-ph]}}.

\bibitem{Delahaye:2007fr}
T.~Delahaye, R.~Lineros, F.~Donato, N.~Fornengo, and P.~Salati, ``{Positrons
  from dark matter annihilation in the galactic halo: Theoretical
  uncertainties},'' \href{http://dx.doi.org/10.1103/PhysRevD.77.063527}{{\em
  Phys.Rev.} {\bf D77} (2008)  063527},
\href{http://arxiv.org/abs/0712.2312}{{\tt arXiv:0712.2312 [astro-ph]}}.

\bibitem{Perelstein:2010fq}
M.~Perelstein and B.~Shakya, ``{Remarks on calculation of positron flux from
  galactic dark matter},''
  \href{http://dx.doi.org/10.1103/PhysRevD.82.043505}{{\em Phys.Rev.} {\bf D82}
  (2010)  043505},
\href{http://arxiv.org/abs/1002.4588}{{\tt arXiv:1002.4588 [astro-ph.HE]}}.

\bibitem{Bahcall:1980fb}
J.~N. Bahcall and R.~Soneira, ``{The Universe at faint magnetidues. 2. Models
  for the predicted star counts},''
\href{http://dx.doi.org/10.1086/190685}{{\em Astrophys.J.Suppl.} {\bf 44}
  (1980)  73--110}.

\bibitem{Sjostrand:2007gs}
T.~Sjostrand, S.~Mrenna, and P.~Z. Skands, ``{A Brief Introduction to PYTHIA
  8.1},'' \href{http://dx.doi.org/10.1016/j.cpc.2008.01.036}{{\em
  Comput.Phys.Commun.} {\bf 178} (2008)  852--867},
\href{http://arxiv.org/abs/0710.3820}{{\tt arXiv:0710.3820 [hep-ph]}}.

\bibitem{Atwood:2009ez}
{\bf LAT Collaboration} Collaboration, W.~Atwood {\em et al.}, ``{The Large
  Area Telescope on the Fermi Gamma-ray Space Telescope Mission},''
  \href{http://dx.doi.org/10.1088/0004-637X/697/2/1071}{{\em Astrophys.J.} {\bf
  697} (2009)  1071--1102},
\href{http://arxiv.org/abs/0902.1089}{{\tt arXiv:0902.1089 [astro-ph.IM]}}.

\bibitem{Cirelli:2009vg}
M.~Cirelli and P.~Panci, ``{Inverse Compton constraints on the Dark Matter e+e-
  excesses},'' \href{http://dx.doi.org/10.1016/j.nuclphysb.2009.06.034}{{\em
  Nucl.Phys.} {\bf B821} (2009)  399--416},
\href{http://arxiv.org/abs/0904.3830}{{\tt arXiv:0904.3830 [astro-ph.CO]}}.

\bibitem{Bernal:2010ip}
N.~Bernal and S.~Palomares-Ruiz, ``{Constraining Dark Matter Properties with
  Gamma-Rays from the Galactic Center with Fermi-LAT},''
  \href{http://dx.doi.org/10.1016/j.nuclphysb.2011.12.016}{{\em Nucl.Phys.}
  {\bf B857} (2012)  380--410},
\href{http://arxiv.org/abs/1006.0477}{{\tt arXiv:1006.0477 [astro-ph.HE]}}.

\bibitem{Abbasi:2012ws}
{\bf IceCube collaboration} Collaboration, R.~Abbasi {\em et al.}, ``{Search
  for Neutrinos from Annihilating Dark Matter in the Direction of the Galactic
  Center with the 40-String IceCube Neutrino Observatory},''
\href{http://arxiv.org/abs/1210.3557}{{\tt arXiv:1210.3557 [hep-ex]}}.

\bibitem{IceCube:2011aj}
{\bf IceCube Collaboration} Collaboration, R.~Abbasi {\em et al.},
  ``{Multi-year search for dark matter annihilations in the Sun with the
  AMANDA-II and IceCube detectors},''
  \href{http://dx.doi.org/10.1103/PhysRevD.85.042002}{{\em Phys.Rev.} {\bf D85}
  (2012)  042002},
\href{http://arxiv.org/abs/1112.1840}{{\tt arXiv:1112.1840 [astro-ph.HE]}}.

\bibitem{BartelsThesis}
C.~Bartels, ``{WIMP Search and a Cherenkov Detector Prototype for ILC
  Polarimetry},''
  \href{http://arxiv.org/abs/https://pubdb.desy.de/fulltext/getfulltext.php?\\%
lid=17403$\&$fid=46030}{{\tt
  https://pubdb.desy.de/fulltext/getfulltext.php?\\lid=17403$\&$fid=46030}}.

\bibitem{Phinney:2007gp}
N.~Phinney, N.~Toge, and N.~Walker, ``{LC Reference Design Report Volume 3 -
  Accelerator},''
\href{http://arxiv.org/abs/0712.2361}{{\tt arXiv:0712.2361 [physics.acc-ph]}}.

\bibitem{Pukhov:1999gg}
A.~Pukhov, E.~Boos, M.~Dubinin, V.~Edneral, V.~Ilyin, {\em et al.}, ``{CompHEP:
  A Package for evaluation of Feynman diagrams and integration over
  multiparticle phase space},''
\href{http://arxiv.org/abs/hep-ph/9908288}{{\tt arXiv:hep-ph/9908288
  [hep-ph]}}.

\bibitem{Abe:2010aa}
{\bf ILD Concept Group - Linear Collider Collaboration} Collaboration, T.~Abe
  {\em et al.}, ``{The International Large Detector: Letter of Intent},''
\href{http://arxiv.org/abs/1006.3396}{{\tt arXiv:1006.3396 [hep-ex]}}.

\bibitem{Rolke:2004mj}
W.~A. Rolke, A.~M. Lopez, and J.~Conrad, ``{Limits and confidence intervals in
  the presence of nuisance parameters},''
  \href{http://dx.doi.org/10.1016/j.nima.2005.05.068}{{\em Nucl.Instrum.Meth.}
  {\bf A551} (2005)  493--503},
\href{http://arxiv.org/abs/physics/0403059}{{\tt arXiv:physics/0403059
  [physics]}}.

\bibitem{Helebrant:2008qz}
C.~Helebrant, D.~Kafer, and J.~List, ``{Precision Polarimetry at the
  International Linear Collider},''
\href{http://arxiv.org/abs/0809.4485}{{\tt arXiv:0809.4485 [physics.ins-det]}}.

\bibitem{Cornwall:1974km}
J.~M. Cornwall, D.~N. Levin, and G.~Tiktopoulos, ``{Derivation of Gauge
  Invariance from High-Energy Unitarity Bounds on the s Matrix},''
\href{http://dx.doi.org/10.1103/PhysRevD.10.1145, 10.1103/PhysRevD.11.972}{{\em
  Phys.Rev.} {\bf D10} (1974)  1145}.

\bibitem{fnumbers}
J.~R. Ellis, K.~A. Olive, and C.~Savage, ``{Hadronic Uncertainties in the
  Elastic Scattering of Supersymmetric Dark Matter},''
  \href{http://dx.doi.org/10.1103/PhysRevD.77.065026}{{\em Phys.Rev.} {\bf D77}
  (2008)  065026},
\href{http://arxiv.org/abs/0801.3656}{{\tt arXiv:0801.3656 [hep-ph]}}.

\bibitem{deltanumbers}
G.~Belanger, F.~Boudjema, A.~Pukhov, and A.~Semenov, ``{Dark matter direct
  detection rate in a generic model with micrOMEGAs 2.2},''
  \href{http://dx.doi.org/10.1016/j.cpc.2008.11.019}{{\em Comput.Phys.Commun.}
  {\bf 180} (2009)  747--767},
\href{http://arxiv.org/abs/0803.2360}{{\tt arXiv:0803.2360 [hep-ph]}}.

\end{thebibliography}\endgroup
